# Water-enhanced interdiffusion of major elements between natural shoshonite and high-K rhyolite melts

Diego González-García[1], Harald Behrens[2], Maurizio Petrelli[1], Francesco Vetere[1], Daniele Morgavi[1], Chao Zhang[2], Diego Perugini[1]

[1] *Dipartimento di Fisica e Geologia, Università degli Studi di Perugia, Piazza Universitá, 1, 06123 Perugia, Italy.*

[2] *Institut für Mineralogie, Leibniz Universität Hannover. Callinstrasse, 3, 30167 Hannover, Germany.*

Corresponding author: **Diego González-García**

e-mail: diego.gonzalez@studenti.unipg.it

Tel.: +39 075 585 2652

Fax: +39 075 585 2603

**Abstract**

The interdiffusion of six major elements (Si, Ti, Fe, Mg, Ca, K) between natural shoshonite and a high-K calc-alkaline rhyolite (Vulcano island, Aeolian archipelago, Italy) has been experimentally measured by the diffusion couple technique at 1200°C, pressures from 50 to 500 MPa and water contents from 0.3 ('nominally dry') to 2 wt%. The experiments were carried out in an internally heated pressure vessel, and major element profiles were later acquired by electron probe microanalysis. The concentration-distance profiles are evaluated using a concentration-dependent diffusivity approach. Effective binary diffusion coefficients for four intermediate silica contents are obtained by the Sauer-Freise modified Boltzmann-Matano method. At the experimental temperature and pressures, the diffusivity of all studied elements notably increases with dissolved $H_2O$ content. Particularly, diffusion is up to 1.4 orders of magnitude faster in a melt containing 2 wt.% $H_2O$ than in nominally dry melts. This effect is slightly enhanced in the more mafic compositions. Uphill diffusion was observed for Al, while all other elements can be described by the concept of effective binary interdiffusion. Ti is the slowest diffusing element through all experimental conditions and compositions, followed by Si. Fe, Mg, Ca and K diffuse at similar rates but always more rapidly than Si and Ti. This trend suggests a strong coupling between melt components. Since effects of composition (including water content) are dominant, a pressure effect on diffusion cannot be clearly resolved in the experimental pressure range.

## 1. Introduction

Diffusive exchange of chemical elements plays a key role in many magmatic processes. As an example, it governs crystal growth and dissolution, bubble nucleation and growth and the compositional modulation during magma mixing processes. In particular, magma mixing is widely considered as one of the basic processes responsible for the generation of the wide compositional variety of igneous rocks on the Earth, within both plutonic and volcanic environments (Anderson, 1976). In addition, it has been recognized as a major trigger mechanism for some highly explosive volcanic eruptions (Sparks et al., 1977; Leonard et al., 2002). The time and space evolution of magma mixing between two chemically distinct magmas is controlled by both physical and chemical processes where diffusion plays a significant role (Perugini et al., 2006; Perugini et al, 2008). Hence, as a time-dependent process, diffusion has the potential to be used as a chronometer of such magma mixing processes (e.g. Perugini et al., 2015). However, in order to fully achieve this goal, a detailed understanding of the diffusion process in natural melts is mandatory.

Although a large number of diffusion data in silicate melts has been published in recent years (for a review of published data, see Zhang et al., 2010), the dataset is not systematic and data for natural systems are scarce. Furthermore, specific literature on the effect of water in diffusivity of major and trace elements is limited (Baker, 1991; Baker & Bossányi, 1994; Mungall et al., 1998; Behrens & Hahn, 2009).

The aim of the present work is to systematically fill this gap by gathering a consistent dataset of diffusion data and relative diffusion coefficients that are applicable to geologically relevant scenarios. We use melts of rhyolitic and shoshonitic composition sampled at Vulcano Island (Aeolian archipelago, Italy) to probe timescales and compositional signatures of diffusion. In detail, we aim at determining the effective binary diffusion coefficients (EBDC hereafter) of six major elements (Si, Ti, Fe, Mg, Ca, K) between the two natural end-members at 1200 °C and at pressures of 50, 100, 300 and 500 MPa, relevant to relatively shallow magmatic systems. The influence of water in the enhancement of the diffusion processes is also considered in detail.

A problem in the use of EBDCs is that they are not generally transferable to other compositions and other conditions. Additionally, non-Fickian effects such as uphill diffusion cannot be described. Complete description of multi-component diffusion in silicate melts requires the determination of the diffusion matrix, which considers the interaction of all melt components (Liang, 2010). In practice, such diffusion matrix can only be determined for melts containing a few components, but not in complex natural systems where the number of components is far larger. In such cases the concept of effective binary diffusion coefficients, when being critically applied, is a suitable approach to get insights into the fundamental mechanisms driving the diffusion process.

## 2. Materials and methods

*2.1 Starting materials and glass synthesis*

Rocks out-cropping on the island of Vulcano range in composition from mafic to silicic with high-K calc-alkaline to shoshonitic and potassic alkaline affinities (Peccerillo, 2005; Keller, 1980; Gioncada et al., 2003). Two natural end-member compositions, sampled during a field campaign (Oct. 2014), have been selected for the present study (**Fig. 1**). The least evolved end-member is a shoshonite sampled at the Vulcanello lava platform (Vetere et al., 2007; Davì et al., 2009). The most evolved is a high-K rhyolitic obsidian from the Pietre Cotte lava flow, belonging to the La Fossa cone (Vetere et al., 2015; De Astis et al., 1997; Clocchiatti et al., 1994; Piochi et al., 2009). Both end-members are products of historic activity within the island. The Vulcanello platform was formed during a long-lasting eruptive period between A.D. 1000 and A.D. 1250 (Arrighi et al., 2006; Vetere et al., 2007). The Pietre Cotte lava flow was emplaced during the 1736 eruption of the La Fossa cone (Keller, 1980; Frazzetta et al., 1983); evidence of magma mixing has been recognized in both systems (Clocchiatti et al., 1994; Aparicio and Frazzetta, 2006; Vetere et al., 2015). Water content in the magmas from Vulcano was found to be lower than 2 wt.%, and usually between 1.0 and 1.5 wt.% (Clocchatti et al., 1994).

Alteration-free rock samples have been selected and, after a deep cleaning with distilled water, they were crushed in an agate mortar to obtain a fine-grained powder. End-member powders were subsequently homogenized through two cycles of melting at 1600°C for 4 hours in a high temperature furnace (Nabertherm® HT 04/17) at ambient pressure using a platinum crucible, followed by crushing and fine powdering in the agate mortar. During this process, the material loses most volatiles that were originally present. Alkali loss was also checked by comparing the re-melted material with literature data on the same rocks (Vetere et al., 2007; Davì et al., 2009) and results indicated that it can be considered negligible. Measured compositions of end-member glasses are reported in **Table 1** and **Fig. 1**. In agreement with Peccerillo and Taylor (1976), the studied compositions define a shoshonitic trend.

Glass with defined water contents were synthesized using $Au_{80}Pd_{20}$ alloy tubes with 5 mm diameter and 4 to 5 cm length. The tubes were welded at one end and filled step by step with rock powder and compacting by a metal rod after each filling step; when necessary, distilled water was added. Welding was carried out with a

Lampert PUK04 pulse arc welder in Ar atmosphere. In the wet experiments, added water was introduced by using a micrometric syringe and checked with a balance with precision of 0.1 μg. Added water contents were 1 wt.% and 2 wt.%, chosen to possibly match that of magmas erupted at Vulcano. Moreover, a water-free ('nominally dry', hereafter ND) capsule for each end-member was prepared. Capsules were filled and weighed before and after the welding procedure. Finally, the capsule was tested for leakage by annealing at 100°C for 12 to 24h.

Glass syntheses were performed in an internally heated pressure vessel (IHPV) (Holloway, 1971; Berndt et al., 2002) at 1200°C and 300 MPa for 24h and isobarically quenched to room temperature by switching off the heating power of the furnace. During experimental runs, the temperature variation within the capsule was less than 10 °C. Initial quench rates were in the order of 200ºC per minute. Subsequent lower quenching rates (100ºC/min at the glass transition range) avoided formation of internal stresses within the glasses during cooling. Glass chips from the end tips of the hydrous glass cylinders were analyzed for water content by pyrolysis and subsequent Karl-Fischer titration (Behrens, 1995, Behrens et al., 1996) to provide a quality control for synthesis procedure. Finally, synthesized glass cylinders were cut with a diamond saw to obtain pieces with length of 3 to 4 mm, and one end of each cylinder was polished.

In addition, the $Fe^{2+}/Fe^{3+}$ ratio was measured by a wet-chemical method in synthesized glasses at high-pressure in the IHPV as described by Schuessler et al. (2008). An advantage of this method is that ferrous iron content and total iron content are determined on the same base solution so that the $Fe^{2+}/Fe^{3+}$ ratio can be determined with high precision using a Shimadzu UV 1800 spectrometer. From each quenched sample, two pieces of glass from the top and bottom of the capsule were analyzed.

Table 2 shows the measured iron oxidation state in both dry and hydrous starting glasses. A relationship between water content of the melt and the oxidation state of iron is evident: higher water contents result in lower the $Fe^{2+}$ due to the dependence of oxygen fugacity on water fugacity (Vetere et al., 2014).

*2.2 Diffusion couple experiments*

Diffusion experiments were performed using the diffusion couple technique (Baker, 1989; Baker, 1990; Nowak and Behrens, 1997). Couples of rhyolite and shoshonite glass cylinders with the same nominal water content (dry, 1 wt.% and 2 wt.%) were placed in contact through their polished end inside a 5 mm diameter $Au_{80}Pd_{20}$ capsule, which was sealed as previously described. The higher density shoshonite glass was located in the bottom side of the capsule in order to avoid gravitationally induced mixing. Tightness of the sealed capsule was checked by weighing it before and after compression in a Cold Seal Pressure Vessel (CSPV) and observing the adaptation of the capsule to the sample shape.

Experiments were run in an IHPV using heating ramps of 30/50/20 °C/min from ambient temperature to 100, 1150 and 1200°C, respectively. Temperature was kept constant at 1200 ± 5 °C for all experiments, and run pressures were 50 MPa, 100 MPa and 300 MPa. Experimental times were 1 h and 4 h in the 300 MPa experiments, and 4 h in the 50 and 100 MPa experiments. An additional experiment was run at 500 MPa, 2 wt.% $H_2O$ with a duration of 4 h. After each run was completed, a rapid quench was applied by letting the capsule fall from the hot region of the vessel to the cold part (Berndt et al., 2002). A zero-time experiment (i.e. an experiment rapidly quenched soon after reaching 1200°C) was performed at 300 MPa to quantify the effect of heating/cooling on diffusion profiles and to demonstrate that quench phases, which might form after synthesis in the shoshonitic glass, were absent when reaching the target temperature. **Table 3** summarizes experimental conditions.

The IHPV vessel used for syntheses and diffusion couple experiments was pressurized with argon, and experiments were performed at the intrinsic hydrogen fugacity of the vessel. The oxygen fugacity within the capsules was controlled by the fugacity of $H_2$ and $H_2O$ through the equilibrium reaction $H_2 + ½\ O_2 = H_2O$. Under intrinsic conditions in the IHPV, the oxygen fugacity in capsules containing $H_2O$-saturated melts (pure $H_2O$ fluid) was found to be close to the $MnO–Mn_3O_4$ buffer (i.e. 3.7 log units higher than the Ni/NiO buffer; Berndt et al., 2002).

## 2.3 Analytical procedure

### 2.3.1 Sample preparation for analyses

Analyses were carried out on doubly polished glass sections (with thickness of ca. 200 μm) from each of the experimental products. Fourier transform infrared spectroscopy (FTIR) was used for the determination of $H_2O$ contents, and an electron probe micro-analyzer (EPMA) for determination of major element concentrations ($SiO_2$, $TiO_2$, $Al_2O_3$, $FeO_{tot}$, $MgO$, $MnO$, $CaO$, $Na_2O$, $K_2O$ and $P_2O_5$; $FeO_{tot}$ refers to total iron content).

### 2.3.2 Fourier transform infrared spectroscopy

Each experiment was analyzed for $H_2O$ abundances by Fourier Transform Infrared Spectroscopy (FTIR). Absorption spectra were collected in the near-infrared (NIR) using an FTIR spectrometer Bruker IFS88 coupled with an IR-ScopeII microscope (operation conditions: MCT narrow range detector for NIR; tungsten lamp and $CaF_2$ beam splitter). Spectral resolution was 4 cm$^{-1}$, and 100 scans were accumulated to obtain the final spectrum in each measured point. Beam size was a rectangle with dimensions of 50 x 100 μm in the focus plane. Measurements were carried out along a transect centered at the interface of each diffusion couple, with 100 μm spacing, and extending 1 mm into each component, to match the EMPA profiles (details reported below).

The absorption peaks of molecular water ($H_2O_m$) and hydroxyl ($OH^-$) species at 5200 and 4500 cm$^{-1}$, respectively, were used for quantification of hydrous species. Their amplitude was measured with the Bruker OPUS software after subtraction of linear baselines. Water species abundances were obtained by application of the modified Beer-Lambert law as given by Stolper (1982):

$$C_i = 100 \cdot \frac{M_i \cdot A_j}{d \cdot \rho \cdot \varepsilon_j} \tag{1}$$

where $C_i$ is the concentration in wt.% of water dissolved as molecules or as OH groups, $A_j$ is the absorptivity at the corresponding wavenumber, denoted as $j$. $M_i$ is the molecular weight of the species (here 18.02 g/mol for $H_2O$); $\rho$ is the calculated density of the glass; $d$ is the thickness of the sample and $\varepsilon_i$ is the linear molar absorption coefficient for the band of interest. The baseline used was a tangent line to both sides of the absorption peak (Ohlhorst et al., 2001). Shoshonite and rhyolite glass densities in g/L were calculated using the equations proposed by Vetere et al. (2011) (1) and Ohlhorst et al. (2001) (2), respectively:

$$\rho_{sho} = (2658 - 19.7 \cdot w) \cdot (1 + 0.00002 \cdot P) \tag{2}$$

$$\rho_{rhy} = 2390 - 17 \cdot w \tag{3}$$

where *w* is the total water content in wt.% and *P* is the pressure in MPa. Absorption coefficients for the end-member composition were taken from Withers and Behrens (1999) for rhyolite ($\varepsilon_{4500}$ = (1.41 ± 0.07) L mol$^{-1}$ cm$^{-1}$; $\varepsilon_{5200}$ = (1.66 ± 0.05) L mol$^{-1}$ cm$^{-1}$) and from Vetere et al. (2011) for shoshonite ($\varepsilon_{4500}$ = (0.80 ± 0.06) L mol$^{-1}$ cm$^{-1}$; $\varepsilon_{5200}$ = (1.03 ± 0.03) L mol$^{-1}$ cm$^{-1}$). Densities and absorption coefficients for intermediate compositions were interpolated using the end-member data. The total water content is the sum of water dissolved as OH ($C_{OH}$, band at 4500 cm$^{-1}$) and water dissolved in molecular form $C_{H2O}$, band at 5200 cm$^{-1}$).

*2.3.3 Electron microprobe analysis*

Major element concentration profiles were obtained with a Cameca SX-100 electron microprobe at the Institute of Mineralogy of the Leibniz University of Hannover. Operating conditions were an accelerating potential of 15 kV, a beam current of 4 nA and a defocused beam diameter of 10 μm, to minimize alkali loss in the glass analysis. Profiles extended 1 mm to each side of the interface, meaning a total length of 2 mm. Spacing of analyses in the central region of the profiles was kept as small as possible (typically 12-13 μm), and was widened to 30 μm in the end plateau regions. Precision and accuracy were determined by measuring VG-568 (rhyolite) and VG-2 (basalt) reference glasses. Obtained analytical errors vary from 4% to 10% for all measured elements. In addition, whenever possible, a second, offset profile was acquired in the same experiment as double check for reproducibility and convection effects.

**3. Results and discussion**

*3.1 Experimental products*

The experimental products, including the nominally dry experiments, were always crystal-free glasses, demonstrating the effectiveness of the rapid quench device. Glass cylinders experienced only minor changes in shape during the runs, and the interface between the two melts showed a mild concave form towards the rhyolite glass as a result of the different surface tension of both melts and the capsule wall (**Fig. 2**). The 300 MPa runs revealed bubble-free glasses, while the 100 MPa and 50 MPa experiments contained bubbles near the interface position and in the rhyolitic melt, in particular for higher water content (runs P100-H2-4 and P050-H2-4; **Table 3**). A quantification of the bubble size and area has been carried out by image analysis by using the software ImageJ (Schneider et al., 2012) on EPMA BSE images. Results show that bubble diameters are very small (up to 3.5 and 1.2 micron in the rhyolitic and shoshonitic glasses, respectively; and up to 8 microns at the interface). The area occupied by the bubbles in the section surface is less than 2% in the rhyolite and, when present, less than 0.6% in the shoshonite. In the interface position the bubble area is between 6% and 8% (**Fig. S1** in the supplementary materials).

A possible explanation for the bubbles in the experimental products is that they originated from air trapped inside the capsule, either in the glass synthesis step (bubbles inside end-member glasses) or in the diffusion couple run (bubbles found in the interface position, trapped between the two glasses prior to the experiment); in this case, they would contain dominantly nitrogen, given its low solubility in the melt at low pressures. In both melts (and especially in the rhyolite), bubbles of this size would not be able to ascend, and hence they form and accumulate at the interface between the two melts. However, given the small size and fraction of these bubbles and their low mobility, we consider their effect negligible.

All run products were measured for water content using the FTIR technique. Obtained water contents are close to the nominal water contents for most of the samples except for the nominally dry ones. These samples contain water although no water was added before syntheses. This water was formed during synthesis by permeation of hydrogen through the capsule wall and subsequent reaction with ferric iron ($Fe_2O_3 + H_2 = 2\ FeO + H_2O$).

Each of the starting glasses of the couple was prepared individually and thus has slightly different water content. As a consequence, a diffusive flux of water is produced in the initial stages of the runs. However, since water diffusion is much faster than diffusion of cations in silicate melts (Zhang et al., 2010), almost constant water contents along the diffusion profiles were measured (**Fig. 3**). Variations are of the order of 15% except for the zero-time experiment, which shows a diffusion-like $H_2O$ profile between both sides of the couples. This indicates that time was not sufficient for water to equilibrate. For this experiment, two water contents are given, corresponding to each of the two halves of the couple. In addition, experiment P300-H0-4 (300 MPa, ND, 4h,) shows an uneven water concentration in the shoshonitic half of the couple (**Fig. 3 and supplementary Fig. S2**). Major element profiles from this run show irregularities and, therefore, it was considered unsuccessful and not used in the modeling and discussion.

*3.2 Concentration-distance profiles*

Selected concentration profiles are shown in **Fig. 4** (for the complete dataset, the reader is directed to **Table S1** and **Fig. S2** in the supplementary materials). It is notable that all the concentration profiles show an asymmetric distribution, extending deeper into the shoshonitic half than in the rhyolitic half. This is the result of the different diffusion rates occurring in melts with different degree of polymerization (e.g., Liang et al., 1996). In our specific case, elements diffuse faster in the shoshonitic than in the rhyolitic melts. A particular case is Al, which shows an uphill-diffusion-related minimum in the rhyolite side of the couple (**Fig. 4**). The effect of water is readily visible in the concentration-distance profiles (**Fig. 5**), producing notably longer diffusion distances at large water contents.

Convection can be a substantial issue in diffusion couple experiments with silicate melts (e.g. Behrens and Stelling, 2011). The following observations indicate that, for our particular system, convection played no relevant role because: (1) convection is expected to produce different compositional profiles within the same sample; on the contrary, microprobe data from profiles measured in the center and near the edge of the couples show no significant difference (see **Fig. S2** in supplemental material); (2) concentration-distance profiles from different experiments with distance normalized by the square root of time (**Fig. S3** in the supplemental material) show a good agreement, as expected for a diffusion-controlled process. In fact, after this normalization, if diffusion is the only process, all profiles should collapse to a single profile, as indeed observed for our experiments. The only exception is represented by experiment P300-H0-4 showing anomalous maxima or minima in the shoshonitic side of the couple (**Fig. S2** in the supplementary materials), indicating it was unsuccessful. As a consequence, it was not considered in further analysis and modeling.

*3.3. Evolution of composition along profiles*

A set of Harker variation diagrams has been plotted to evaluate the evolution of the composition along the diffusion profiles with varying experimental conditions (**Fig. 6**). Deviations from a linear mixing trend are observed due to the different diffusivities for different oxides. This effect tends to be larger in the oxides with the highest diffusivity contrast relative to SiO$_2$, as MgO or CaO. The uphill diffusion effect present in the Al$_2$O$_3$ profiles is prominently visible, characterized by a minimum at 70 wt.% SiO$_2$. Moreover, in spite of analytical scatter of Na$_2$O and relatively small differences between the end-members, the Harker diagram shows indication of a non-linear variation along the diffusion profiles. It is notable that, no variation of compositional trends was observed with changing H$_2$O content, pressure or experimental time, as evidenced by superimposition of the data from different experiments. Due to the consistency of the compositional trends, we define four intermediate compositions Lt$_{58}$ (latite), Tr$_{62}$, Tr$_{66}$ (trachytes) and Rh$_{70}$ (rhyolite) where subscripts denote the SiO$_2$ contents (**Table 1**). These compositions were used to monitor the evolution of diffusivity along the shoshonite-rhyolite join.

*3.4. Determination of diffusion coefficients*

The asymmetry of the profiles points to a variation of diffusivity with composition and, hence, a simple error function fit (assuming constant diffusivity) is not suitable to describe the mixing process.

Composition-dependent diffusion coefficients were determined by a three-step procedure (e.g. Crank, 1975; Zhang, 2008): (1) fitting and smoothing the analytical data with a Legendre polynomial; (2) normalization of the compositional range of fitted curve; (3) applying the method of Sauer and Freise (1962) to derive the diffusion coefficients for the element of interest.

Classically, the Boltzmann-Matano method (Boltzmann, 1894; Matano, 1932) has been used to study concentration dependent diffusivity. This method requires the determination of the position of the Matano interface, which is often difficult to identify after completion of the experiments. As a consequence, artifacts (jumps and gaps in plots of diffusion coefficients vs. compositional variables) may occur in the interface region if the Matano interface is not correctly identified. The modified method of Sauer and Freise (1962) is mathematically equivalent and does not require the a priori knowledge of the position of the Matano interface.

A prerequisite for the Sauer-Freise analysis is that the molar volume of the melts does not significantly vary along the diffusion profile, a condition that is met in the studied system. Using the computation scheme of Ochs and Lange (1999), the change of molar volume between mafic and felsic end-member is less than 4%.

**Fig. 7** shows an example of polynomial fitting. It is important to note that polynomial fits cannot be used to obtain diffusivities at the ends of the diffusion profiles, because they do not result in constant concentration at the end tips.

The analytical solution for one-dimension molar volume independent diffusion from Sauer and Freise (1962) is

$$D(x) = \frac{1}{-2t(\partial c/\partial x)}\left[(1-c(x))\int_x^\infty c\,dx + c(x)\int_{-\infty}^x (1-c(x))\,dx\right] \quad (4)$$

In equation (4), $D(x)$ is the diffusivity at $x$; $c(x)$ is the normalized composition of the diffusing component, where $c(x) = 1$ at $x = -\infty$, and $c(x) = 0$ at $x = \infty$; $t$ is the experimental time in seconds and $x$ is the position (distance in meters) along the profile. It is worth noting that, compared to the classical Boltzmann-Matano analysis, the approach of Sauer and Freise is not sensitive to the origin of the coordinate system. In our case $x = 0$ is arbitrarily chosen to be close to the initial interface between the two halves of the diffusion couples.

Both the polynomial fitting and the Sauer-Freise calculation procedures were implemented in a house-developed script in the Python programming language, from which distance-diffusivity profiles (and hence composition-diffusivity profiles) were obtained.

*3.5. Diffusion data*

The effective binary diffusion (EBD) (Zhang, 2010) coefficients for six elements (Si, Ti, Fe, Mg, Ca and K) obtained by the described procedure are listed in **Table 4**. The concept of EBD assumes that the flux of each component is associated with a counterflux of all other components, resulting effectively in a quasi-binary system (Zhang, 2010). Na is not included here since the difference in sodium content between end-members is too small (**Table 1**) to properly resolve profiles, although some indications of uphill diffusion are present. Al shows uphill diffusion (**Fig. 3**) indicating that both Fickian diffusion and elemental activity are playing a role. A pronounced minimum is observed at $SiO_2$ contents of ca. 70 wt.%, but no maximum is observed in the shoshonite side. This might be the effect of a combination of faster diffusion in the more mafic side and the low compositional gradient of $Al_2O_3$. Uphill diffusion is an evidence of coupling of Al to other melt components, mainly alkaline earth elements diffusing together with Al towards the rhyolite. Such coupling effects were observed by Liang (2010) in synthetic melts of the system $MgO-CaO-Al_2O_3-SiO_2$, and by Guo and Zhang (2016) in a 7-component haplobasalt system ($SiO_2-TiO_2-Al_2O_3-MgO-CaO-Na_2O-K_2O$). A similar effect has been also observed in natural basalt-rhyolite diffusion couple experiments (Koyaguchi, 1989; Richter et al., 2003), but not in rhyolite-dacite couples (Baker, 1990, Baker, 1991). Additionally, no significant Al uphill diffusion was observed in the system $Na_2O-K_2O-Al_2O_3-SiO_2-$

(H$_2$O) (Chakraborty, 1995a; Chakraborty, 1995b; Mungall, 1998). These findings suggest that coupling of Al to other melt components is particularly pronounced in systems containing alkaline earth elements and systems with large contrasts in composition. However, it is worth noting that uphill diffusion can occur almost in any element by means of coupling with other melt components if the appropriate conditions are given (Guo and Zhang, 2016). It is also important to note that diffusivities obtained in natural melts are subject to complex interaction occurring between components of the melt, often resulting in coupling of diffusion fluxes. The diffusivity values determined for such complex systems usually differ from diffusivities derived from simpler, synthetic systems and, as a consequence, results obtained here are only strictly applicable to the studied compositional systems.

Ti and Si are the slowest diffusing components in all experiments and through all the studied compositional spectrum, although Si shows a different behavior in the mafic sides of the couple and displays a larger increase in diffusivity. The lowest value of diffusion coefficient in the calculated dataset is -13.17 ± 0.14 (log $D$ units, with $D$ in m$^2$/s] for Ti in the dry Rh$_{70}$ composition. In the 2 wt.% water-bearing experiments, Si diffusion in the mafic terms often approaches that of Ca and K. On the other hand, the diffusivities of Fe, Mg, Ca and K are commonly very similar at the same experimental conditions and differences fall within error. However, in the majority of experiments Mg and Ca tend to be the fastest diffusing elements, both in the mafic to the felsic compositions. These trends are more readily seen in the experiments with added water (**Fig. 8**). This is in agreement with existing data derived from the CAS, MAS and CMAS synthetic systems, which show coupling between Ca and Mg (Liang, 2010). The observation of complex couplings between elements is also common in other literature data. Watson (1982) found a similar diffusion rate for all melt components in granite-basalt couples, and Zhang et al. (1989) showed similar results in quartz dissolution experiments in andesite.

The plot of field strength (defined as $Z/r^2$, where $Z$ is the atomic number and $r$ is ionic radius; iron is assumed to be $Fe^{2+}$) versus log $D$ for the 300 MPa experiments (**Fig. 8**) shows that diffusivity is not systematically correlated with the ionic field strength of cations. The trend is slightly negative in the most silicic compositions ($Rh_{70}$, $Tr_{66}$), and it tends to disappear towards the mafic compositions ($Tr_{62}$, $Lt_{58}$). This slightly negative trend disappears in the ND experiments, where mainly flat profiles are observed. It is also noticeable that, although Ti shows an anomalous low diffusivity and commonly breaks the trends, in some cases Ca of Fe have a similar or lower diffusivity (typically within the analytical error). These plots confirm the diffusive coupling between Mg, Ca, Fe and K, and the low diffusivity of Ti.

All components show a similar dependence on melt composition at the same water content. At all conditions a systematic increase of diffusivity with decreasing silica content is observed for all elements. The value of log $D$ varies linearly with $SiO_2$ content, although small deviations appear for some elements in the least evolved compositions (**Fig. 9**). The increase of diffusivity is relatively small for nominally dry melts (factor of 1.3 from $Rh_{70}$ to $Lt_{58}$) but increases with water content to factors of 3.5 to 5.5 for melts containing 2 wt.% $H_2O$. The larger increase of diffusivity is observed in experiments with the largest dissolved water contents, with some exceptions, most notably in some elements such as Si, Fe and K, at pressure of 100 MPa. In particular, K does not show any relevant variation with water content.

The effect of pressure in our experiments (**Fig. S4** in the supplementary materials) is much smaller than that of water content and anhydrous composition, and could not be properly resolved. Only in the experiments performed with nominally dry melts there is a very slight positive correlation between pressure and the diffusion of Si, Ti, Mg and Ca, with no visible effects on Fe and K. However, variations barely fall outside estimated error and the inclusion of the 500 MPa experiment in the 2 wt.% dataset does not confirm this trend. In experiments performed with 2 wt.% $H_2O$, trends appear to be dominated by small differences in water content. In an attempt to better distinguish the effect of pressure, the 2 wt.% $H_2O$ experimental dataset has been normalized to the water content of the 500 MPa experiment. Results (see **Fig. S4** in the supplemental material) show that no appreciable pressure effect can be seen for most elements.

*3.5.1 Effect of water and viscosity*

The major factor influencing diffusivity of major elements in our experiments is the dissolved water content. **Fig. 10** shows how diffusion coefficients vary with water content for the 50, 100 and 300 MPa experiments. The evaluation of the concentration-distance profiles shows a variable amount of enhancement of diffusivity in different components and experimental conditions. In the mafic compositions at 50 and 300 MPa, water effect is notable for Si, Fe and Mg, elements that show an enhancement of one order of magnitude in the 2 wt.% $H_2O$ experiments relative to the nominally dry ones (up to 20 times faster for Si in 50 MPa experiments for the $Lt_{58}$ compositional term). The effect of water gradually diminishes towards the felsic members, where wet diffusion coefficients (e.g. Si, Fe, Mg) are 5 to 15 times faster than the dry ones. However, other elements (notably K) do not show this trend. Similarly, the 100 MPa experiment show mainly parallel trends for the intermediate compositions, and the observed enhancements in diffusivity are lower than in the 50 and 300 MPa datasets. The dependence of diffusivity, taken as log $D$, on water content (wt.%) is apparently linear in the investigated range (**Fig. 10**). Comparison with literature data is difficult, since a few data are available for hydrous melts. Van der Laan et al. (1994) and Mungall et al. (1999) found water-induced enhancements in Mg diffusivity in felsic melts similar to those presented here, but equivalent data from mafic melts are almost absent.

The diffusivities of the different elements are coupled to different extent to melt relaxation and viscosity, a feature that has been observed in trace element diffusion data (e.g. Mungall, 2002; Behrens and Hahn, 2009). The Eyring equation

$$D = \frac{k_B * T}{\lambda * \eta} \tag{5}$$

where $k_B$ is the Boltzmann constant, $\lambda$ is the interatomic jump distance and $\eta$ is the viscosity of the melt, can be used to relate diffusion of network former elements to melt viscosity (Zhang, 2010).

An attempt has been made to relate the Si diffusion coefficients obtained in this work to the Eyring diffusivity. The viscosity of all four intermediate compositions has been calculated by using the Giordano et al. (2008) general viscosity model. Results are presented in **Fig. 11** and show that it is only possible to fit Si diffusion in the most mafic end-member using jump distances of c.a. 0.30 nm (the value of the Si-O-Si distance) for experiments performed at either 50, 100 or 300 MPa pressure. In the most silicic compositions the distance between calculated and experimental values increases, with calculated values being up to one order of magnitude lower than measured values. It is remarkable that trends within one composition varying from dry to hydrous are linear and roughly parallel to the best-fit trend. Deviations from the Eyring diffusivity are interpreted as a decoupling of element diffusivity from network relaxation. It is well known that alkali cations are only weakly affected by network relaxation, but high field strength elements are strongly coupled to network dynamics

(Mungall, 2002). We interpret the observed systematic deviation as a further consequence of the strong coupling already observed and discussed before.

*3.5.2. Petrological and volcanological implications*

The first and most important implication of the results presented here is the quantification of the effects caused by the presence of dissolved water in natural melts on element diffusion. Although a number of previous works have described the diffusion in both dry and hydrous melts (e.g. Baker, 1991; Baker & Bossányi, 1994; Mungall et al., 1998; Behrens & Hahn, 2009), this is the first work using natural end-member compositions from a specific volcanic system (Vulcano island, Aeolian archipelago, Italy). Our results clearly indicate that the diffusion of all major elements is notably enhanced even by relatively low amounts of dissolved water. Previous works (e.g. Perugini et al., 2010; Perugini et al., 2015, Morgavi et al., 2015) have shown that the signature of diffusion between compositionally different melts is recognizable in volcanic rocks that recorded pre-eruptive magma mixing events. Consequently, the study of mass exchanges between magmas by diffusion can provide important insights on the timescales of magma mixing and its potential role as an eruption trigger (e.g. Perugini et al., 2015). Up to now this approach has been essentially applied considering dry magmatic compositions. However, as demonstrated in this study, water content can strongly enhance diffusivity, thus providing the opportunity to better refine the existing models in order to derive more precise timescales. For example, the empirical relationships obtained in this work between water content and diffusivity can be used in numerical models in which the mixing process develops at different pressures (i.e. different water contents), from deep magmatic reservoirs to volcanic conduits. Element diffusivities are expected to vary because of water exsolution from the ascending magmas. In light of the experiments presented here, in such a scenario, diffusion coefficients for chemical elements cannot be considered constant during magma mixing. It follows that including our experimental results in future models might allow understanding in greater detail the effects of mass transfer processes between magma during their travel towards the Earth's surface.

Results presented here also contribute to shed new light on the complexity of the diffusive interactions occurring in complex, multicomponent natural melts. In particular, during mixing processes, a fractionation of chemical elements is expected due to their different diffusion coefficients (e.g. Perugini et al., 2006). Although our results might be valid only for the shoshonitic to high-K calc-alkaline system (i.e. the natural melts considered in the experiments), they highlight that there is a strong coupling among Fe, Mg, Ca and K (and sometimes Si) possibly preventing important diffusive fractionation processes for this group of elements. On the contrary, in the same timescale, Ti (because of its low diffusivity) and Al (because of the presence of uphill diffusion) may experience such fractionation processes to a greater extent. Further work on different magmatic compositions might help in understanding if this behavior of chemical elements is peculiar for the system studied here, or it is a generalized feature that characterizes most natural silicate systems.

Finally, the natural end-member melts used in the experiments belong to the active volcanic system of the island of Vulcano (Aeolian archipelago, Italy). In particular, they represent potential end-member compositions that interacted in this natural system to generate the wide compositional spectrum that has been erupted by this volcano in the last centuries. The evidence of magma mixing processes is widespread in the system (banded pumices, enclaves, etc.), especially in the last eruptive activity

(i.e. 1888-1890; Clocchiatti et al., 1994; Vetere et al., 2015). It is therefore crucial to understand how mass exchange processes between magmas develop via chemical diffusion, as they might impact upon the rheological properties of magmas, the amount and composition of the crystallizing minerals and, ultimately, on eruption style and dynamics. Experimental results presented here can aid in developing more robust petrological models for the compositional evolution of the plumbing system of this volcanic system by including in the models realistic relationships between the water content of magmas and the ability of chemical element to diffuse in the molten mass.

## 4. Conclusions

A set of isothermal diffusion coefficients has been experimentally determined for six major elements (Si, Ti, Fe, Mg, Ca and K) to study the interdiffusion at high pressure and temperature between natural high-K rhyolite and shoshonite melts, both in dry and wet conditions. The presence of uphill diffusion in Al and the analytical scatter for Na prevented the measurement of their diffusivities. The diffusion of measured elements results in compositional evolutions that slightly differ from the linear mixing trends. Evidence of coupled diffusivities among melt components is observed, as evidenced by uphill diffusion and similarity of the diffusivities of Fe, Mg, Ca and K. Dissolved water is the major influencing parameter on diffusivities of all elements. In general, diffusion is up to 1.4 orders of magnitude faster in samples containing up to 2 wt.% $H_2O$ than in nominally dry melts, although the greater effects are observed in the mafic compositions. No clear indication of pressure dependence is observed in the investigated pressure range (50 to 500 MPa).

Our experimental results have implications on the study of magma mixing events and the rate of homogenization and chemical exchange between mixing melts. In particular, the complexity of the diffusion mechanism in multicomponent natural melts observed in our study indicates that in order to fully understand mass exchange processes during magma mixing, the mobility of chemical elements in different melt compositions must be systematically studied and well understood. Further systematic studies are needed in order to generate consistent datasets of diffusion coefficients for different end-member couples. These data are essential to evaluate the potential of mass transfer processes during magma mixing in generating the wide compositional spectrum that is often observed in mixed rocks. At present, due of the lack of systematic data on diffusivities between compositionally different melts, this crucial aspect of magma differentiation still remains elusive.

**Acknowledgements**

This research was funded by the European Research Council Consolidator Grant ERC-2013-CoG (No. 612776 – CHRONOS) to D. Perugini; and by the MIUR-DAAD Joint Mobility Project (grant number 57262582) to F. Vetere and H. Behrens. D. Morgavi thanks the project AEOLUS (No. MORGABASE2015) funded by the Fondo di Ricerca di Base of Department of Physics and Geology, University of Perugia. D. González-García wishes to acknowledge the kind help received from R. Balzer while performing the experiments, R. Almeev during the microprobe analysis, and J. Feige for the careful preparation of experiment sections, all of them at the Institute of Mineralogy, Hannover. In addition, we acknowledge A. Zezza for completing the first experiment under her M.Sc. thesis work. We are grateful to Y. Zhang and an anonymous reviewer, whose constructive reviews significantly improved the quality of this manuscript. We acknowledge R. Astbury as well for the review of the English in the manuscript.

**Figure captions**

**Fig. 1**: Compositions of end-members and intermediate compositions for which diffusivities have been extracted, plotted in (A) a total alkali-silica diagram (TAS) and (B) a $SiO_2$ vs. $K_2O$ diagram (Peccerillo and Taylor, 1976). Compositions are normalized on an anhydrous basis. SHO and RHY refer to the end-members shoshonite and rhyolite, respectively. $Lt_{58}$, $Tr_{62}$, $Tr_{66}$ and $Rh_{70}$ indicate average compositions along the diffusion profiles (subscripts indicate the $SiO_2$ content).

**Fig. 2**: Examples of experimental diffusion couples. (A) microphotograph of an experiment section (run P500-H2-4, 4 h, 2 wt.% $H_2O$, 500 MPa) in transmitted light with the two measured analytical profiles (yellow lines): one centered and one offset. (B) interface area of a 300 MPa, bubble-free experiment (run P300-H2-1, 1 h, 2 wt.% $H_2O$). (C) EPMA back scattered electron (BSE) image of another 300 MPa experiment (run P300-H1-4, 1 wt.% $H_2O$, 4h,). (D) interface area of a bubble-bearing

experiment at 50 MPa (run P050-H1-4, 1 h, 1 wt.% $H_2O$). (E) BSE image of the interface region of the same experiment.

**Fig. 3**: Water concentration along the analytical profiles for each of the measured experiments, except P050-H0-4. Data for experiments at different water contents are plotted in different panels. The vertical dashed line represents the inferred position of the interface between the two melts. Rhyolite end is in the left side; shoshonite end is in the right side. For experiment P050-H0-4 (star), only two data points are available at distances from the interface larger than 1 mm.

**Fig. 4**: Example of concentration-distance profiles of major elements (run P300-H2-4; 2 wt.% $H_2O$, 300 MPa, 1200 °C, 4 h). Vertical dashed line represents the inferred position of the interface between melts. Rhyolite end is in the left side; shoshonite end is in the right side. For the complete set of concentration-distance profiles, refer to **Fig. S2** in the supplemental materials.

**Fig. 5**: Variation of $SiO_2$ composition-distance profiles as a function of different water contents at 50, 100 and 300 MPa.

**Fig. 6**. Harker variation diagrams for all experimental products, differentiated by pressure and water content. Thick line represents a linear mixing trend between the end-members. Vertical grey bands are the silica contents of the four intermediate compositions ($Lt_{58}$ to $Rh_{70}$).

**Fig. 7**. Example of polynomial fits (solid curves) to composition-distance EPMA profiles (expressed in atomic percent) for experiment P300-H2-1 (2 wt.% $H_2O$, 300 MPa, 1 h). Vertical dashed line represents the inferred position of the interface between melts. Error bars represent the analytical error associated with EPMA analyses.

**Fig. 8:** Diffusivities obtained for the 300 MPa series of experiments plotted against element field strength.

**Fig. 9**: Relationship between Si and Mg diffusion coefficients and the $SiO_2$ content. All points obtained from the application of the Sauer-Freise procedure are plotted. A linear fit is also provided as a guide. Data corresponds to run P300-H2-1.

**Fig. 10.** Experimental diffusivities of six major elements as a function of water content at 50, 100, and 300 MPa. Linear fits are added as visual guide (for the equation parameters, refer to **Table S2** in the supplemental materials).

**Fig. 11**. Comparison between experimentally measured Si diffusivities and Eyring diffusivities calculated from viscosity using a jump distance ($\lambda$) of 0.30 nm. Central solid line represents the equivalence between measured and calculated diffusivities; dashed lines represent a difference of plus or minus one log unit. Coloured lines are linear fits to the data of the intermediate compositions. Viscosities are calculated using the model from Giordano et al. (2008). Nominal water content and composition are coded by shape and colour of symbols as follows. Circles: 2 wt.% $H_2O$ MPa;

triangles: 1 wt.% H$_2$O MPa; squares: nominally dry. Red: Lt$_{58}$; orange: Tr$_{62}$; green: Tr$_{66}$; blue: Rh$_{70}$.

**Table captions**

**Table 1**. Composition of end-members and intermediate compositions for which diffusivities have been obtained normalized to a water-free basis. Iron is given as total ferrous iron oxide (FeO$_t$). End-member data are the average composition based on 10 microprobe data points measured on the starting materials melted in air. Intermediate compositions are the average of all data points where diffusivities have been determined. Standard deviation of data is given in parenthesis.

**Table 2**: Fe oxidation state in synthetized glasses measured by UV-VIS spectroscopy. Water contents were measured by the Karl Fischer titration (KFT) method.

**Table 3**. Experimental conditions. Experiment P300-H2-0 (zero-time) shows an uneven water profile, and water concentrations are given for each end-member. Experiment P300-H2-4 is discussed in the text but discarded because of uneven water concentration profile. Temperature was kept constant at 1200 °C for all experiments.

**Table 4**. Diffusion coefficients obtained from the evaluation of concentration-distance profiles using the method of Sauer and Freise (1962). Diffusion coefficients

are reported as log $D$, with $D$ in m$^2$/s. Water contents are given in wt.%. ND stands for nominally dry. Values in italics are estimated errors.

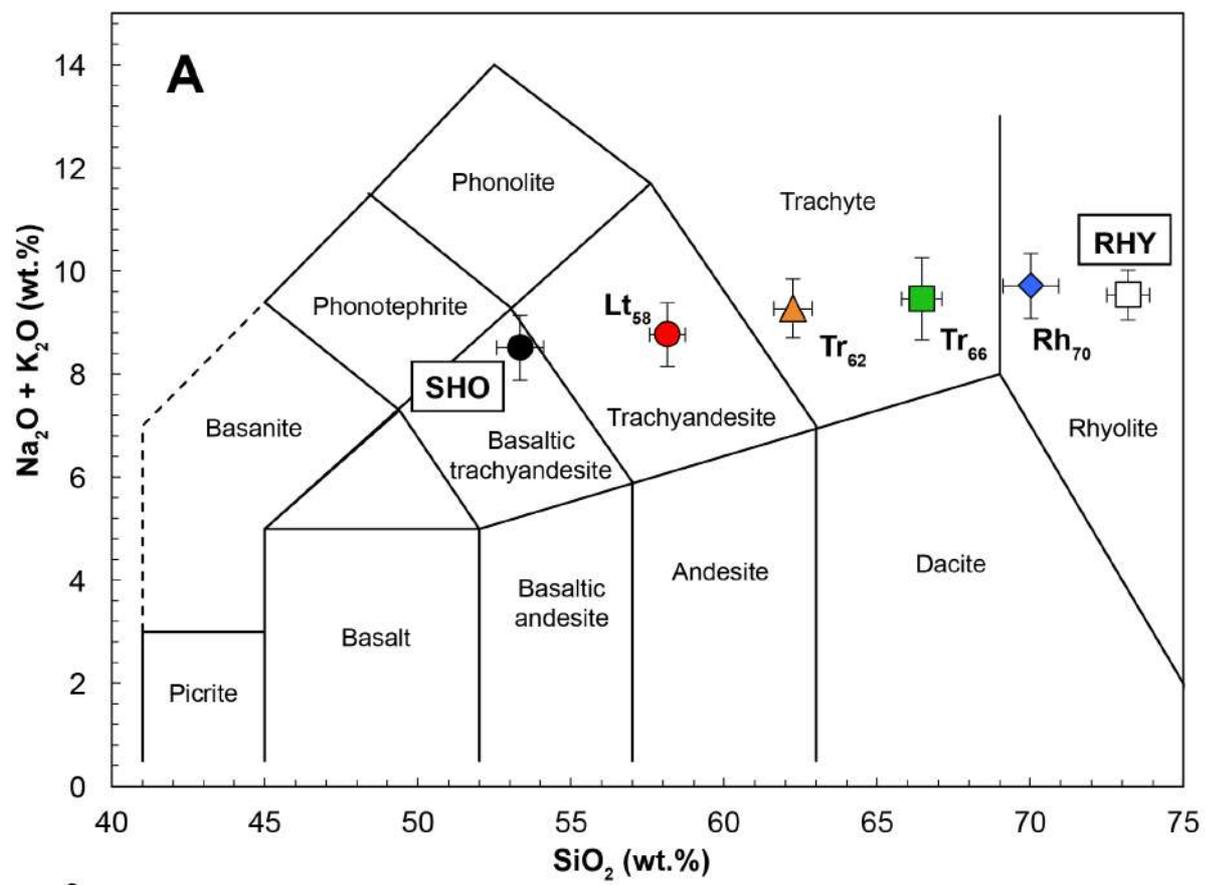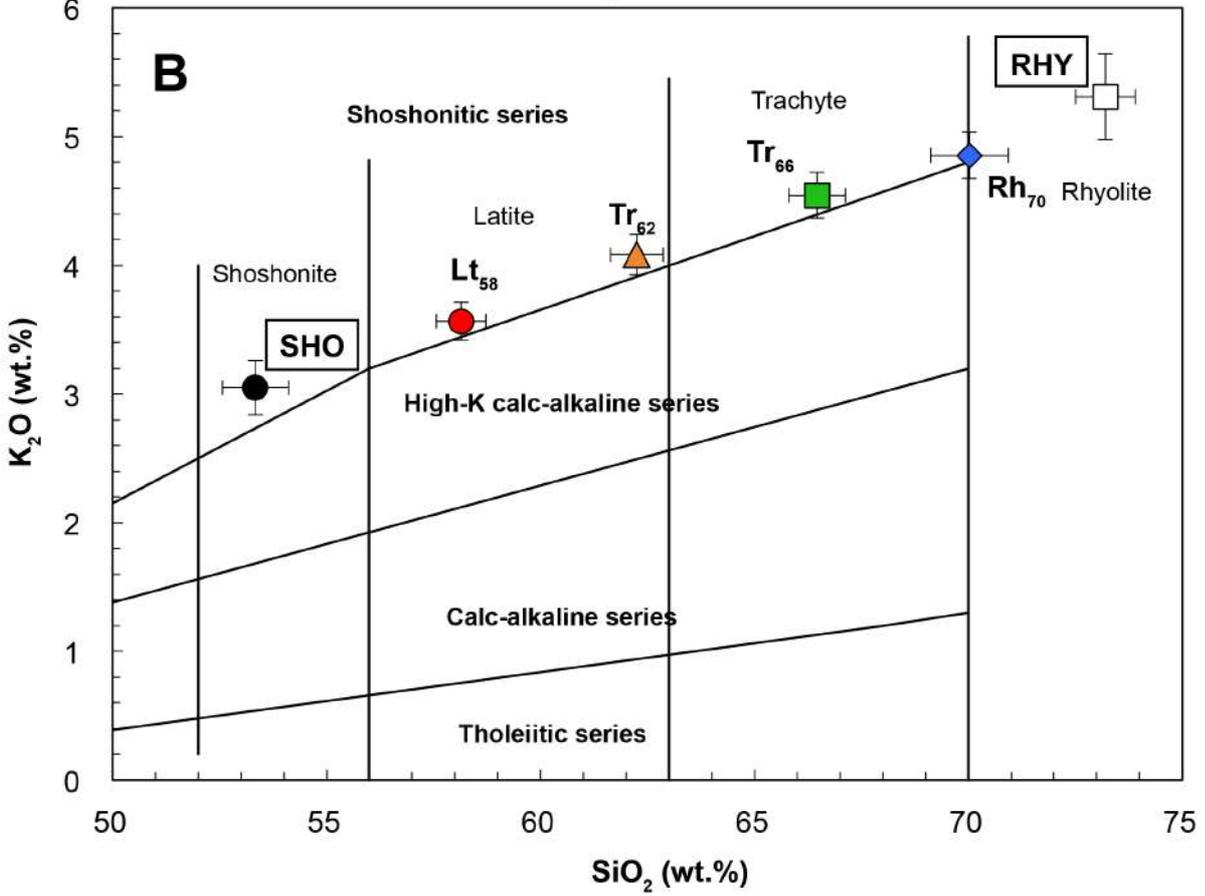

**Figure 1**

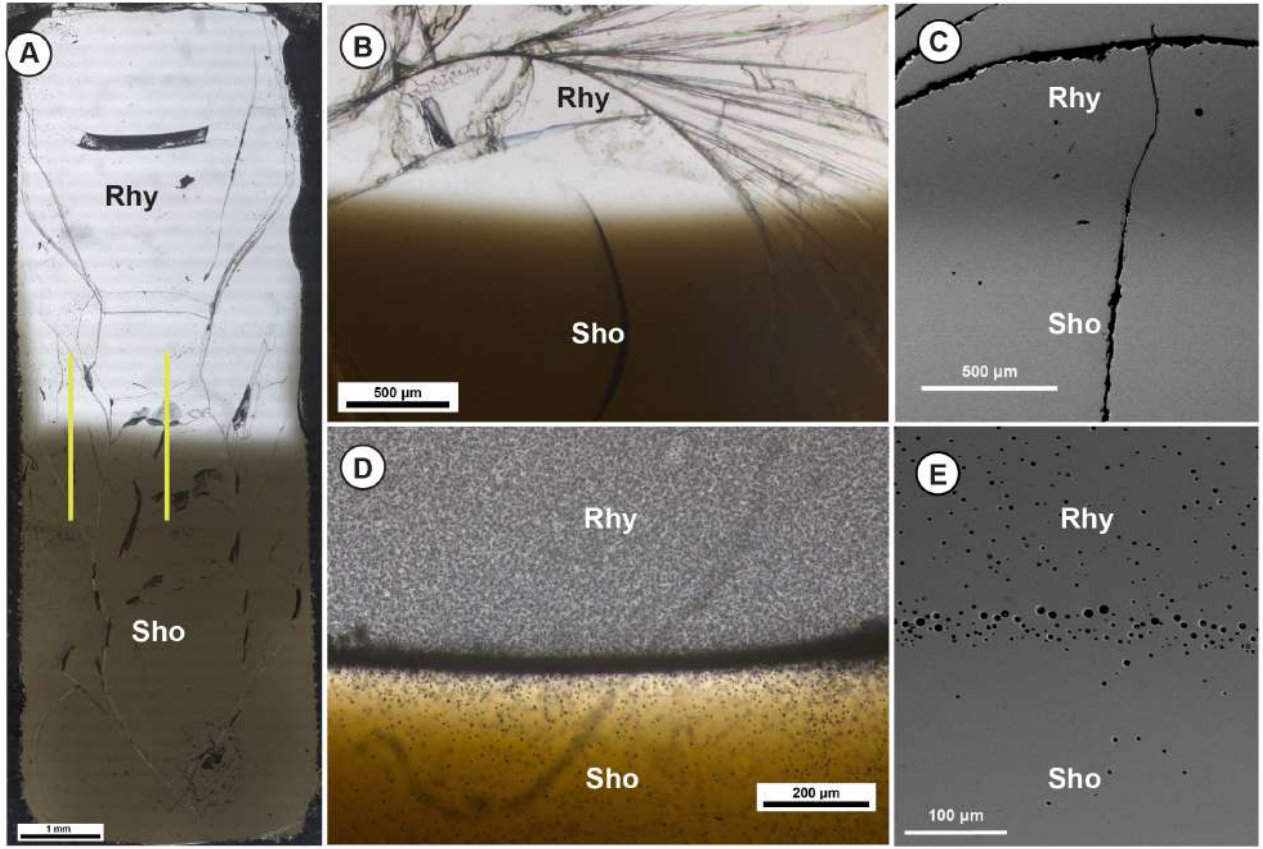

**Figure 2**

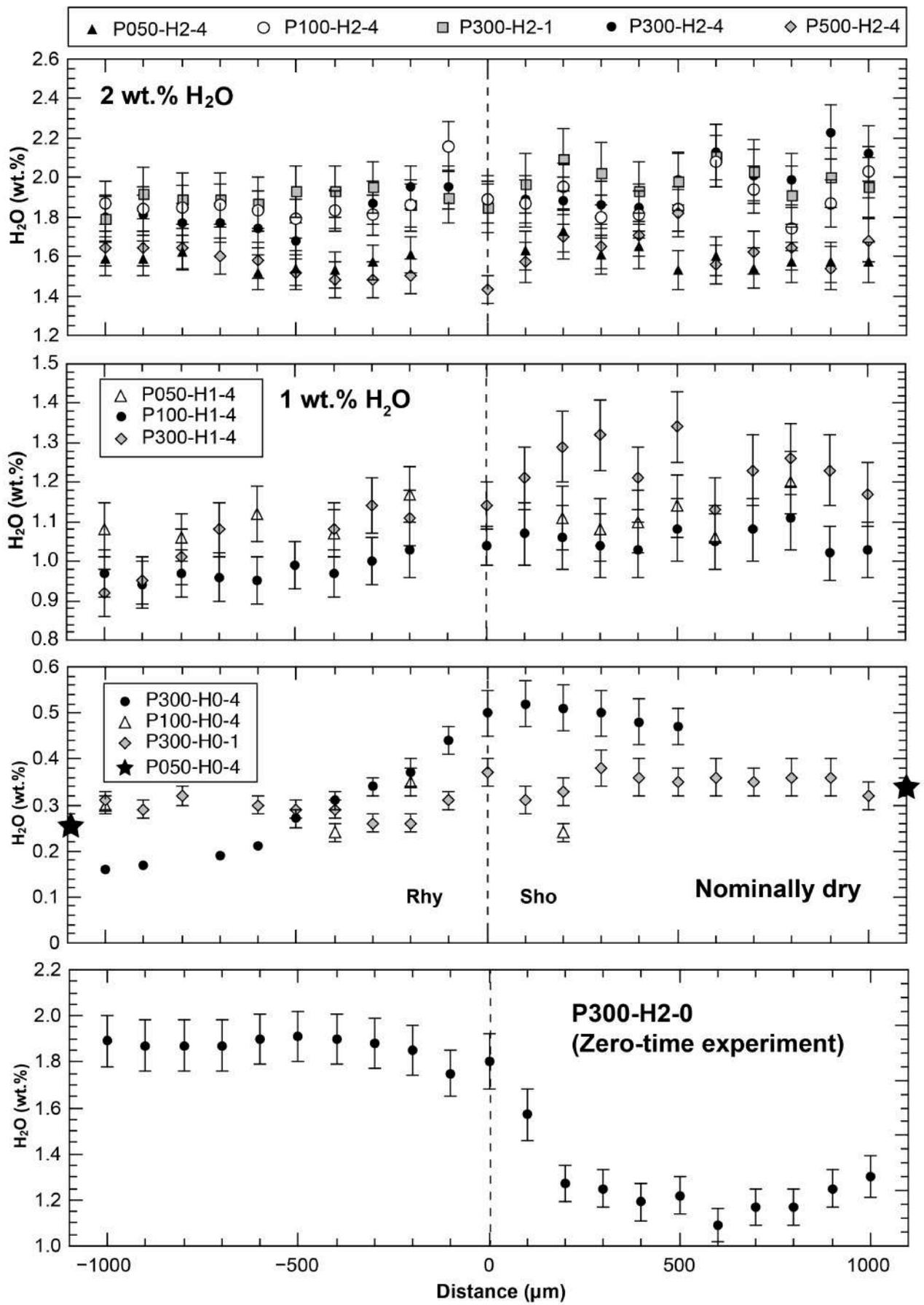

**Figure 3**

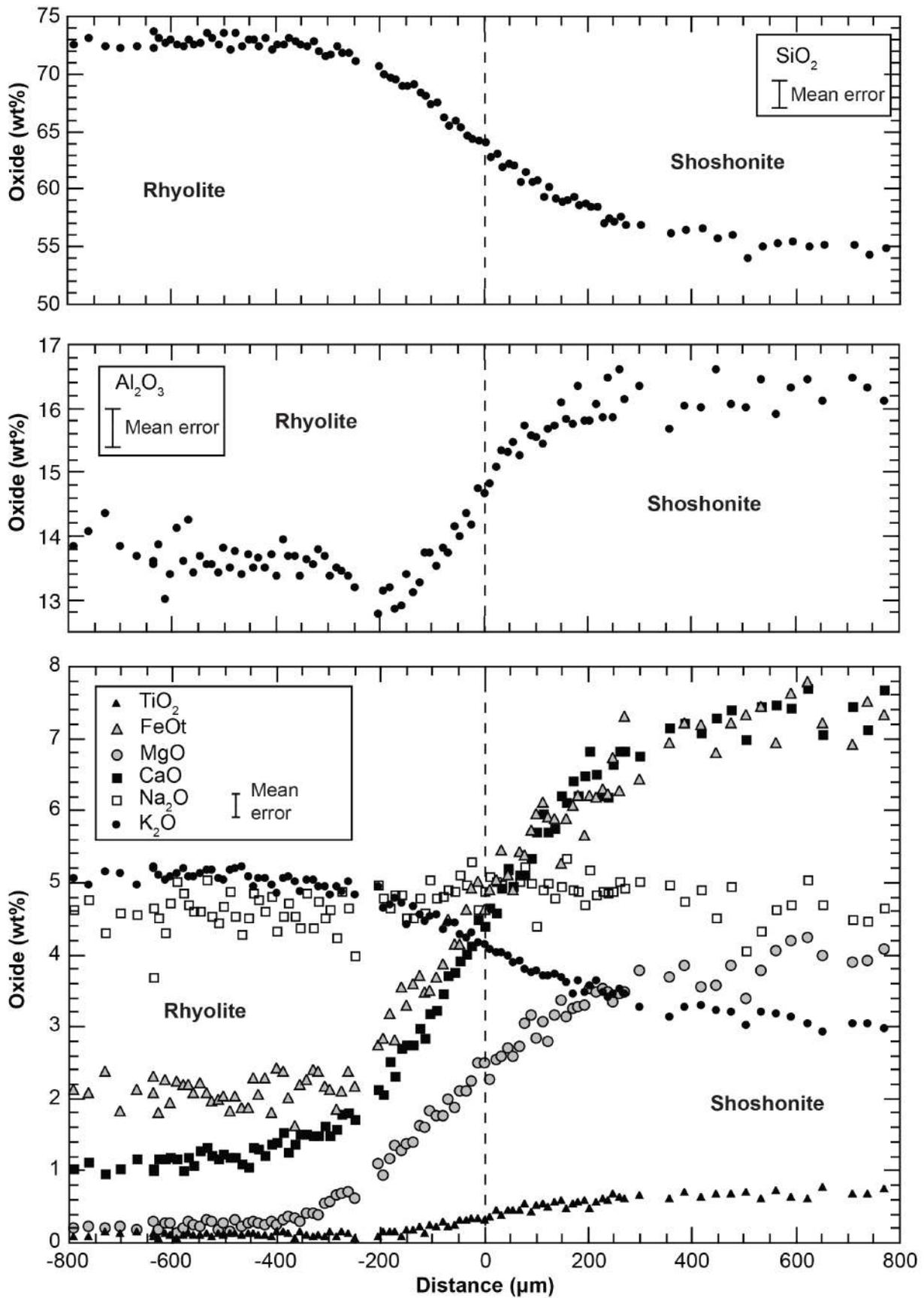

**Figure 4**

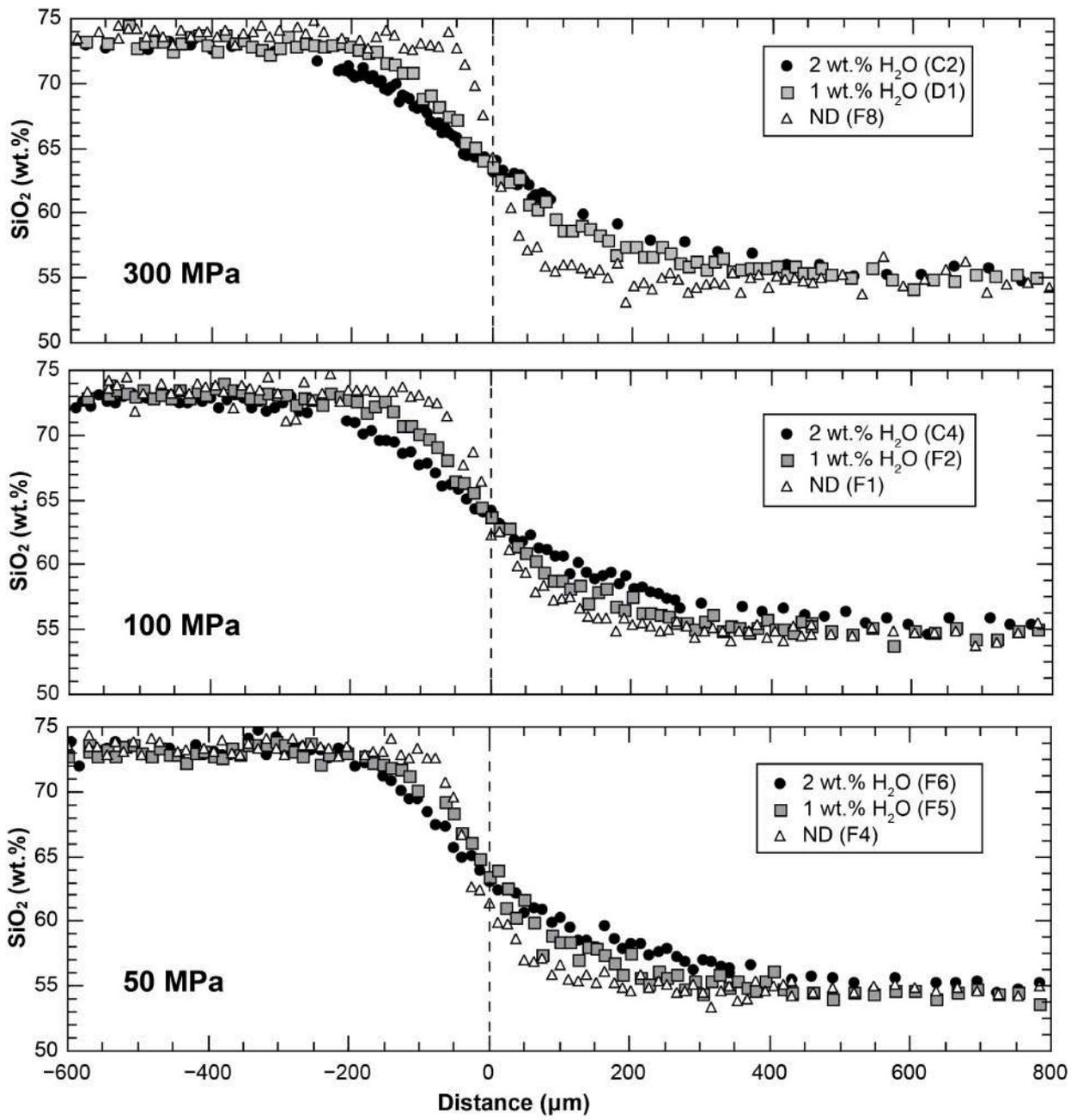

**Figure 5**

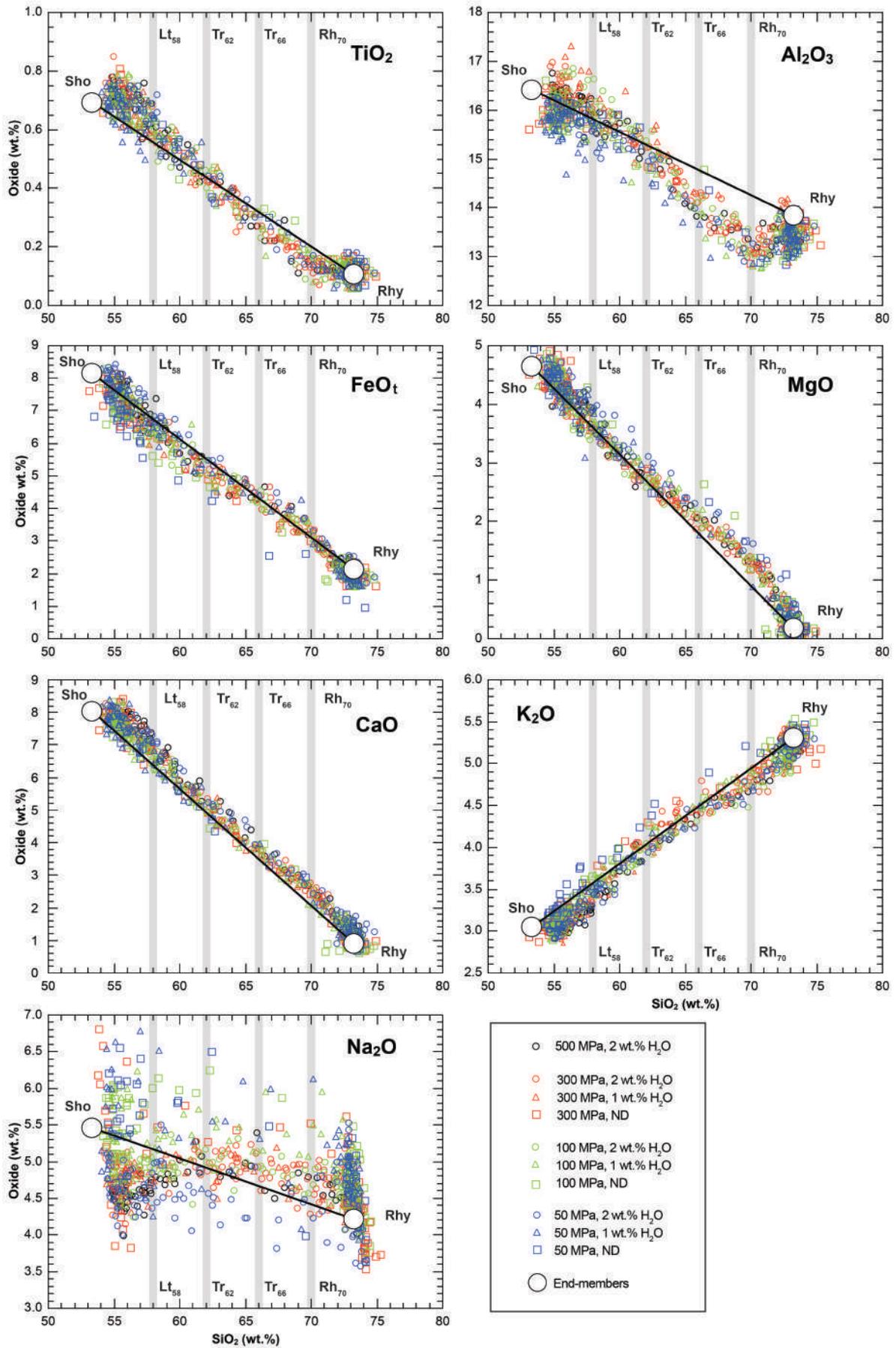

**Figure 6**

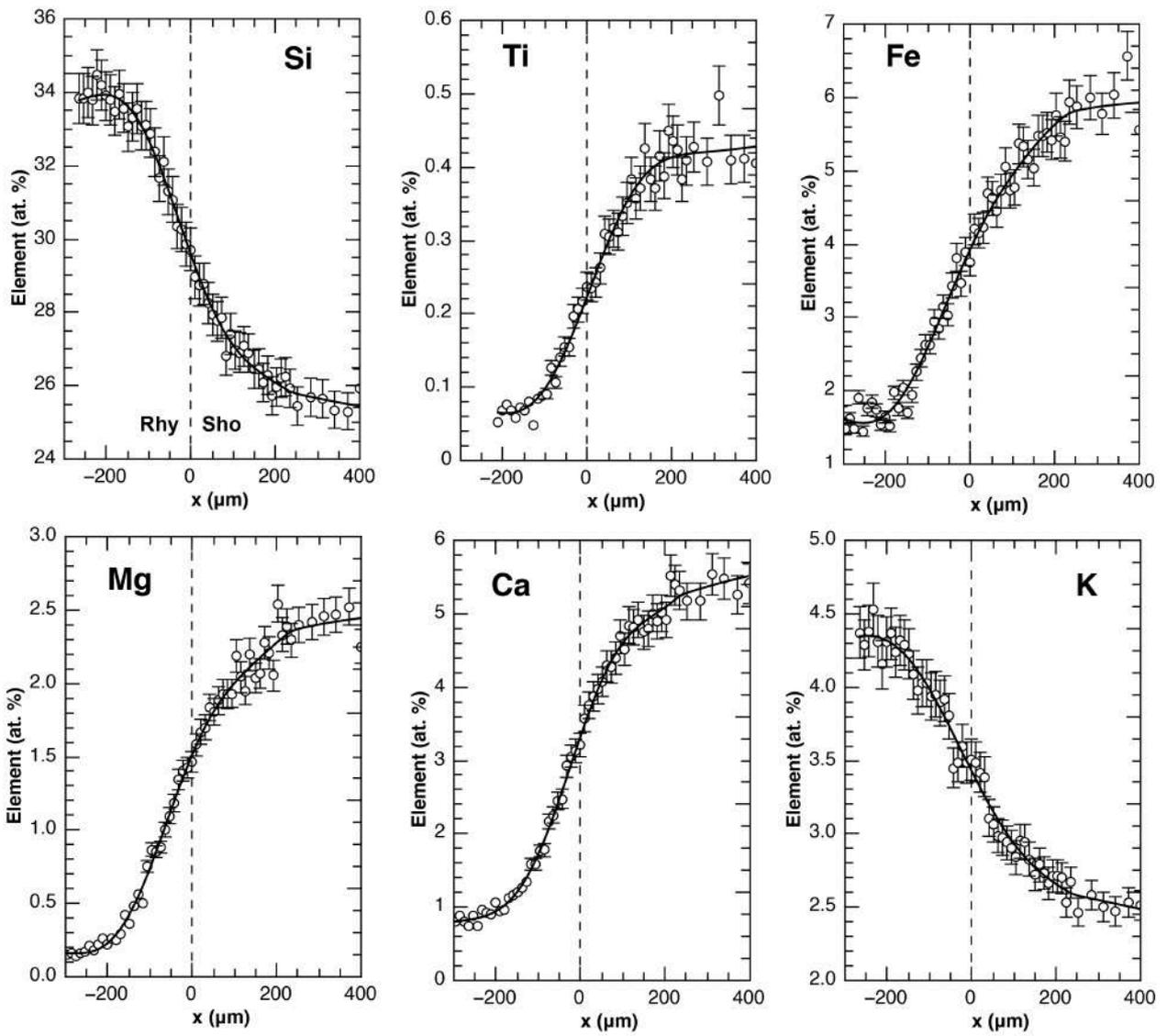

**Figure 7**

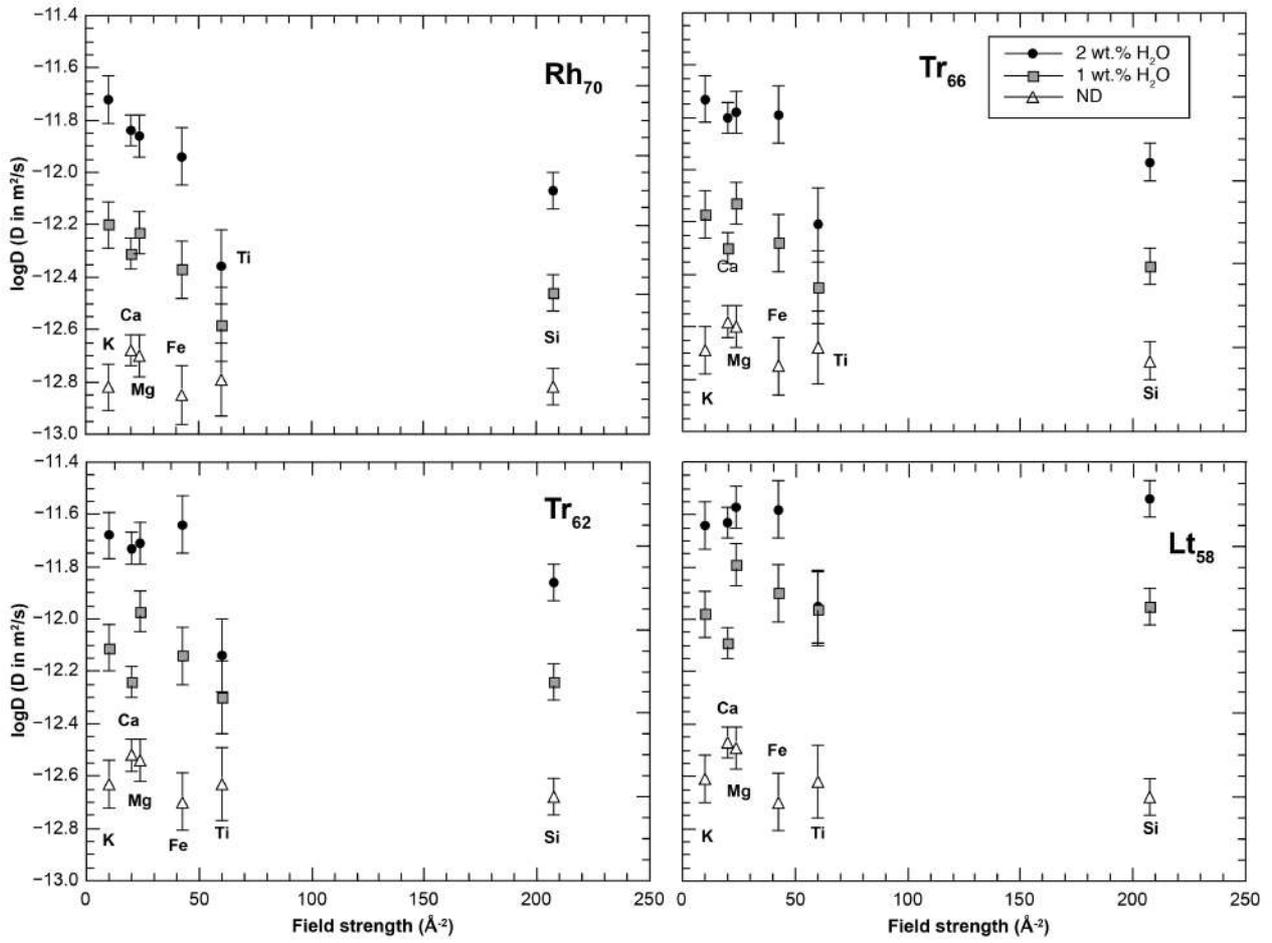

**Figure 8**

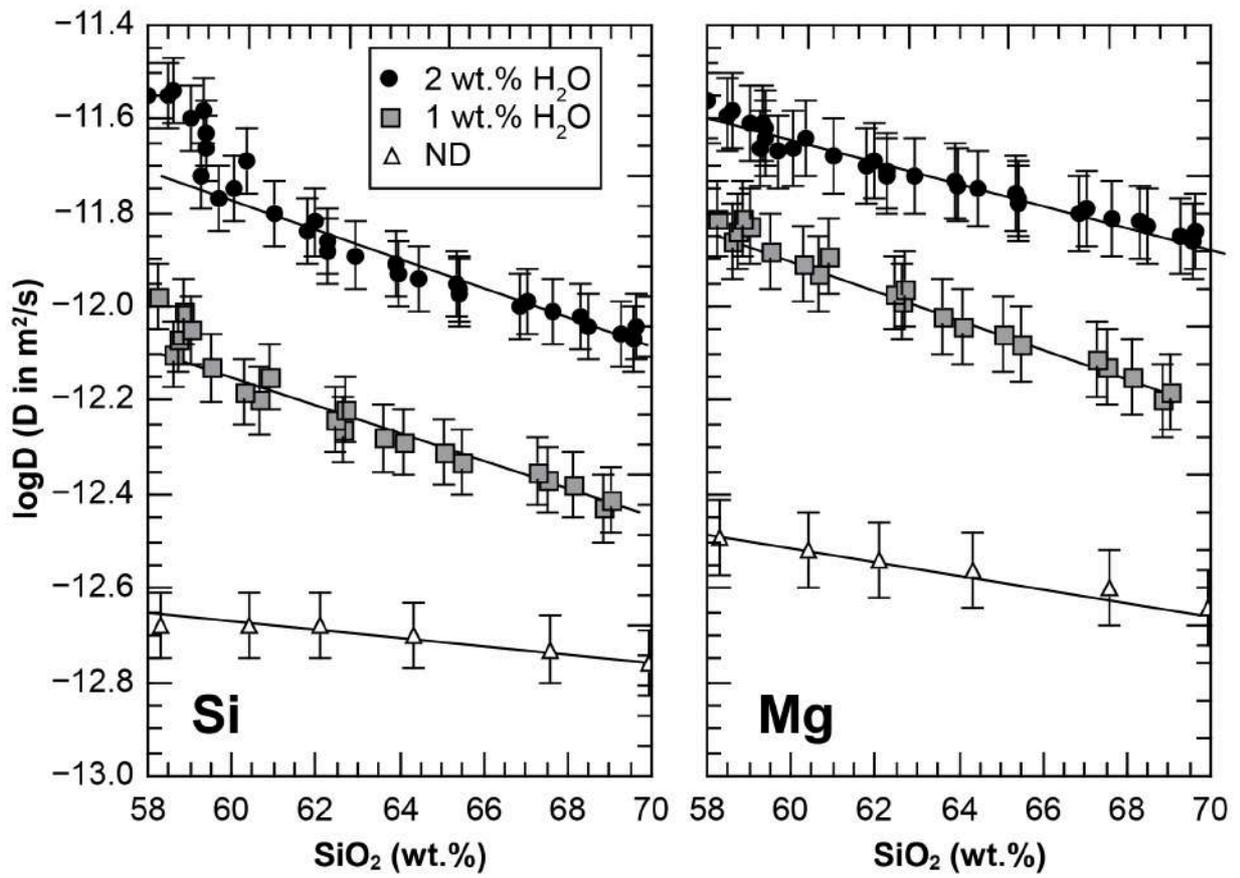

**Figure 9**

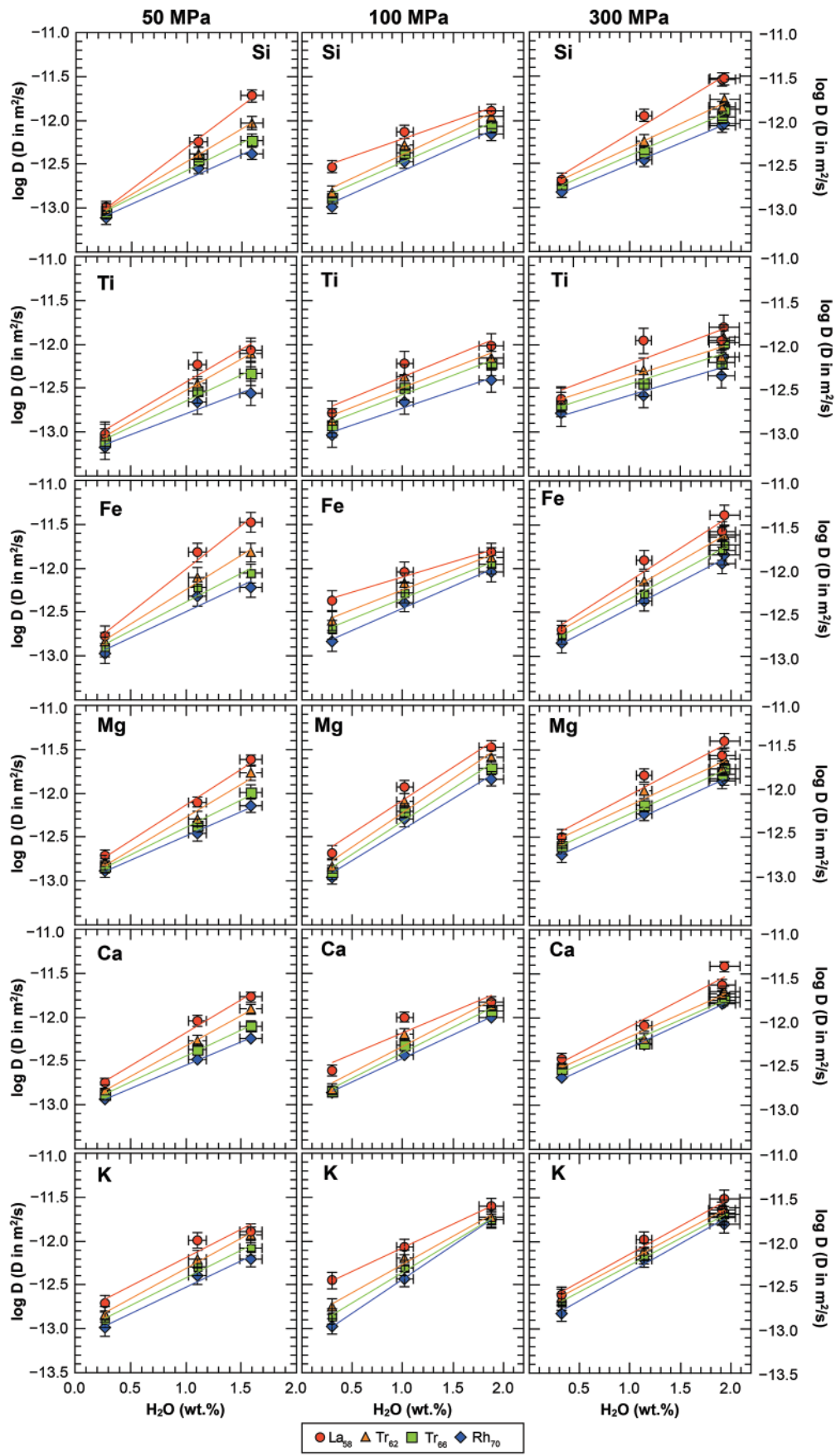

**Figure 10**

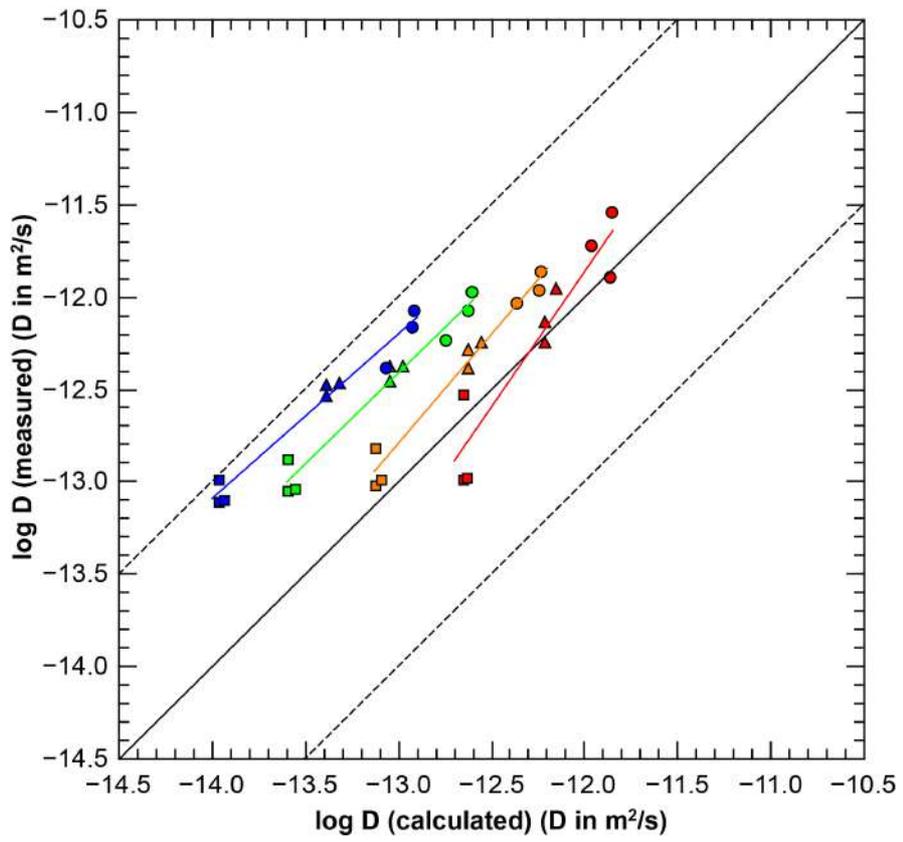

**Figure 11**

**Table 1**

|  | End members | | Intermediate terms | | | |
|---|---|---|---|---|---|---|
|  | Shoshonite | Rhyolite | $Lt_{58}$ Latite | $Tr_{62}$ Trachyte | $Tr_{66}$ Trachyte | $Rh_{70}$ Rhyolite |
| $SiO_2$ | 53.34 *(0.77)* | 73.20 *(0.67)* | 58.15 *(0.58)* | 62.25 *(0.62)* | 66.45 *(0.66)* | 70.02 *(0.91)* |
| $TiO_2$ | 0.69 *(0.04)* | 0.11 *(0.03)* | 0.61 *(0.03)* | 0.42 *(0.02)* | 0.27 *(0.03)* | 0.15 *(0.04)* |
| $Al_2O_3$ | 16.42 *(0.15)* | 13.84 *(0.31)* | 15.74 *(0.26)* | 14.99 *(0.23)* | 14.04 *(0.32)* | 13.11 *(0.23)* |
| $FeO_t$ | 8,14 *(0.28)* | 2.14 *(0.26)* | 6.32 *(0.47)* | 5.02 *(0.34)* | 3.98 *(0.58)* | 3.16 *(0.48)* |
| MgO | 4.64 *(0.10)* | 0.18 *(0.09)* | 3.67 *(0.22)* | 2.79 *(0.23)* | 2.08 *(0.38)* | 1.35 *(0.26)* |
| MnO | 0.21 *(0.13)* | 0.08 *(0.03)* | 0.08 *(0.07)* | 0.10 *(0.08)* | 0.09 *(0.08)* | 0.09 *(0.04)* |
| CaO | 8.04 *(0.18)* | 0.92 *(0.18)* | 6.51 *(0.28)* | 5.08 *(0.23)* | 3.60 *(0.30)* | 2.39 *(0.38)* |
| $Na_2O$ | 5.46 *(0.21)* | 4.22 *(0.20)* | 5.20 *(0.55)* | 5.19 *(0.52)* | 4.91 *(0.74)* | 4.85 *(0.50)* |
| $K_2O$ | 3.05 *(0.07)* | 5.31 *(0.06)* | 3.57 *(0.15)* | 4.08 *(0.16)* | 4.55 *(0.22)* | 4.86 *(0.18)* |

**Table 2**

| End-member | Rhyolite | | | Shoshonite | | |
|---|---|---|---|---|---|---|
| H2O nom. (wt.%) | ND | 1 | 2 | ND | 1 | 2 |
| $H_2O$ (wt%) | 0.30 ± 0.08 | 1.01 ± 0.07 | 1.99 ± 0.07 | 0.75 ± 0.07 | 1.01 ± 0.07 | 1.83 ± 0.07 |
| $Fe^{2+}/Fe_{tot}$ | 0.72 ± 0.04 | 0.62 ± 0.03 | 0.61 ± 0.03 | 0.85 ± 0.05 | 0.81 ± 0.03 | 0.74 ± 0.03 |

**Table 3**

| Run | P (MPa) | t (s) | H$_2$O nom. (wt%) | H$_2$O FTIR (wt%) |
|---|---|---|---|---|
| P050-H0-4 | 50 | 14400 | ND | 0.27 ± 0.03 |
| P050-H1-4 | 50 | 14400 | 1 | 1.11 ± 0.08 |
| P050-H2-4 | 50 | 14400 | 2 | 1.59 ± 0.10 |
| P100-H0-4 | 100 | 14400 | ND | 0.35 ± 0.05 |
| P100-H1-4 | 100 | 14400 | 1 | 1.02 ± 0.08 |
| P100-H2-4 | 100 | 14400 | 2 | 1.88 ± 0.12 |
| P300-H0-4 | 300 | 14400 | ND | 0.38 ± 0.05* |
| P300-H0-1 | 300 | 3600 | ND | 0.32 ± 0.03 |
| P300-H1-4 | 300 | 14400 | 1 | 1.14 ± 0.08 |
| P300-H2-4 | 300 | 14400 | 2 | 1.91 ± 0.13 |
| P300-H2-1 | 300 | 3600 | 2 | 1.95 ± 0.15 |
| P500-H2-4 | 500 | 14400 | 2 | 1.64 ± 0.08 |
| P300-H2-0 | 300 | 0 | 2 | 1.87 ± 0.11 (Rhy)<br>1.26 ± 0.08 (Sho) |

**Table 4**

| Pressure | | 50 MPa | | | | | | 100 MPa | | | | | | 300 MPa | | | | | | | | 500 MPa | |
|---|---|---|---|---|---|---|---|---|---|---|---|---|---|---|---|---|---|---|---|---|---|---|---|
| Experiment time (s) | | 14400 | | 14400 | | 14400 | | 14400 | | 14400 | | 14400 | | 3600 | | 14400 | | 14400 | | 3600 | | 14400 | |
| Nominal H$_2$O | | ND | | 1 wt.% | | 2 wt.% | | ND | | 1 wt.% | | 2 wt.% | | ND | | 1 wt.% | | 2 wt.% | | 2 wt.% | | 2 wt.% | |
| Measured H$_2$O | | 0.27 | 0.03 | 1.11 | 0.08 | 1.59 | 0.10 | 0.30 | 0.03 | 1.02 | 0.08 | 1.88 | 0.12 | 0.32 | 0.03 | 1.14 | 0.08 | 1.91 | 0.13 | 1.94 | 0.15 | 1.64 | 0.08 |
| Comp. | Melt | logD | 1σ | logD | 1σ | logD | 1σ | logD | 1σ | logD | 1σ | logD | 1σ | logD | 1σ | logD | 1σ | logD | 1σ | logD | 1σ | logD | 1σ |
| Si | Lt$_{58}$ | -12.99 | 0.07 | -12.24 | 0.07 | -11.72 | 0.07 | -12.53 | 0.07 | -12.13 | 0.07 | -11.89 | 0.07 | -12.68 | 0.07 | -11.95 | 0.07 | -11.54 | 0.07 | -11.53 | 0.07 | -11.95 | 0.07 |
| | Tr$_{62}$ | -13.02 | 0.07 | -12.38 | 0.07 | -12.03 | 0.07 | -12.82 | 0.07 | -12.28 | 0.07 | -11.96 | 0.07 | -12.68 | 0.07 | -12.24 | 0.07 | -11.86 | 0.07 | -11.77 | 0.07 | -12.11 | 0.07 |
| | Tr$_{66}$ | -13.05 | 0.07 | -12.45 | 0.07 | -12.23 | 0.07 | -12.88 | 0.07 | -12.37 | 0.07 | -12.07 | 0.07 | -12.73 | 0.07 | -12.37 | 0.07 | -11.97 | 0.07 | -11.88 | 0.07 | -12.17 | 0.07 |
| | Rh$_{70}$ | -13.11 | 0.07 | -12.54 | 0.07 | -12.38 | 0.07 | -12.99 | 0.07 | -12.47 | 0.07 | -12.16 | 0.07 | -12.82 | 0.07 | -12.46 | 0.07 | -12.07 | 0.07 | -12.04 | 0.07 | -12.24 | 0.07 |
| Ti | Lt$_{58}$ | -13.02 | 0.14 | -12.23 | 0.14 | -12.07 | 0.14 | -12.78 | 0.14 | -12.22 | 0.14 | -12.02 | 0.14 | -12.62 | 0.14 | -11.96 | 0.14 | -11.95 | 0.14 | -11.81 | 0.14 | -12.10 | 0.14 |
| | Tr$_{62}$ | -13.05 | 0.14 | -12.44 | 0.14 | -12.11 | 0.14 | -12.87 | 0.14 | -12.37 | 0.14 | -12.15 | 0.14 | -12.63 | 0.14 | -12.30 | 0.14 | -12.14 | 0.14 | -11.92 | 0.14 | -12.28 | 0.14 |
| | Tr$_{66}$ | -13.10 | 0.14 | -12.53 | 0.14 | -12.33 | 0.14 | -12.92 | 0.14 | -12.49 | 0.14 | -12.22 | 0.14 | -12.68 | 0.14 | -12.45 | 0.14 | -12.21 | 0.14 | -11.98 | 0.14 | -12.38 | 0.14 |
| | Rh$_{70}$ | -13.17 | 0.14 | -12.66 | 0.14 | -12.56 | 0.14 | -13.03 | 0.14 | -12.66 | 0.14 | -12.41 | 0.14 | -12.79 | 0.14 | -12.58 | 0.14 | -12.36 | 0.14 | -12.14 | 0.14 | -12.54 | 0.14 |
| Fe | Lt$_{58}$ | -12.77 | 0.11 | -11.82 | 0.11 | -11.48 | 0.11 | -12.37 | 0.11 | -12.04 | 0.11 | -11.82 | 0.11 | -12.70 | 0.11 | -11.90 | 0.11 | -11.58 | 0.11 | -11.39 | 0.11 | -11.74 | 0.11 |
| | Tr$_{62}$ | -12.84 | 0.11 | -12.10 | 0.11 | -11.82 | 0.11 | -12.60 | 0.11 | -12.17 | 0.11 | -11.88 | 0.11 | -12.70 | 0.11 | -12.14 | 0.11 | -11.64 | 0.11 | -11.62 | 0.11 | -11.88 | 0.11 |
| | Tr$_{66}$ | -12.90 | 0.11 | -12.22 | 0.11 | -12.06 | 0.11 | -12.70 | 0.11 | -12.28 | 0.11 | -11.96 | 0.11 | -12.75 | 0.11 | -12.28 | 0.11 | -11.79 | 0.11 | -11.73 | 0.11 | -11.94 | 0.11 |
| | Rh$_{70}$ | -12.97 | 0.11 | -12.32 | 0.11 | -12.22 | 0.11 | -12.84 | 0.11 | -12.39 | 0.11 | -12.04 | 0.11 | -12.85 | 0.11 | -12.37 | 0.11 | -11.94 | 0.11 | -11.84 | 0.11 | -12.00 | 0.11 |
| Mg | Lt$_{58}$ | -12.71 | 0.08 | -12.10 | 0.08 | -11.62 | 0.08 | -12.68 | 0.08 | -11.93 | 0.08 | -11.48 | 0.08 | -12.49 | 0.08 | -11.79 | 0.08 | -11.57 | 0.08 | -11.40 | 0.08 | -11.93 | 0.08 |
| | Tr$_{62}$ | -12.79 | 0.08 | -12.29 | 0.08 | -11.77 | 0.08 | -12.09 | 0.08 | -12.83 | 0.08 | -11.59 | 0.08 | -12.54 | 0.08 | -11.97 | 0.08 | -11.71 | 0.08 | -11.60 | 0.08 | -11.99 | 0.08 |
| | Tr$_{66}$ | -12.83 | 0.08 | -12.37 | 0.08 | -11.99 | 0.08 | -12.90 | 0.08 | -12.20 | 0.08 | -11.71 | 0.08 | -12.60 | 0.08 | -12.99 | 0.08 | -11.78 | 0.08 | -11.71 | 0.08 | -12.05 | 0.08 |
| | Rh$_{70}$ | -12.88 | 0.08 | -12.46 | 0.08 | -12.14 | 0.08 | -12.96 | 0.08 | -12.30 | 0.08 | -11.84 | 0.08 | -12.70 | 0.08 | -12.23 | 0.08 | -11.86 | 0.08 | -11.83 | 0.08 | -12.12 | 0.08 |
| Ca | Lt$_{58}$ | -12.75 | 0.06 | -12.04 | 0.06 | -11.77 | 0.06 | -12.61 | 0.06 | -12.00 | 0.06 | -11.83 | 0.06 | -12.47 | 0.06 | -12.09 | 0.06 | -11.63 | 0.06 | -11.42 | 0.06 | -11.86 | 0.06 |
| | Tr$_{62}$ | -12.83 | 0.06 | -12.27 | 0.06 | -11.91 | 0.06 | -12.82 | 0.06 | -12.19 | 0.06 | -11.87 | 0.06 | -12.52 | 0.06 | -12.24 | 0.06 | -11.73 | 0.06 | -11.70 | 0.06 | -12.04 | 0.06 |
| | Tr$_{66}$ | -12.88 | 0.06 | -12.37 | 0.06 | -12.10 | 0.06 | -12.85 | 0.06 | -12.32 | 0.06 | -11.93 | 0.06 | -12.58 | 0.06 | -12.30 | 0.06 | -11.80 | 0.06 | -11.77 | 0.06 | -12.10 | 0.06 |
| | Rh$_{70}$ | -12.94 | 0.06 | -12.48 | 0.06 | -12.24 | 0.06 | -12.86 | 0.06 | -12.43 | 0.06 | -12.01 | 0.06 | -12.68 | 0.06 | -12.31 | 0.06 | -11.84 | 0.06 | -11.83 | 0.06 | -12.16 | 0.06 |
| K | Lt$_{58}$ | -12.71 | 0.09 | -11.99 | 0.09 | -11.89 | 0.09 | -12.45 | 0.09 | -12.07 | 0.09 | -11.60 | 0.09 | -12.61 | 0.09 | -11.98 | 0.09 | -11.64 | 0.09 | -11.51 | 0.09 | -12.04 | 0.09 |
| | Tr$_{62}$ | -12.83 | 0.09 | -12.20 | 0.09 | -11.93 | 0.09 | -12.75 | 0.09 | -12.19 | 0.09 | -11.73 | 0.09 | -12.63 | 0.09 | -12.11 | 0.09 | -11.68 | 0.09 | -11.62 | 0.09 | -12.10 | 0.09 |
| | Tr$_{66}$ | -12.90 | 0.09 | -12.30 | 0.09 | -12.08 | 0.09 | -12.86 | 0.09 | -12.31 | 0.09 | -11.76 | 0.09 | -12.69 | 0.09 | -12.17 | 0.09 | -11.73 | 0.09 | -11.69 | 0.09 | -12.14 | 0.09 |
| | Rh$_{70}$ | -12.99 | 0.09 | -12.40 | 0.09 | -12.21 | 0.09 | -12.97 | 0.09 | -12.43 | 0.09 | -11.75 | 0.09 | -12.82 | 0.09 | -12.20 | 0.09 | -11.72 | 0.09 | -11.81 | 0.09 | -12.19 | 0.09 |

**Supplementary materials**

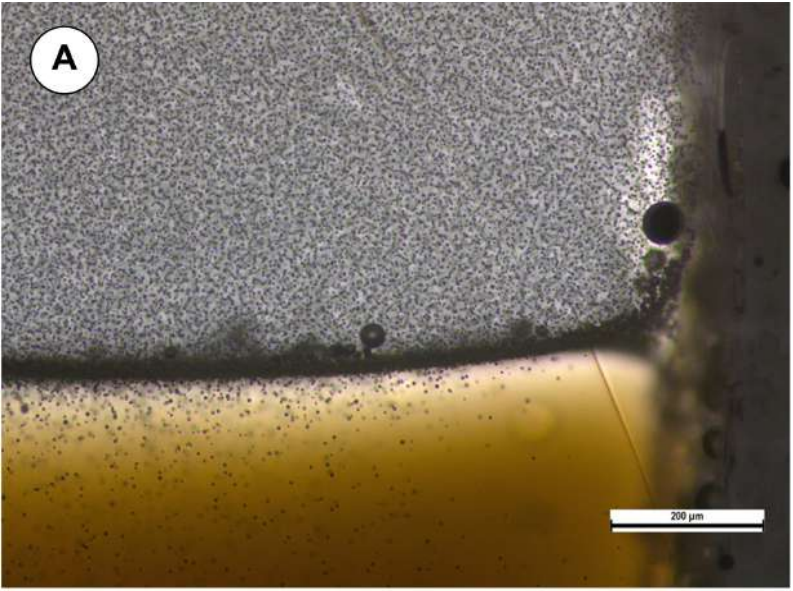

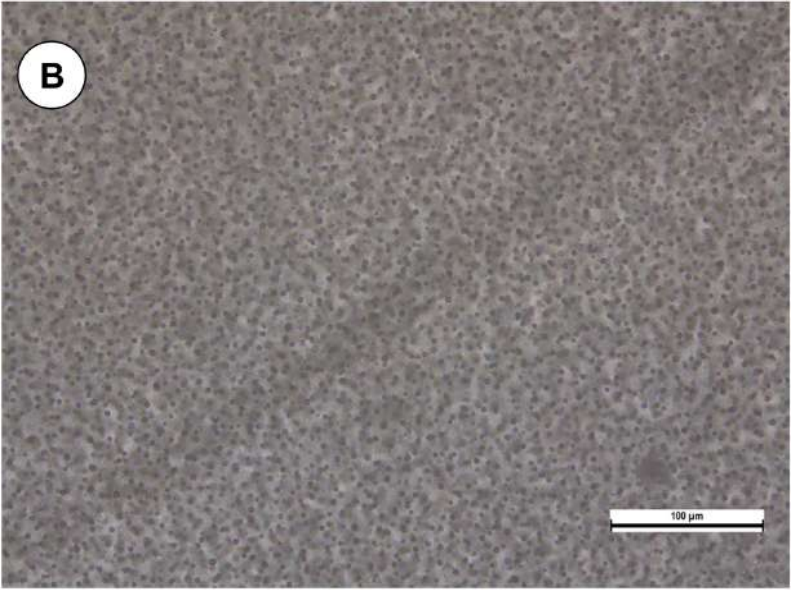

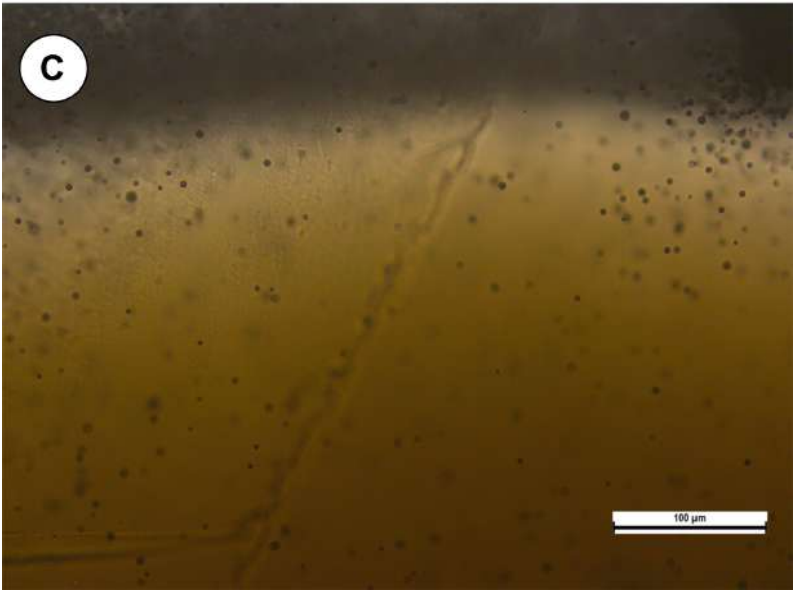

**Fig. S1.** Bubble distribution in low pressure experiment (P050-H1- 4; 50 Mpa, 1 wt.% $H_2O$ and 4 h.). (A) Microphotograph of the interfacial region of the experiment (thickness ca. 200 microns). (B) and (C) details of the bubble content of the rhyolitic and shoshonitic glasses, respectively.

**Fig. S2:** Concentration-distance profiles of all experiments measured by EPMA. The rhyolitic end-member is represented in the left half (negative distance values). Filled circles represent central profiles; empty circles represent offset profiles (when available). All analyses were normalized to a water-free basis. Zero distance is the estimated position of the interface.

**Experiment P050-H0-4**

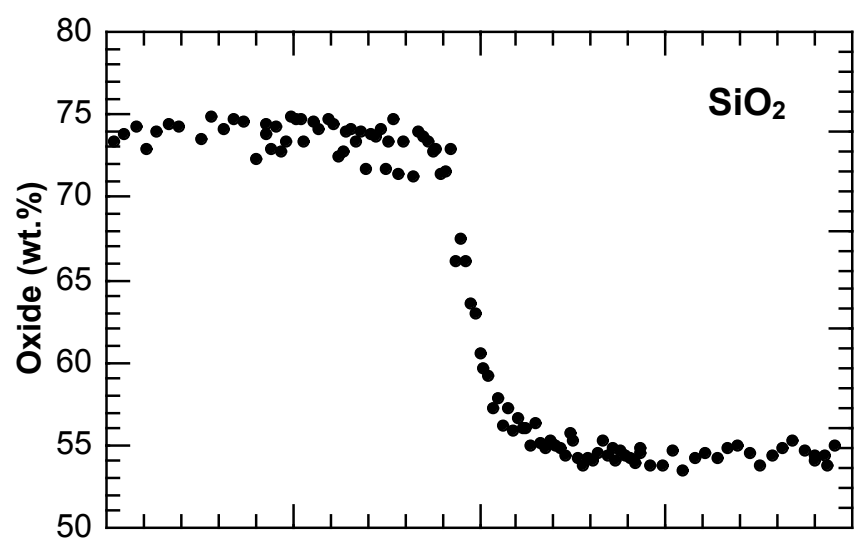
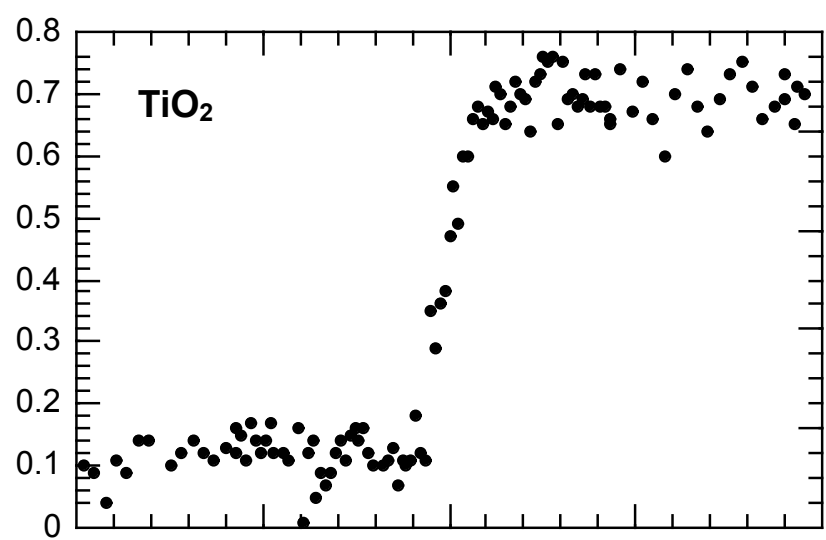
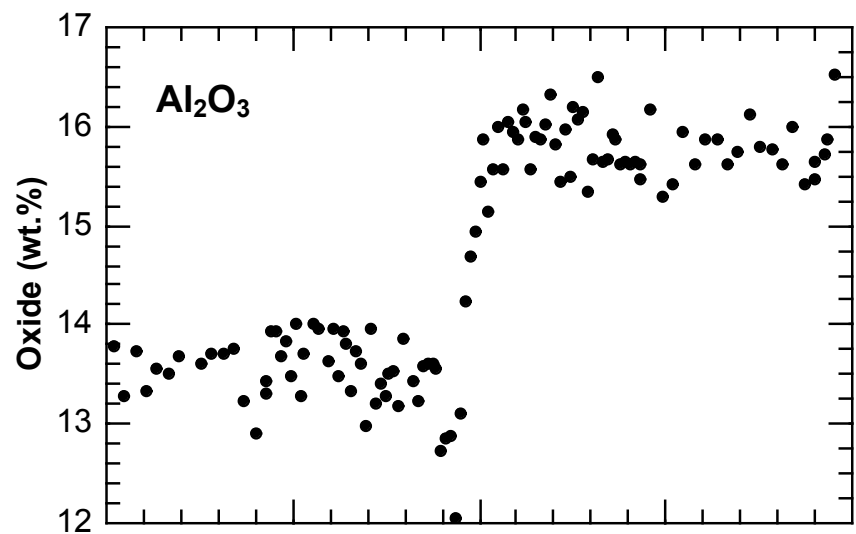
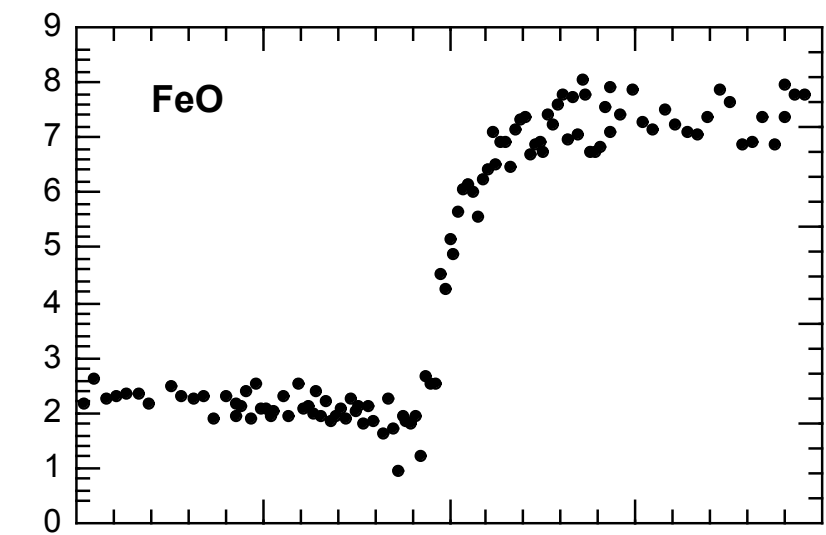
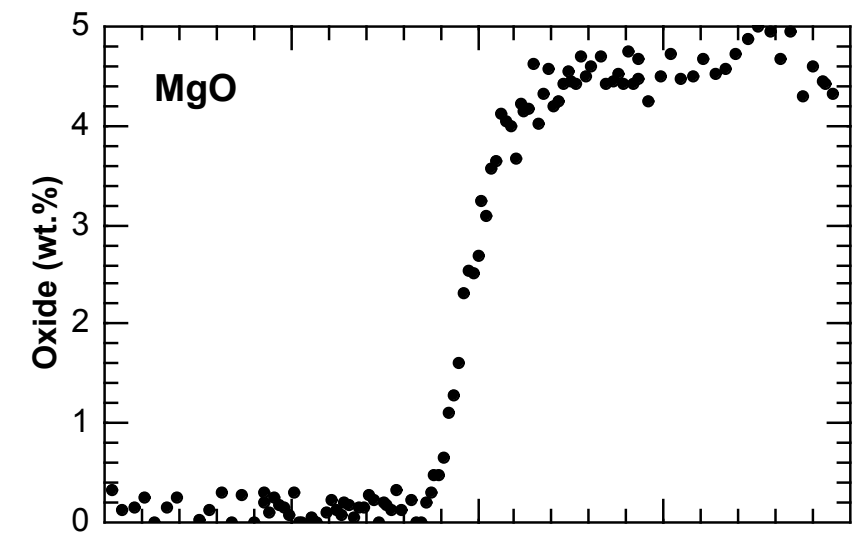
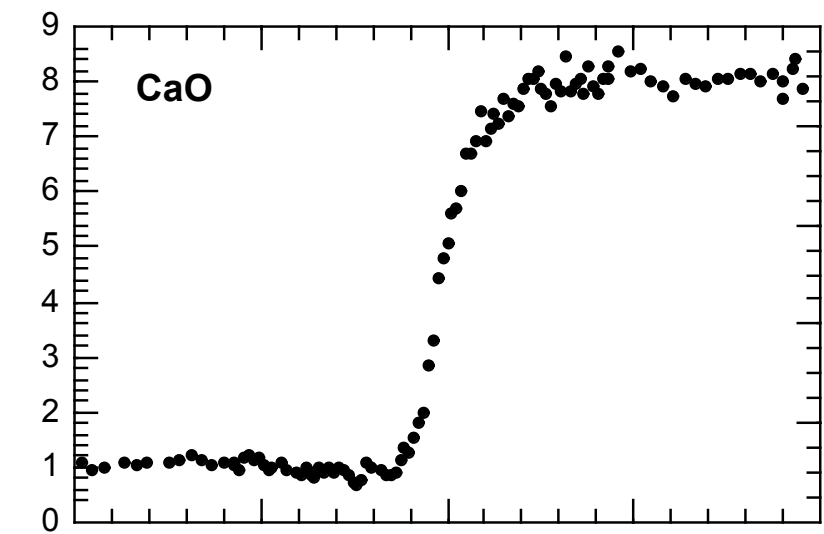
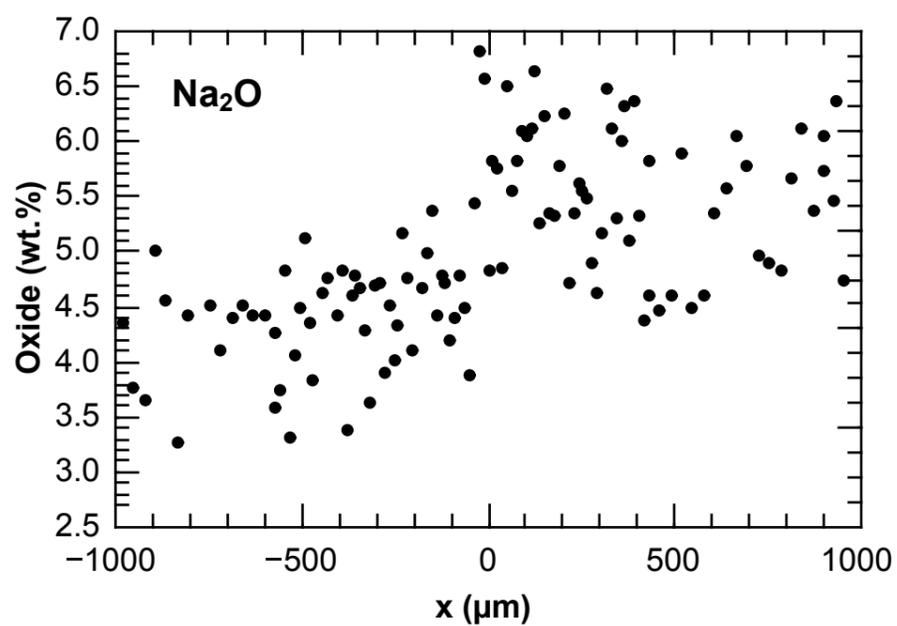
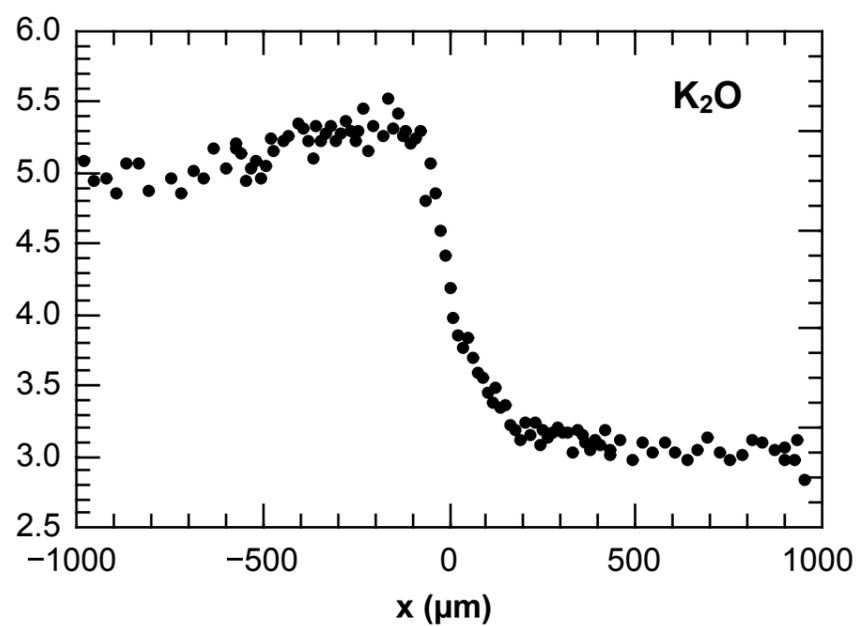

**Experiment P050-H1-4**

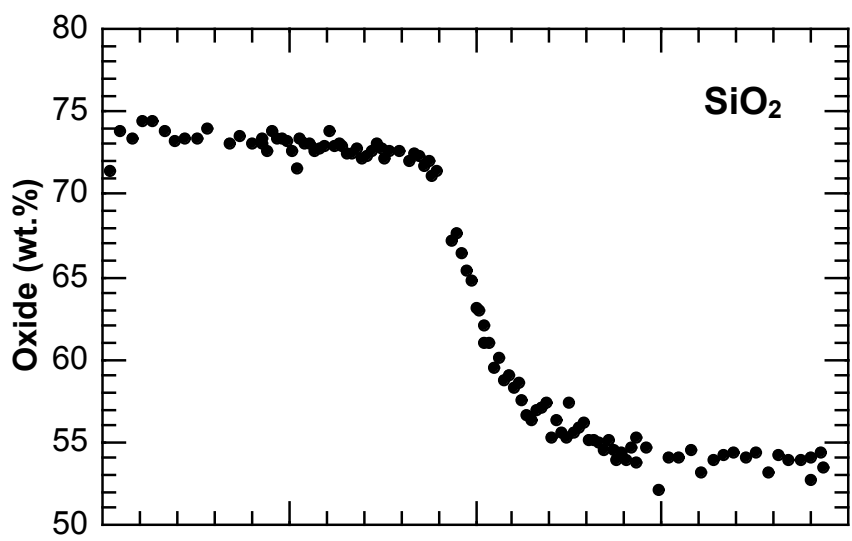
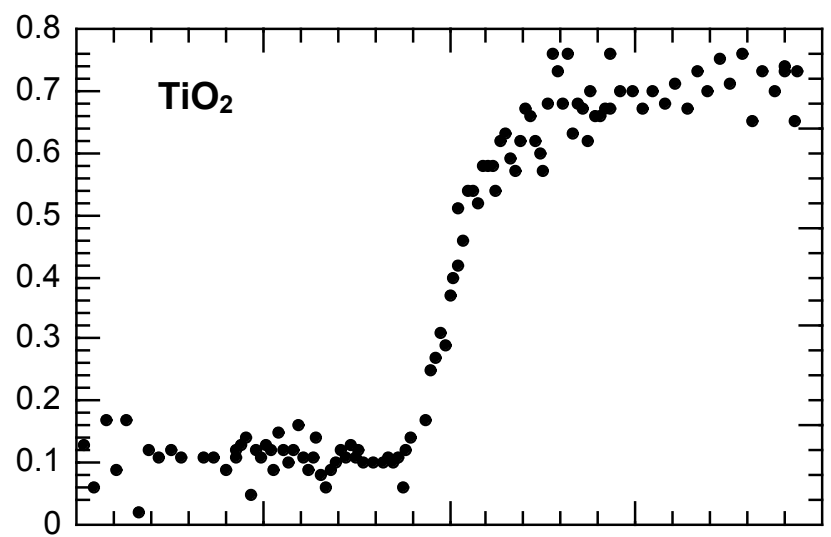
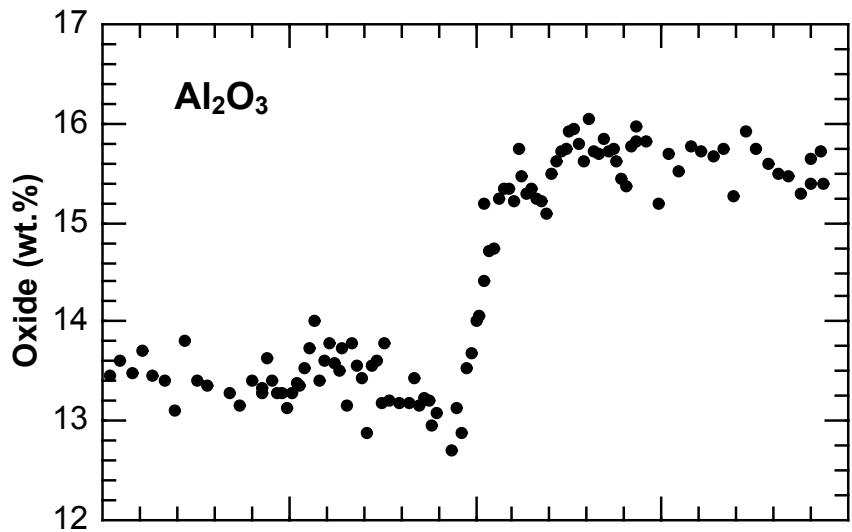
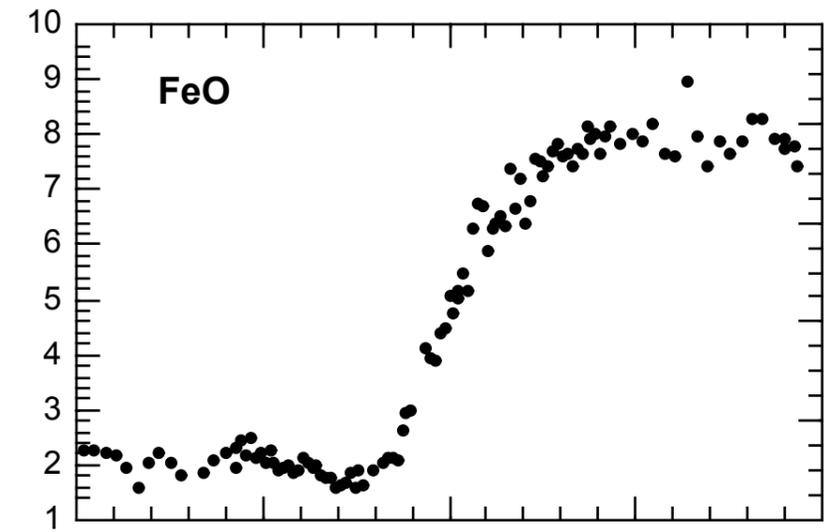
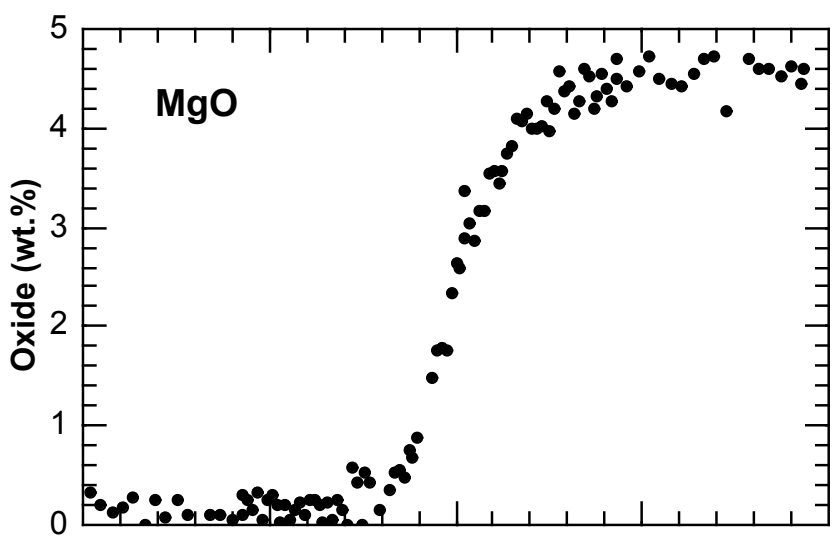
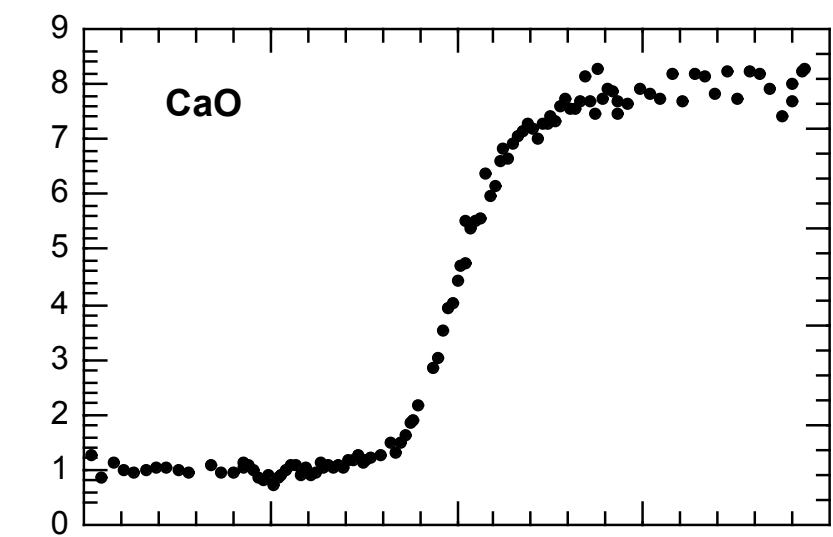
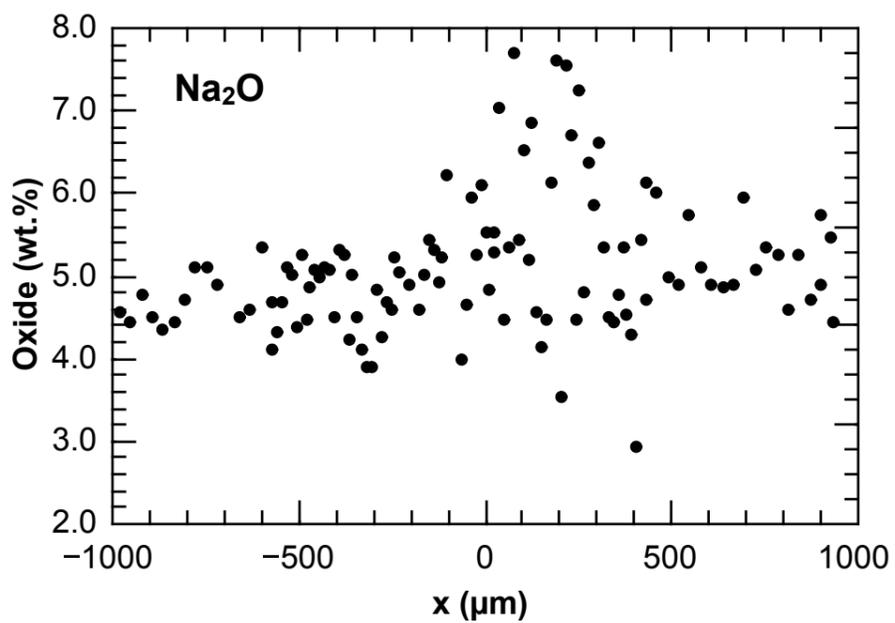
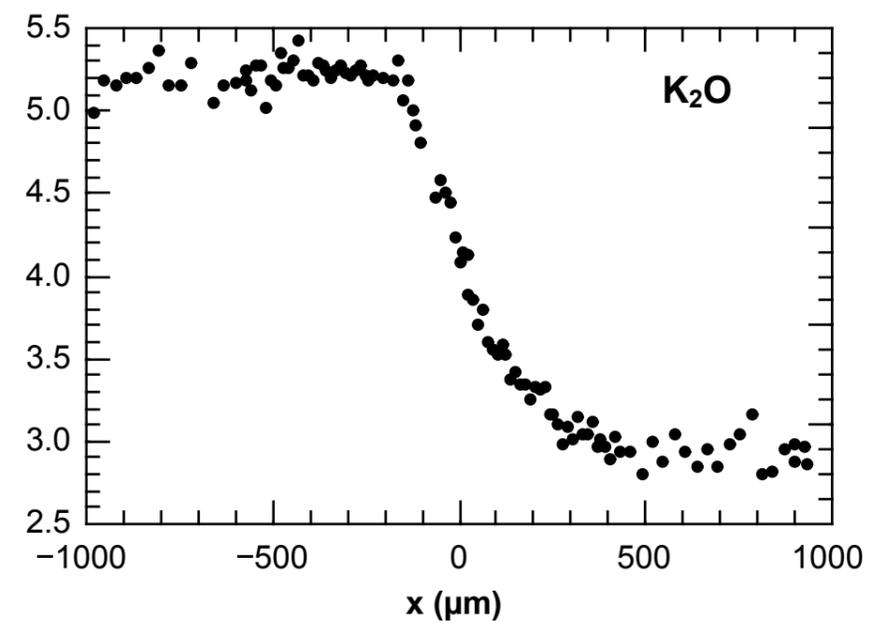

**Experiment P050-H2-4**

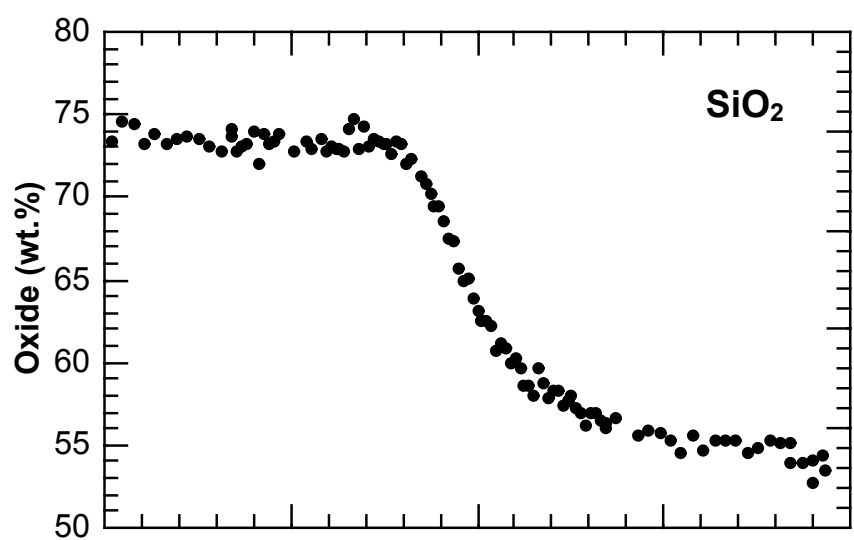
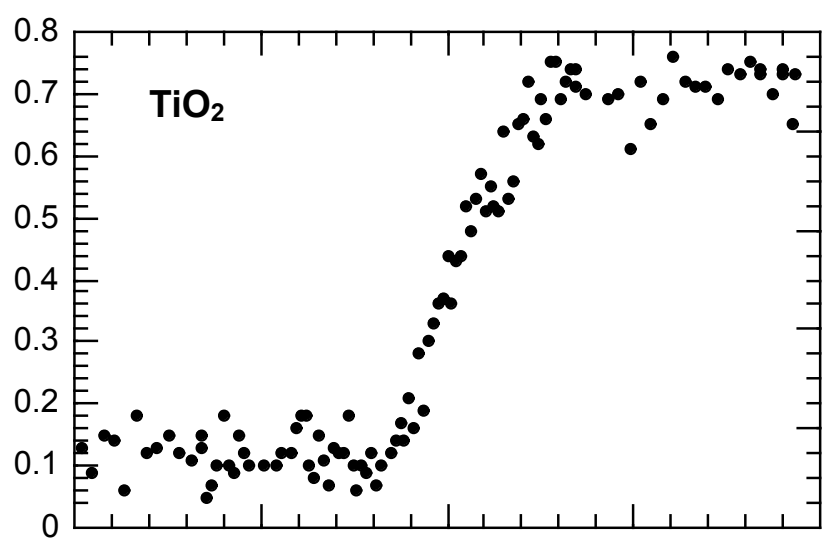
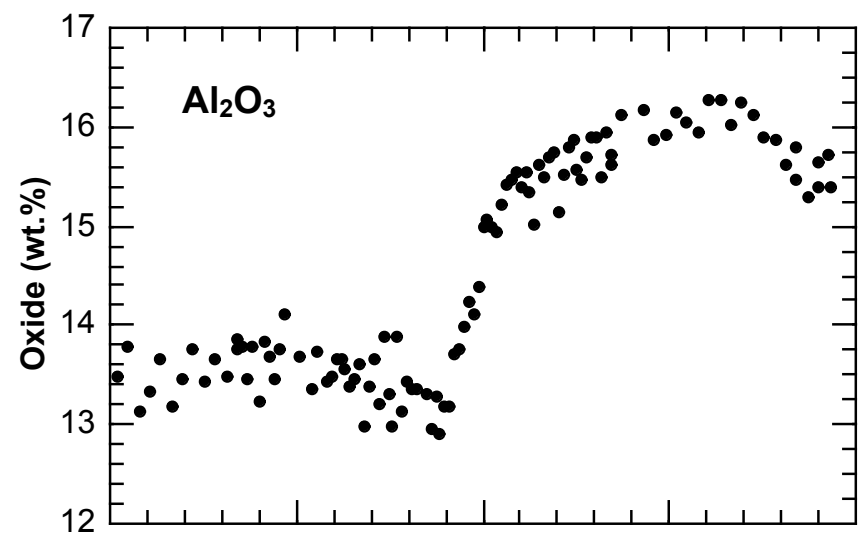
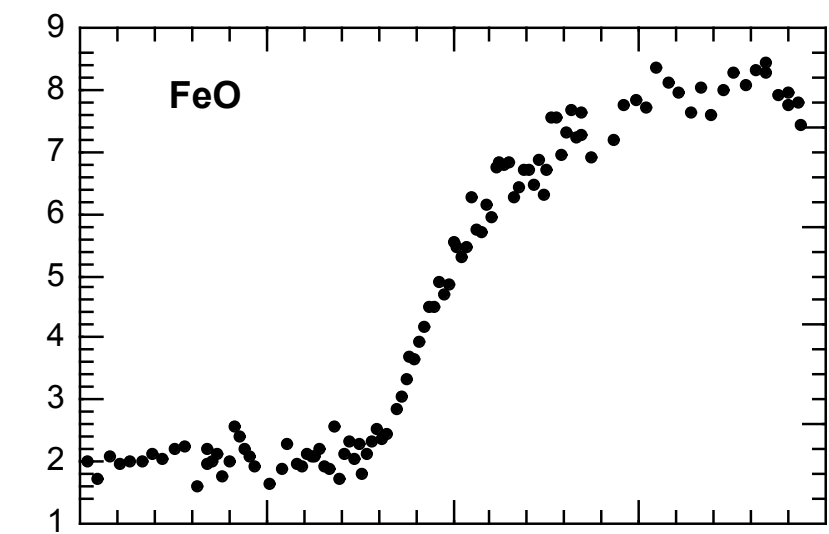
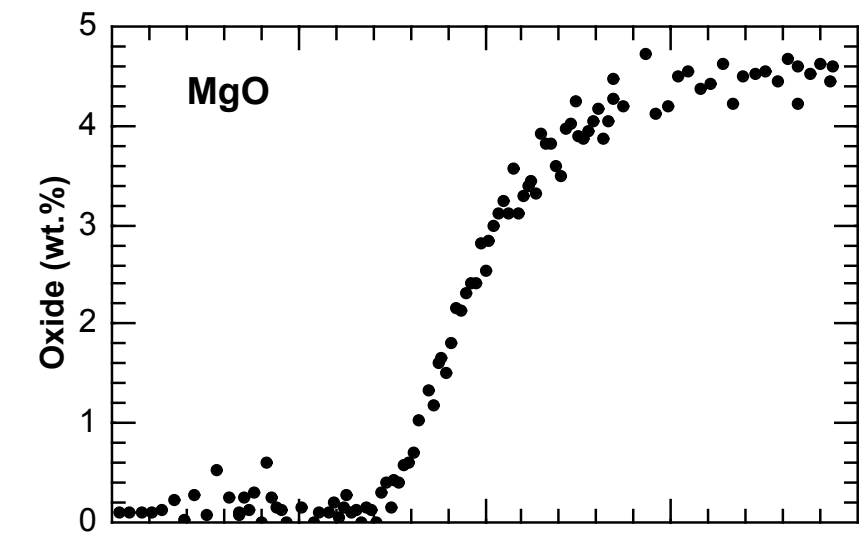
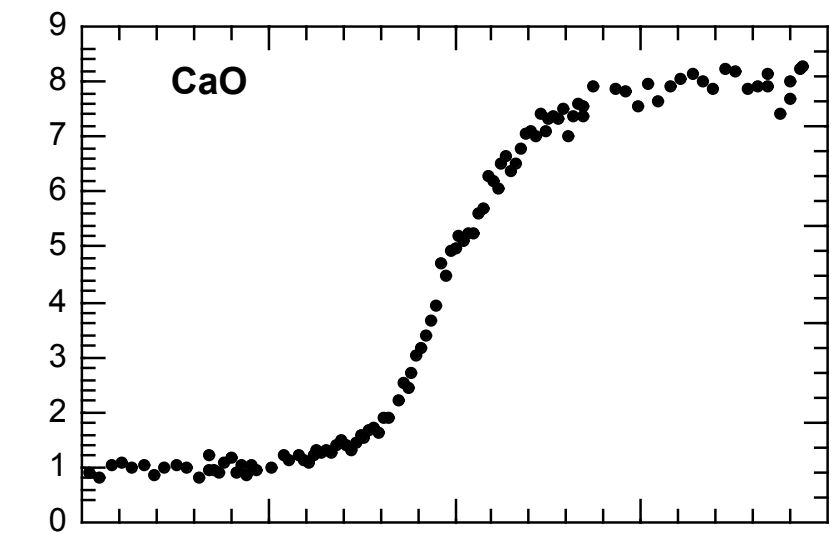
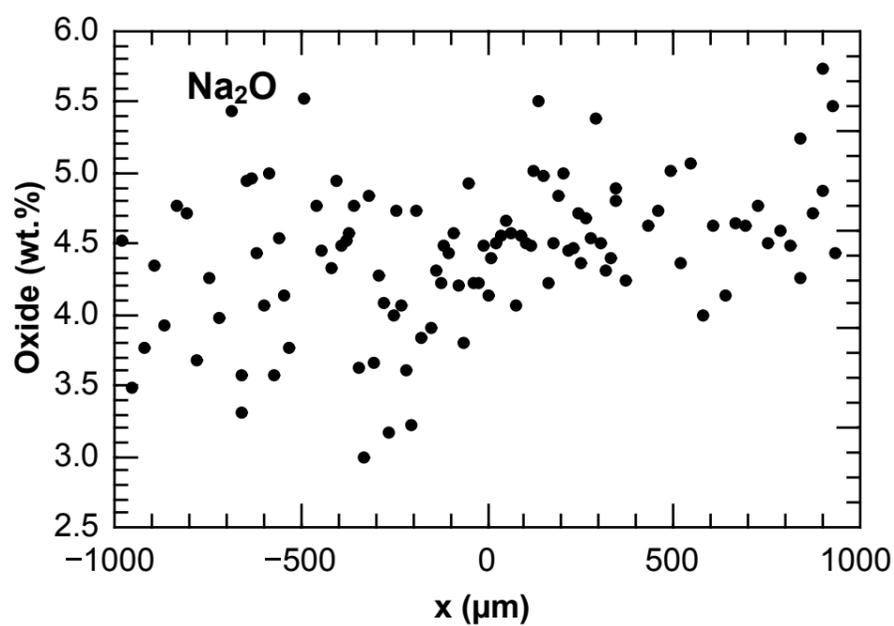
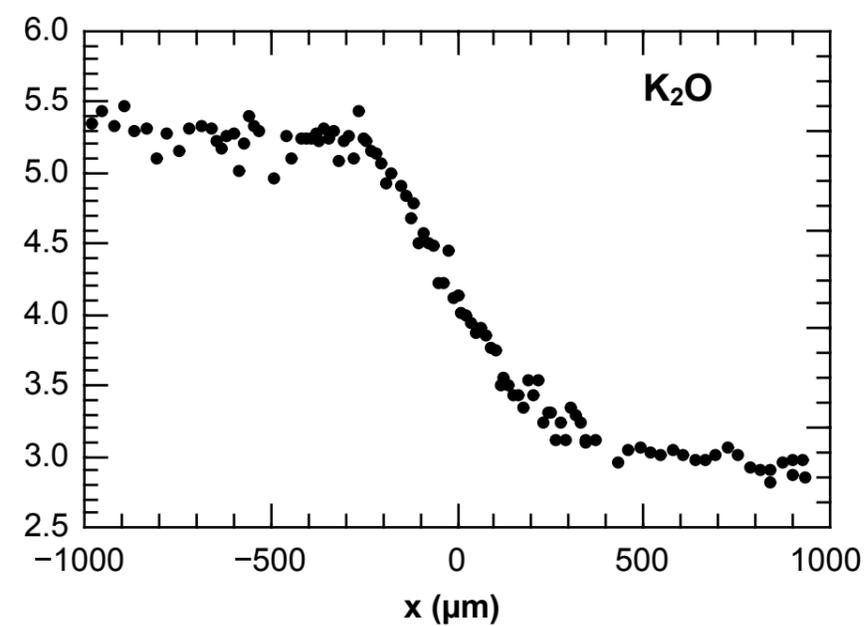

**Experiment P100-H0-4**

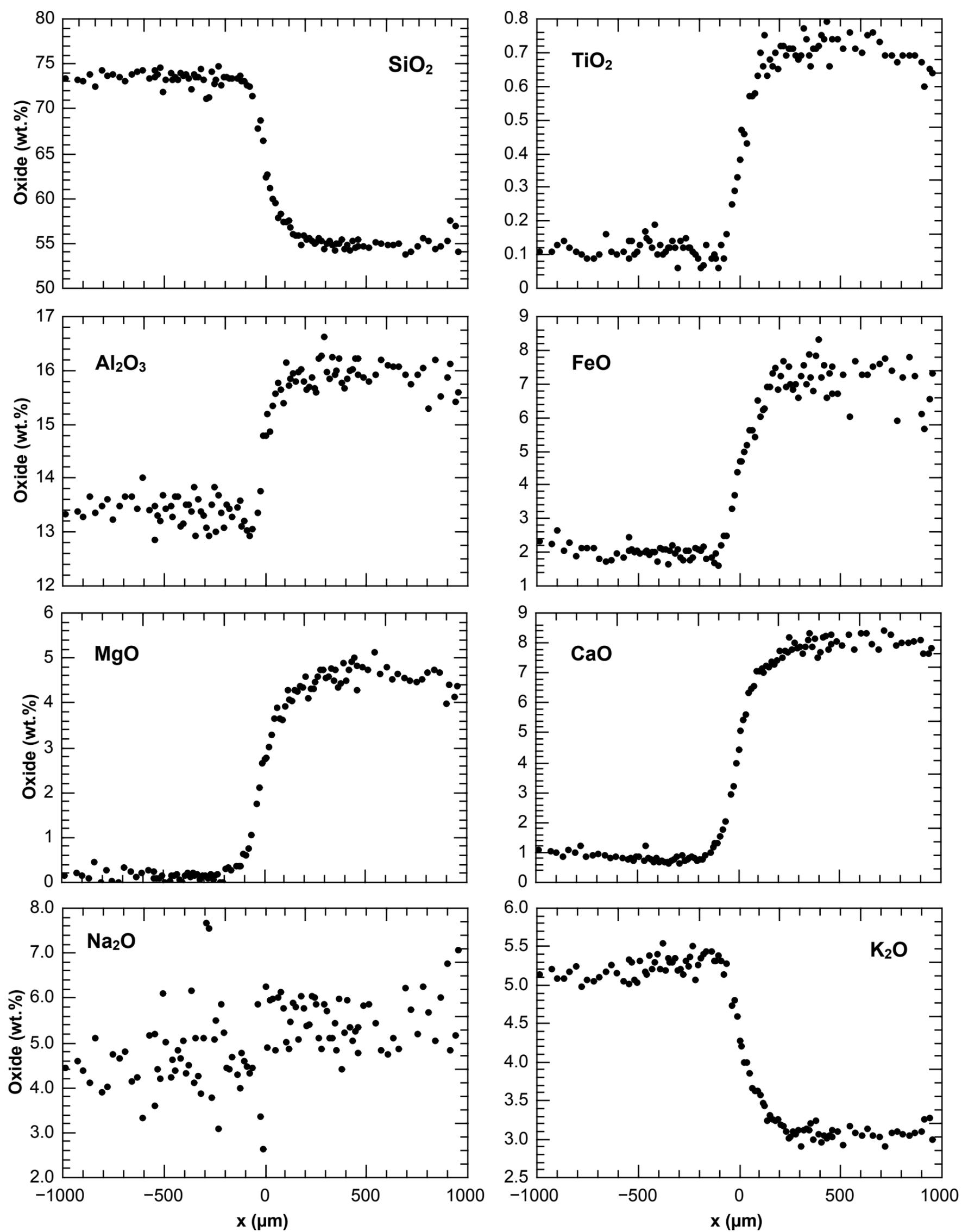

**Experiment P100-H1-4**

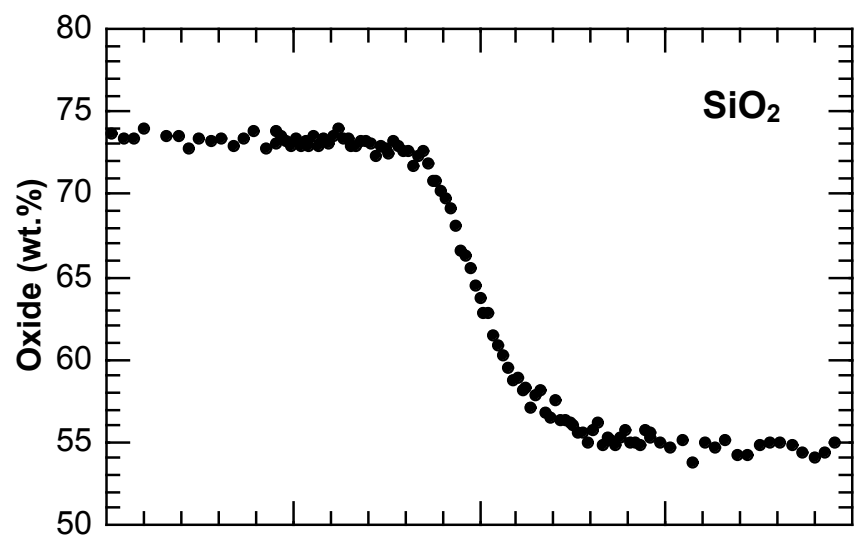
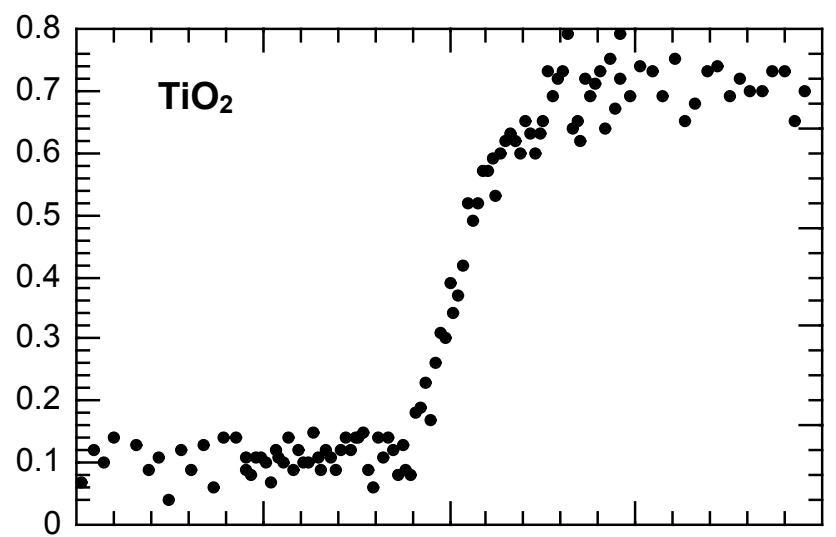
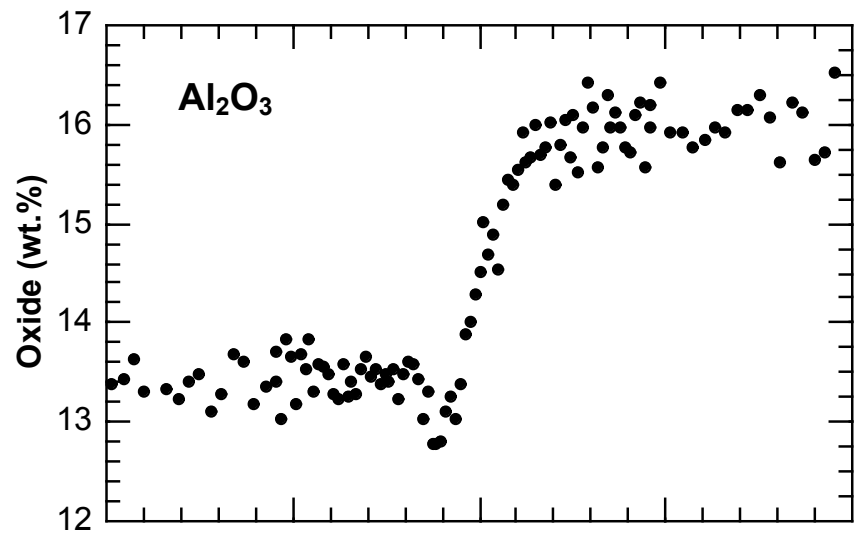
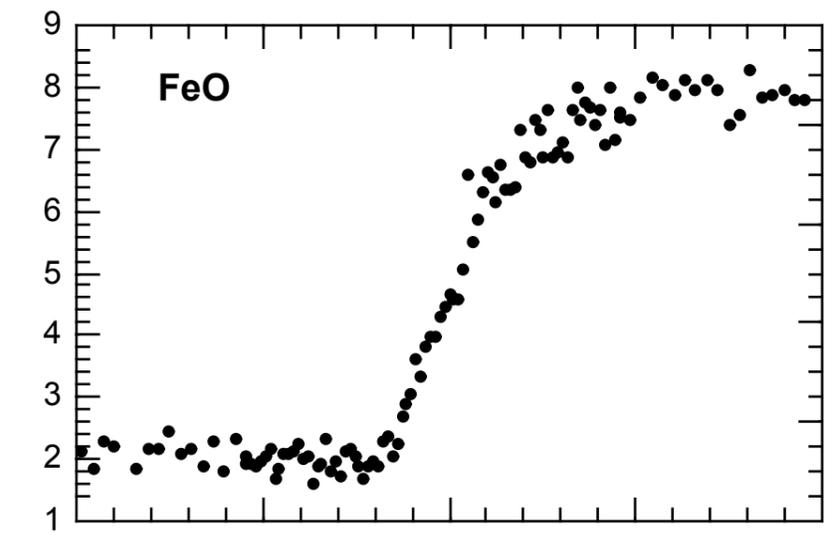
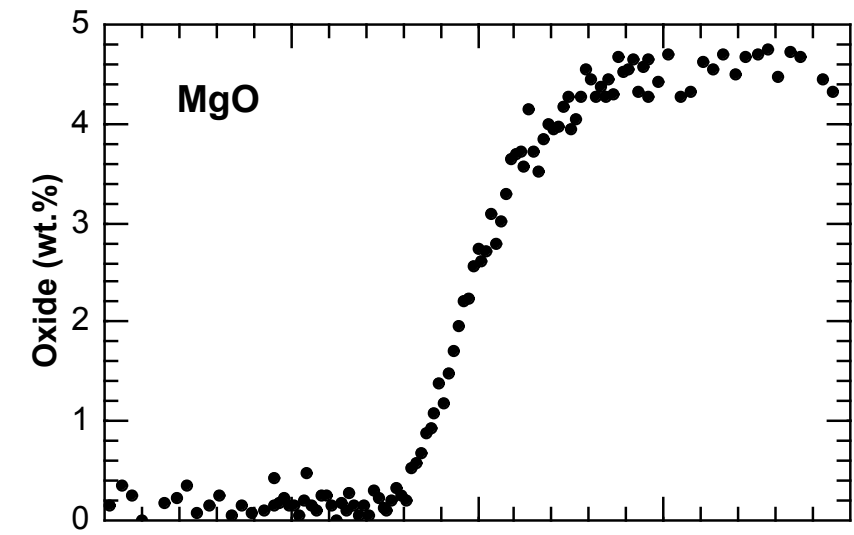
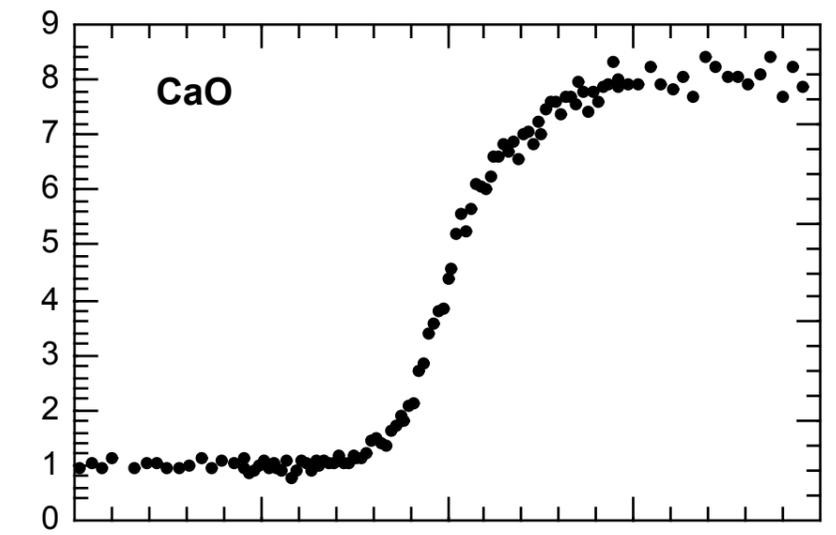
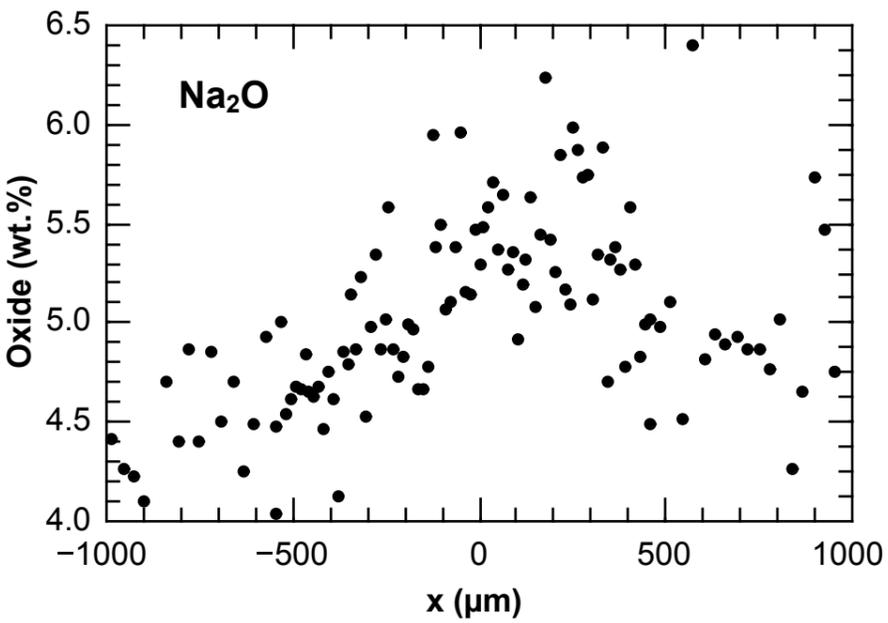
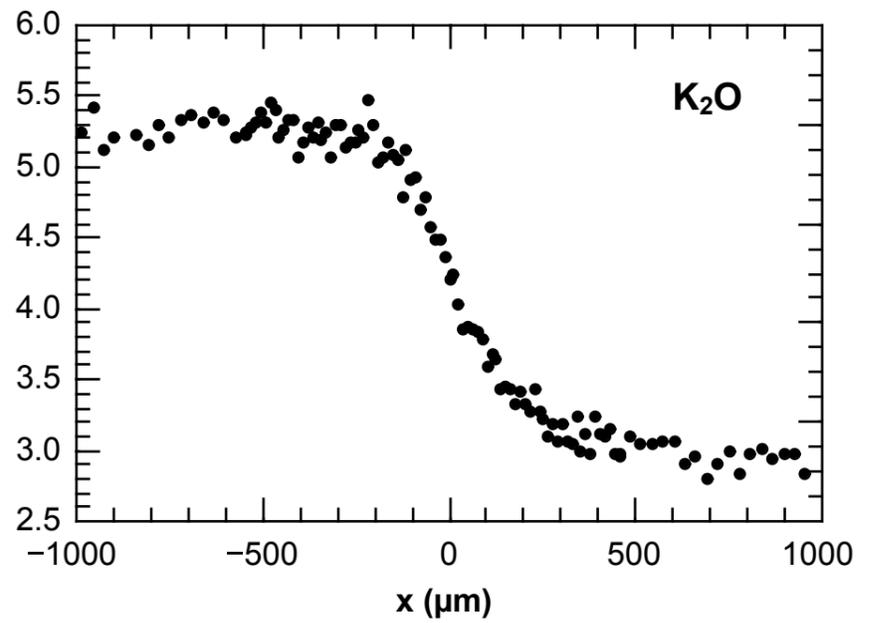

# Experiment P100-H2-4

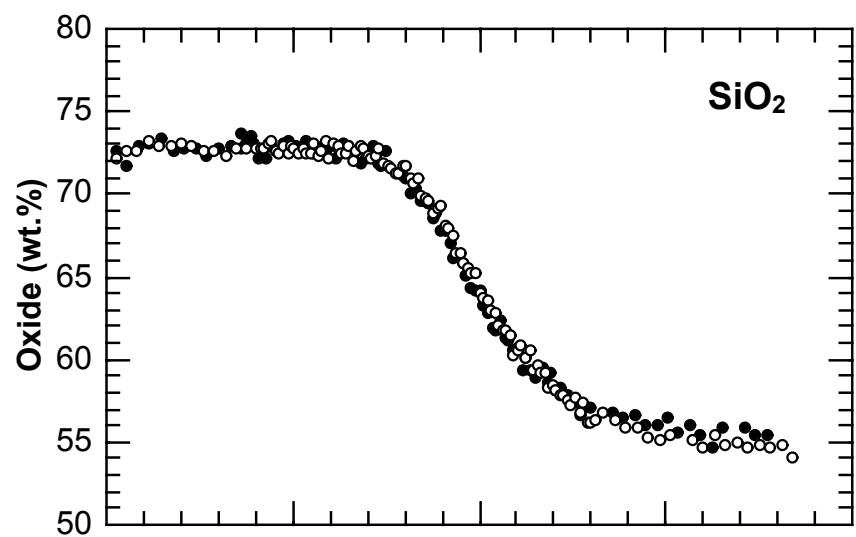
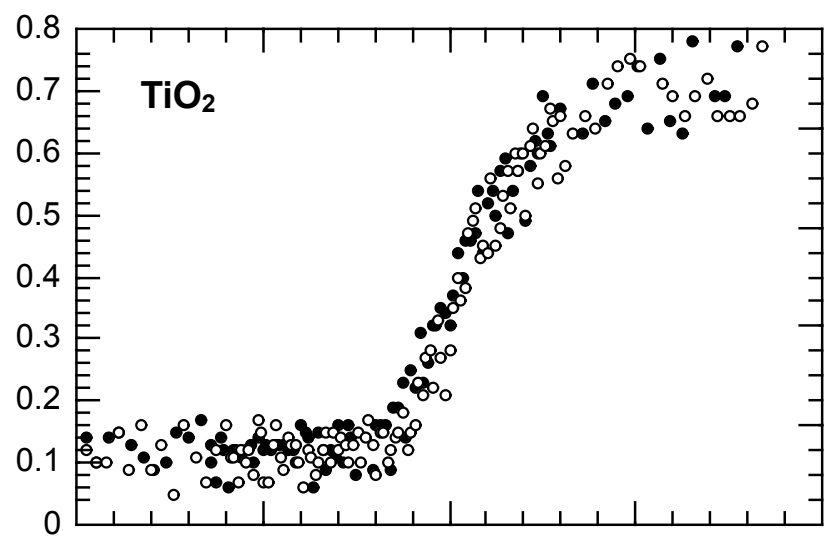
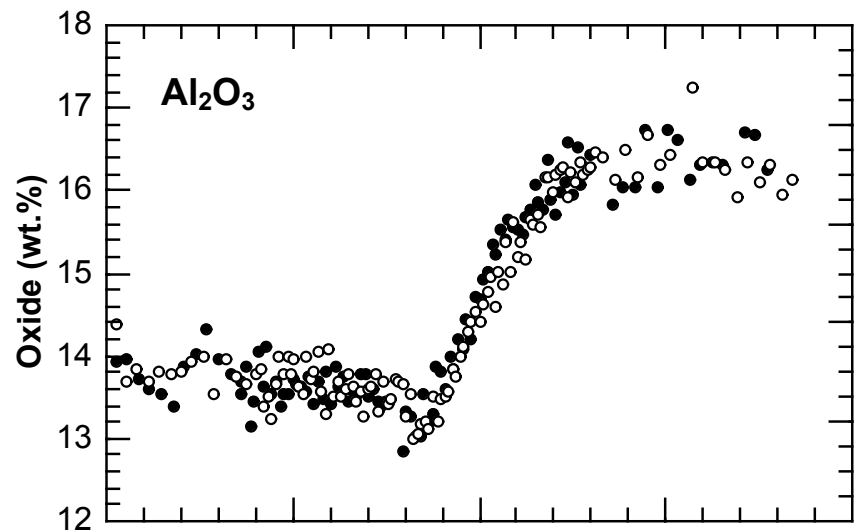
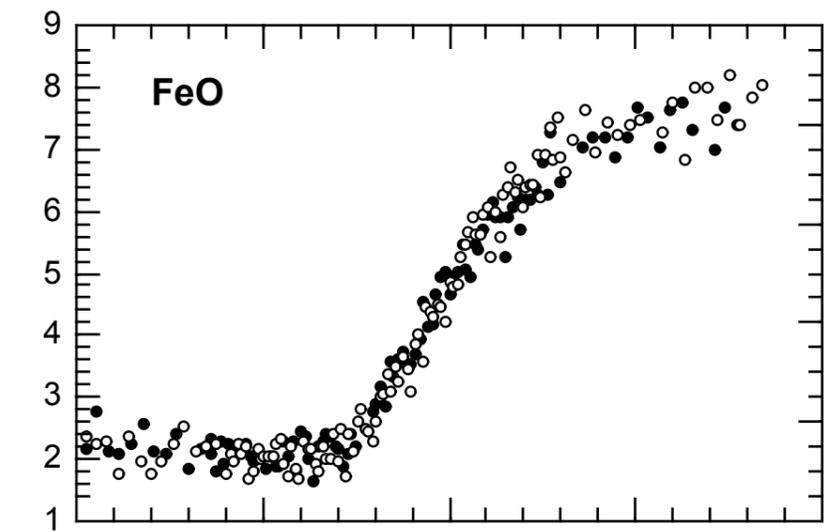
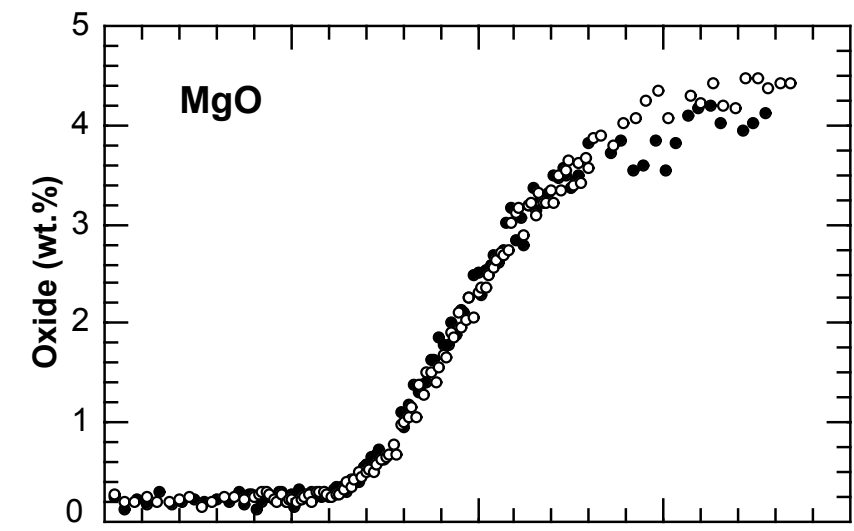
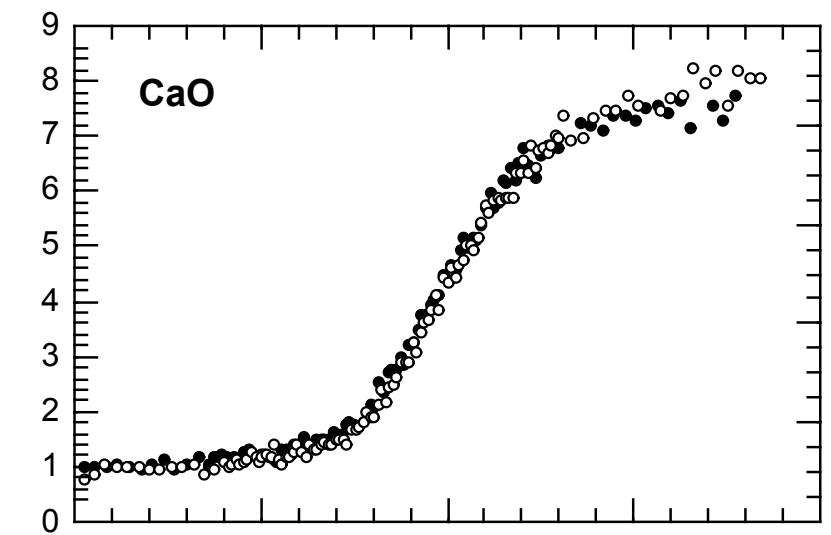
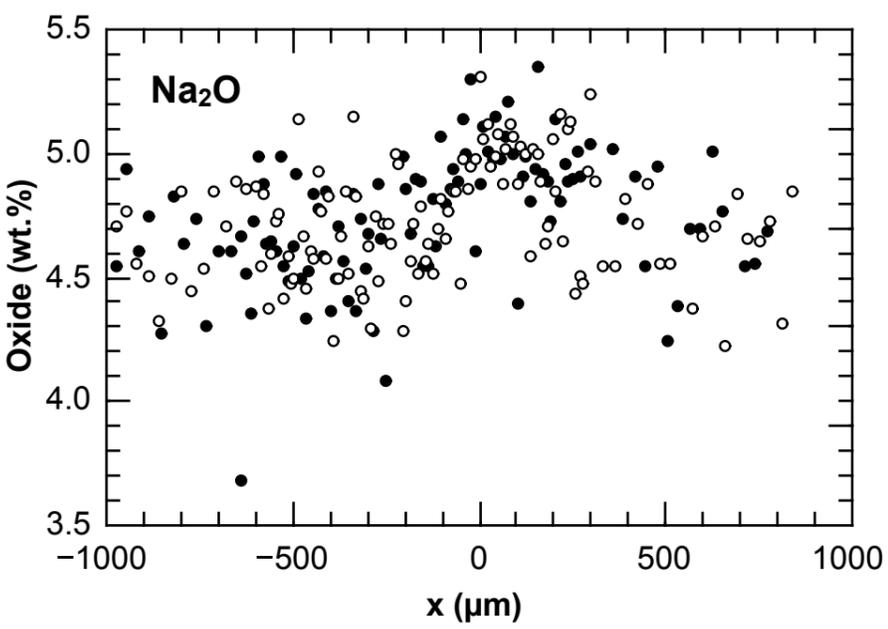
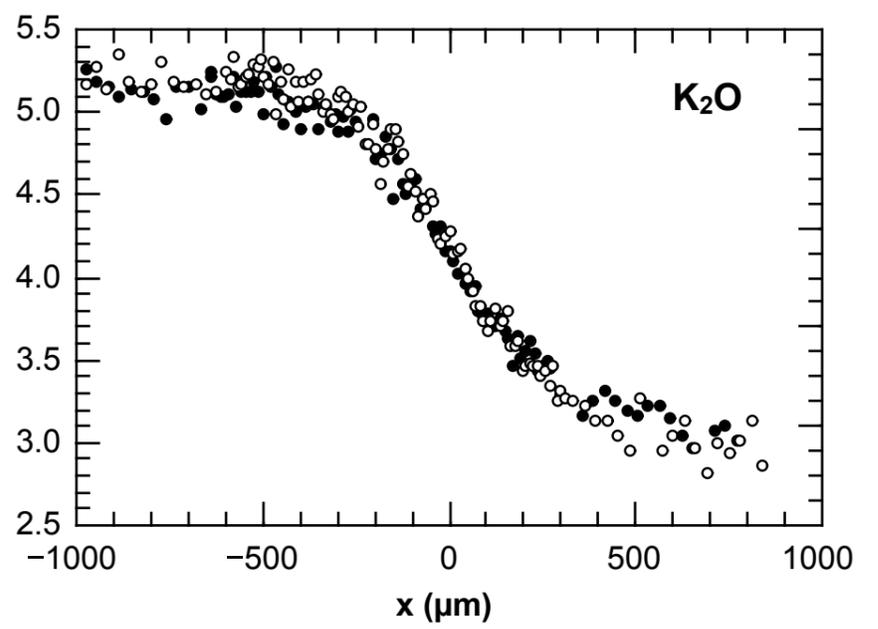

**Experiment P300-H0-1**

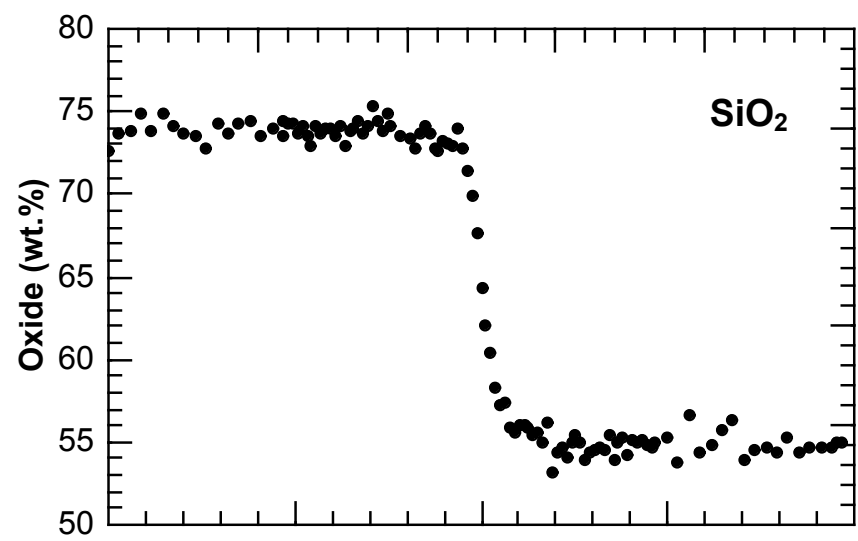
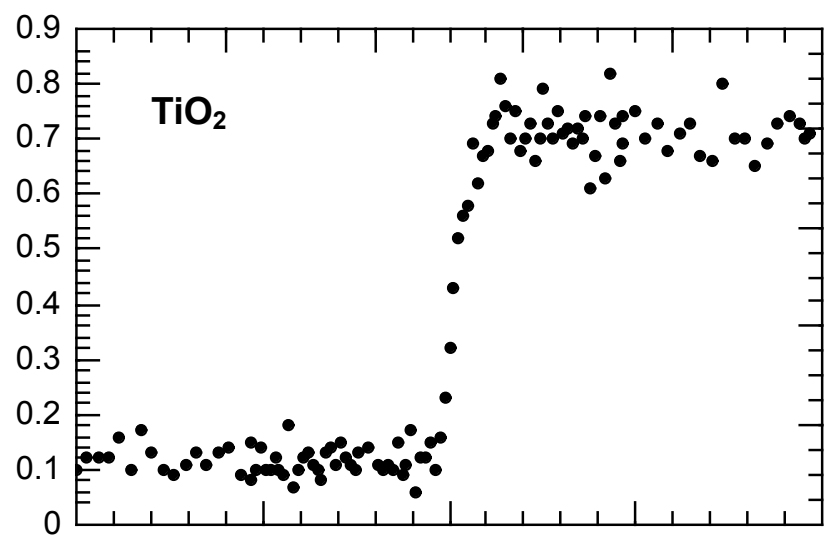
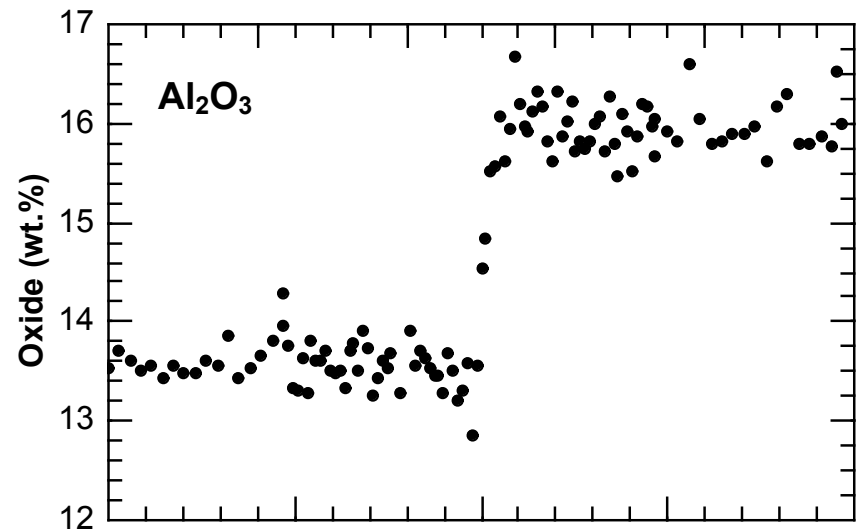
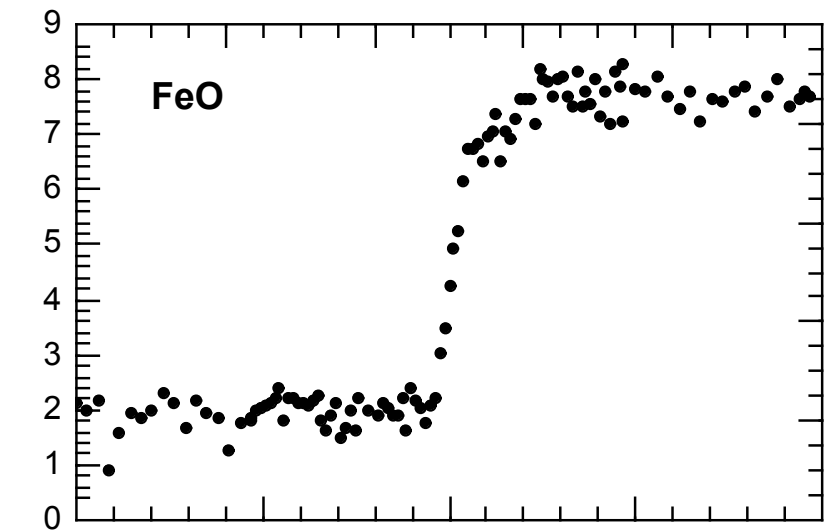
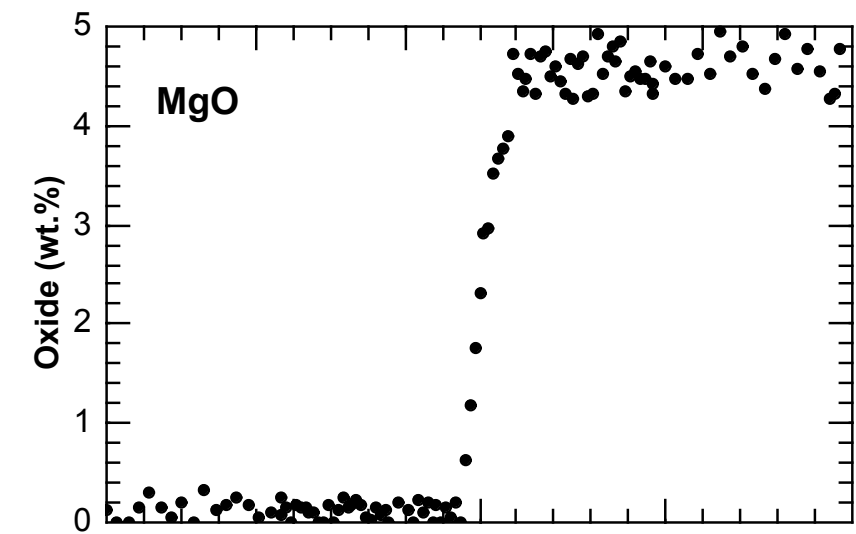
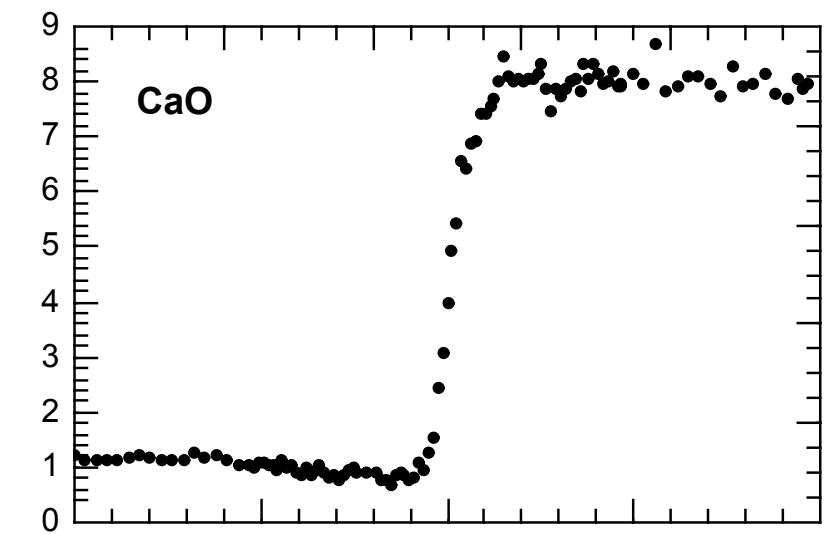
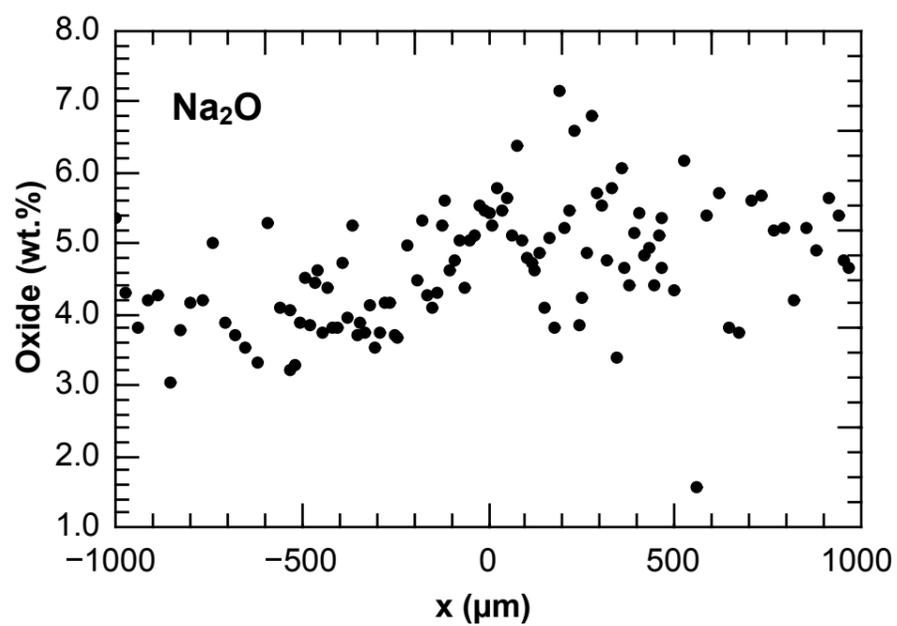
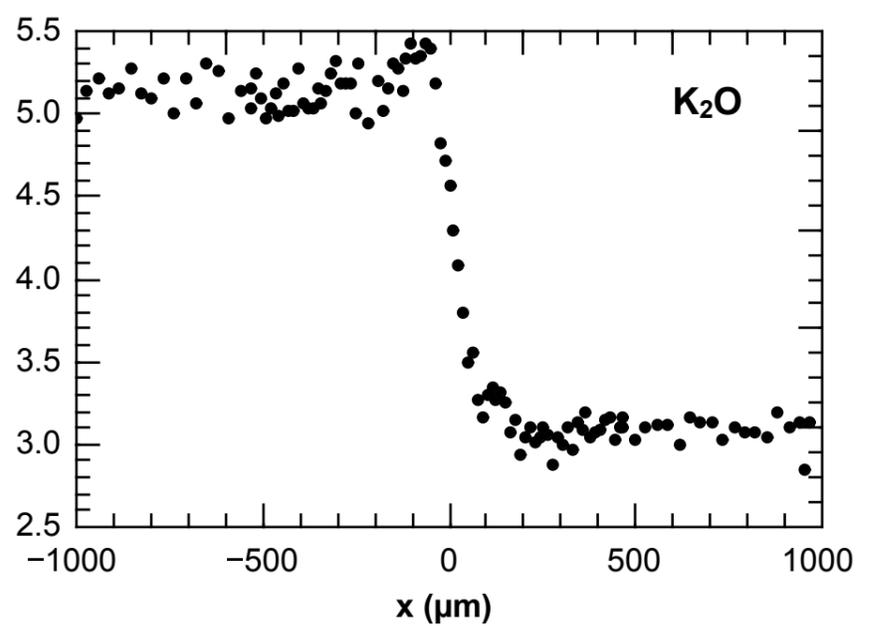

**Experiment P300-H0-4**

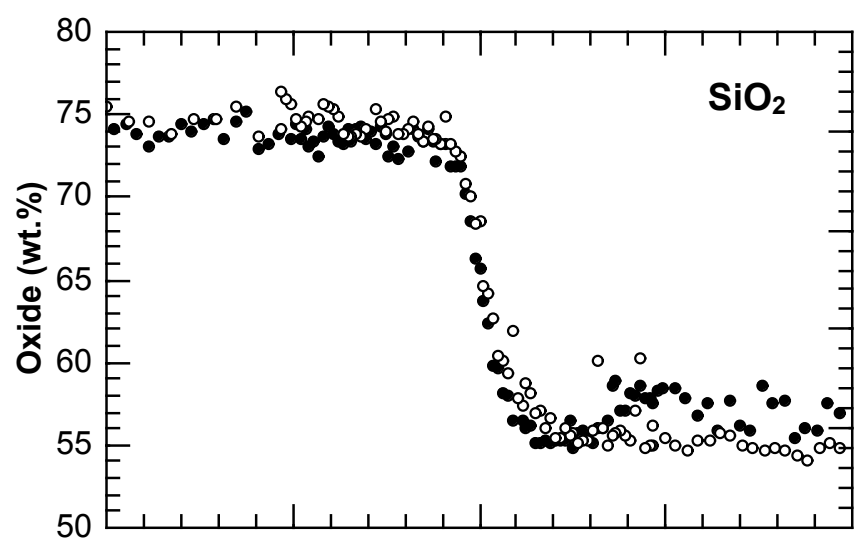
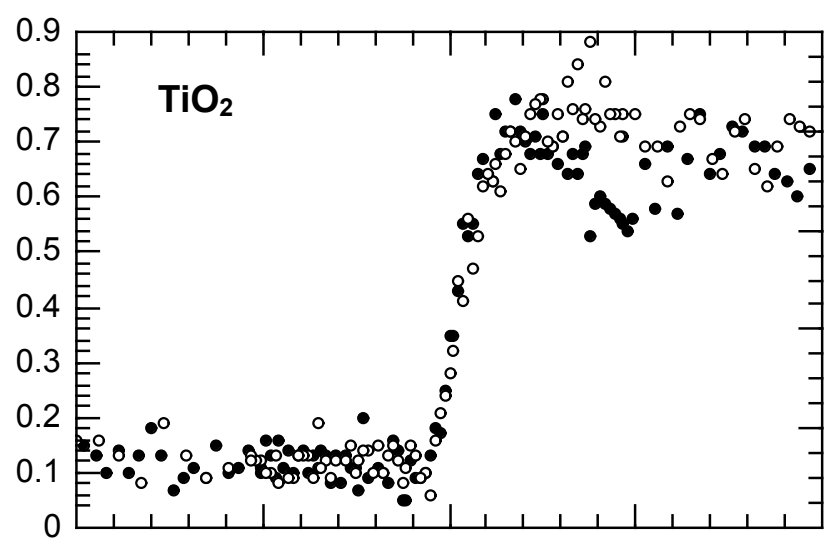
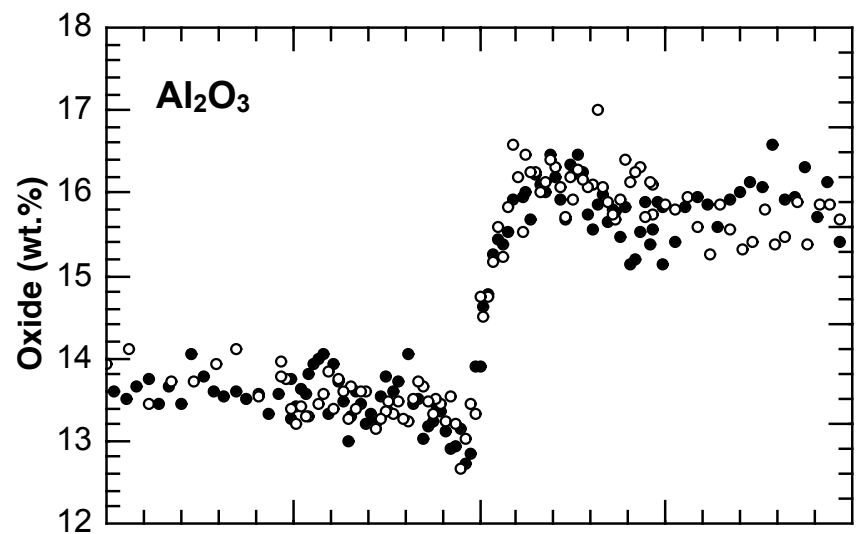
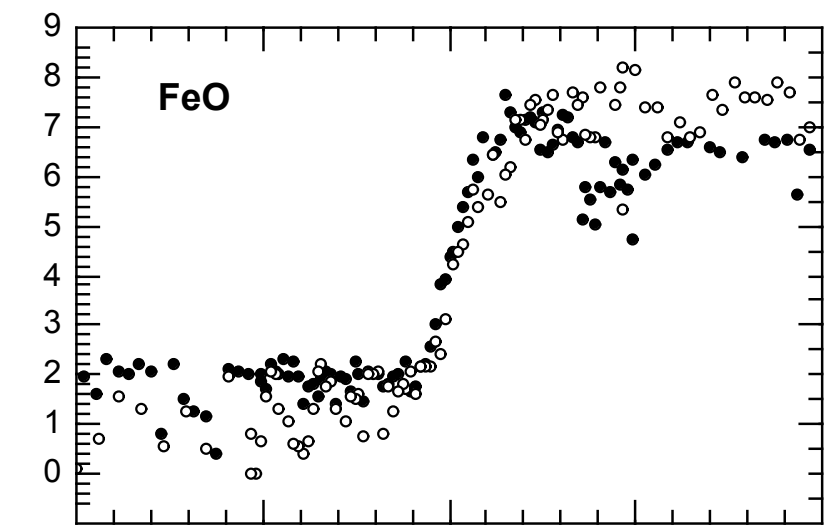
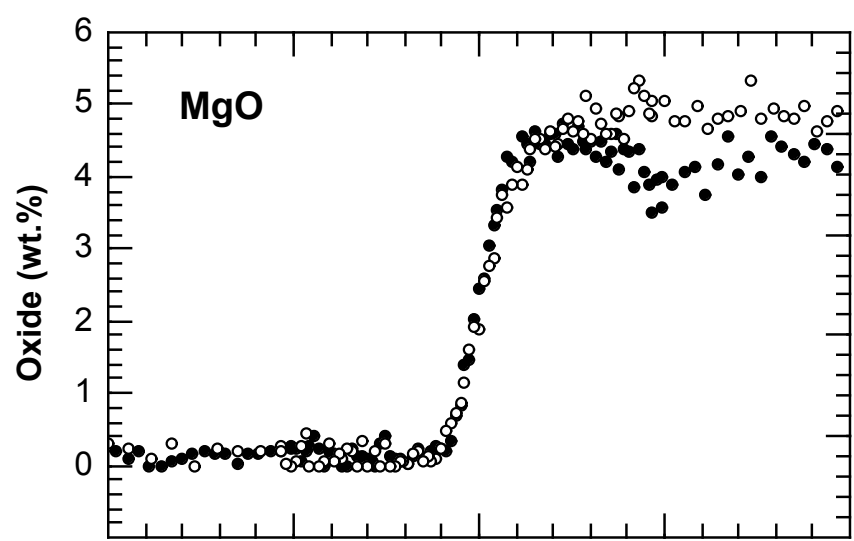
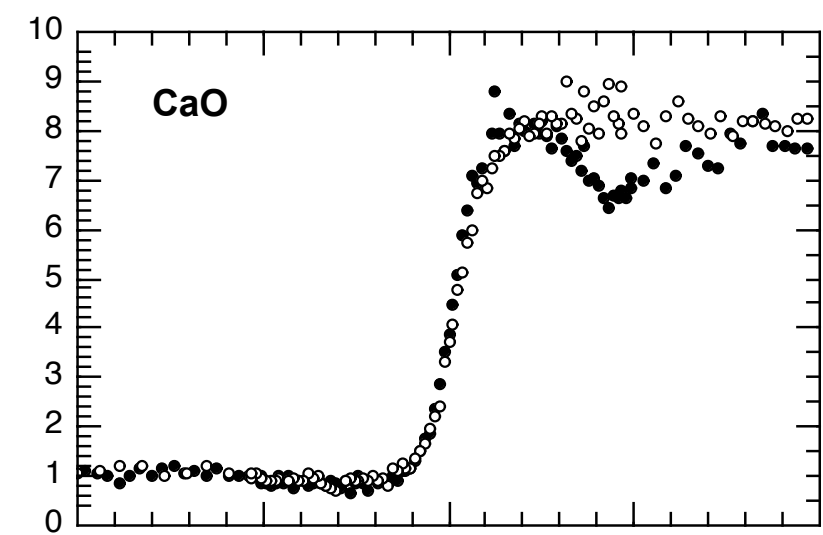
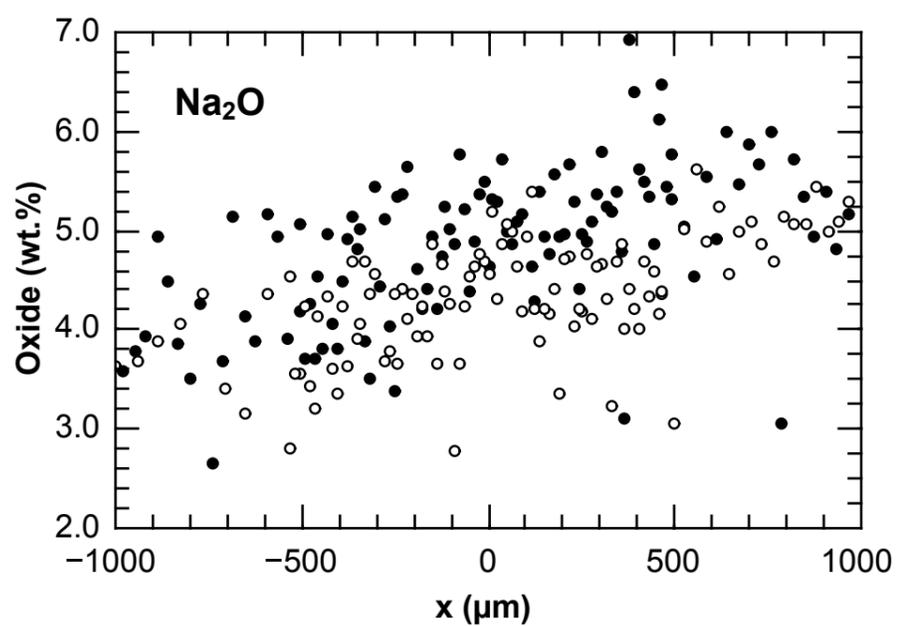
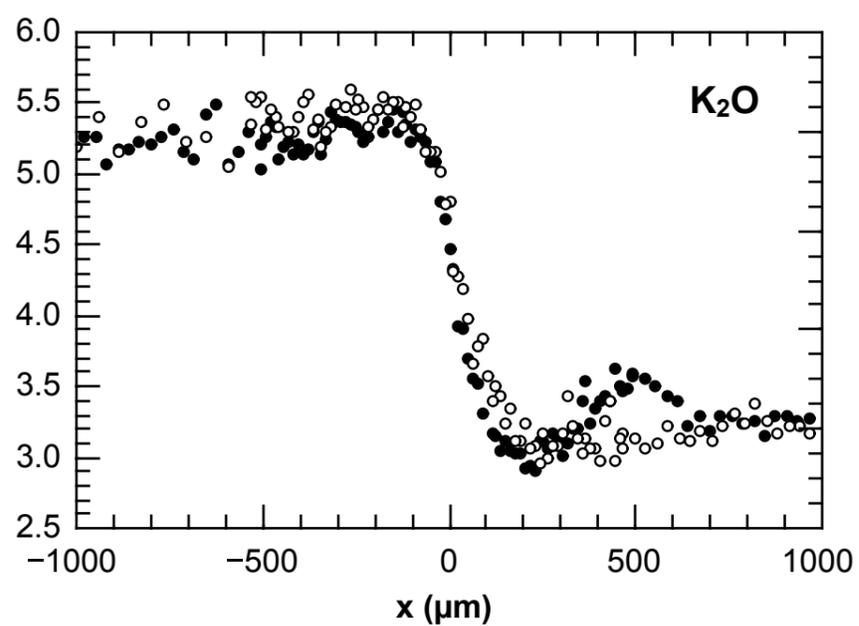

**Experiment P300-H1-4**

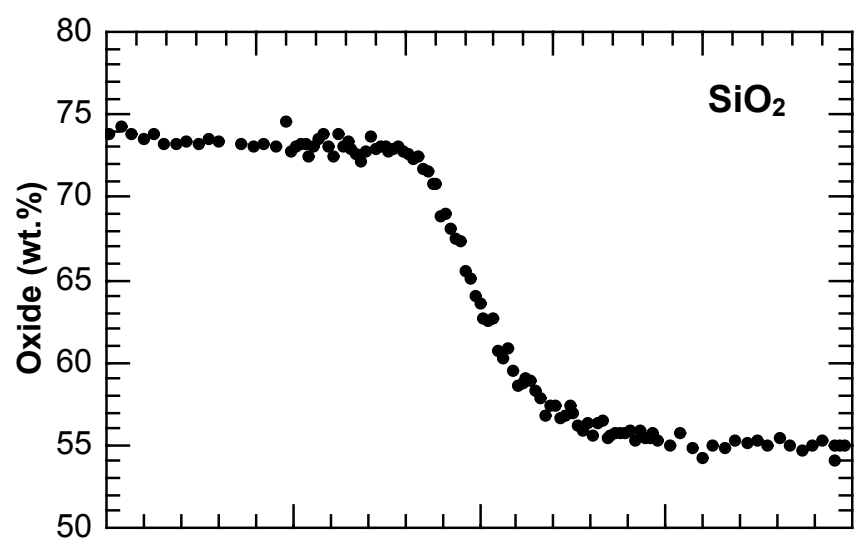
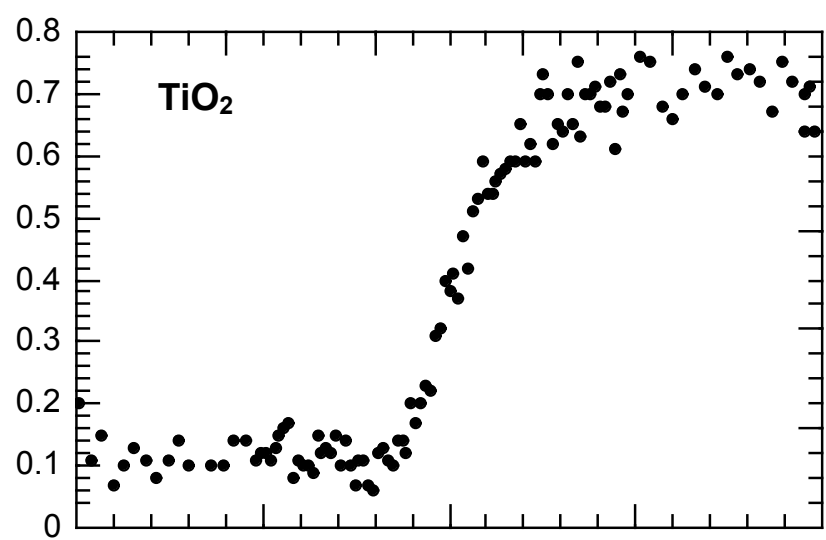
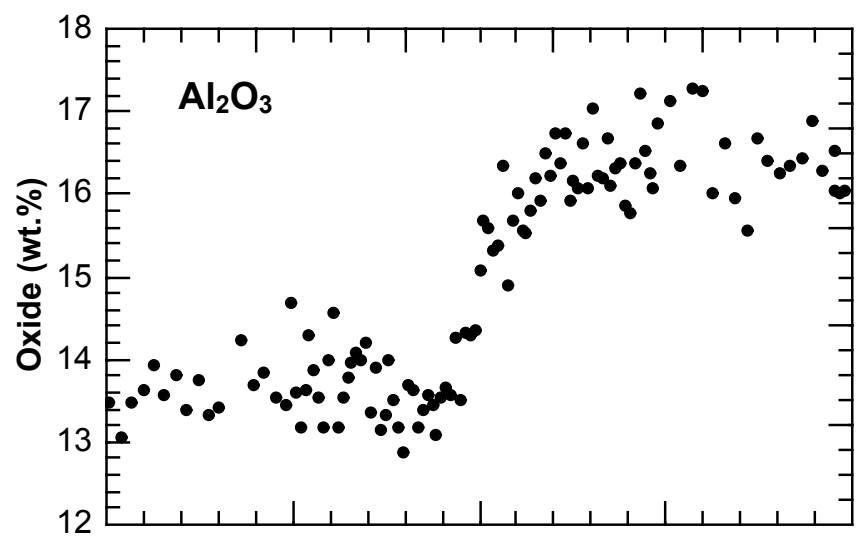
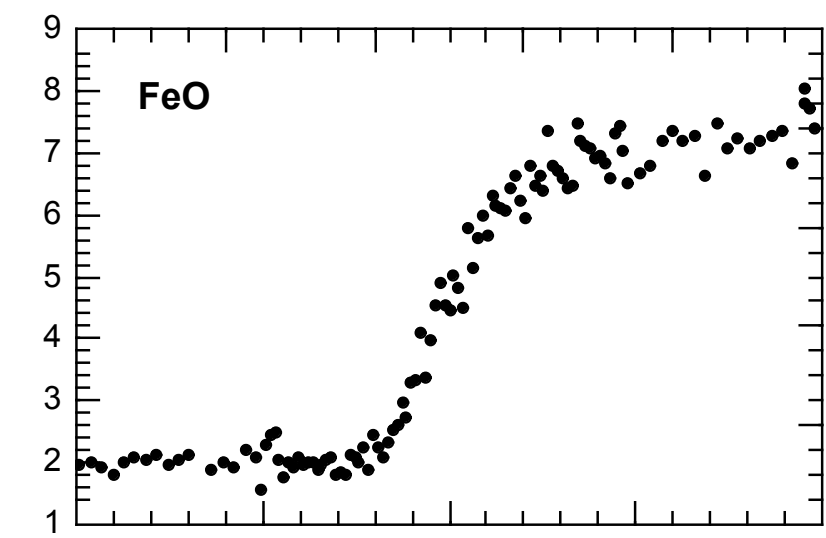
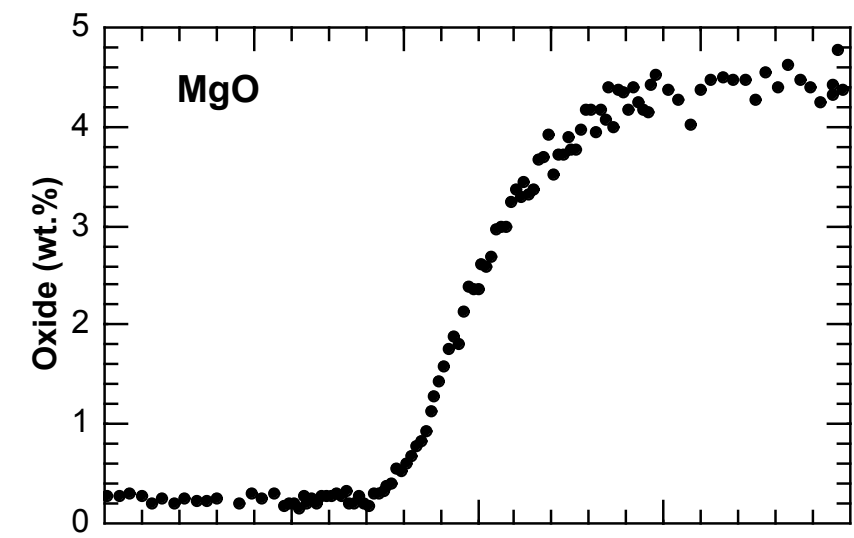
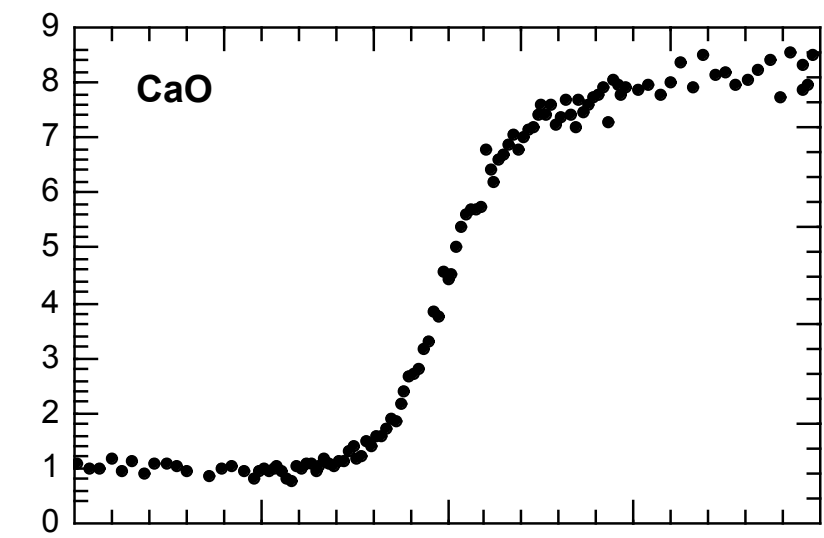
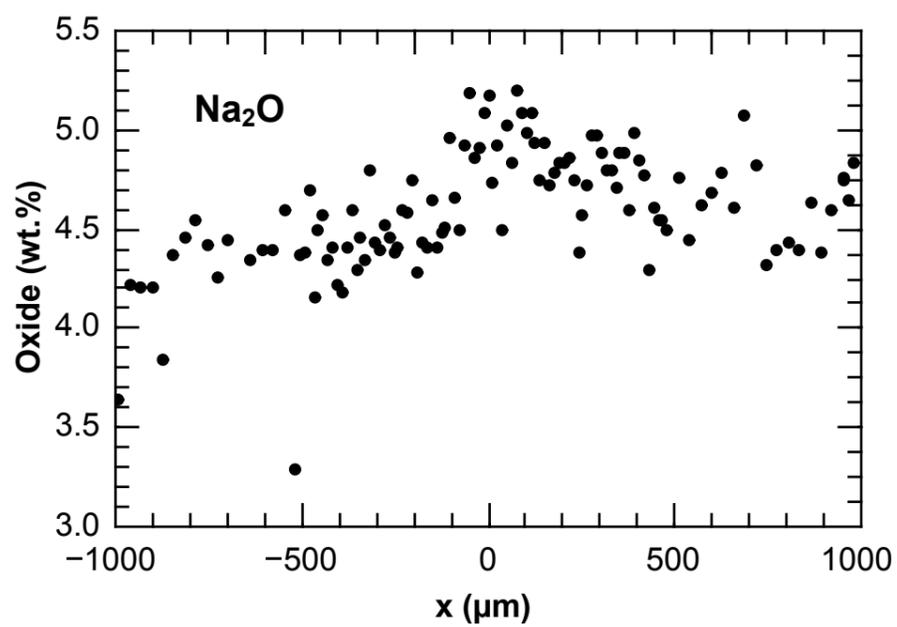
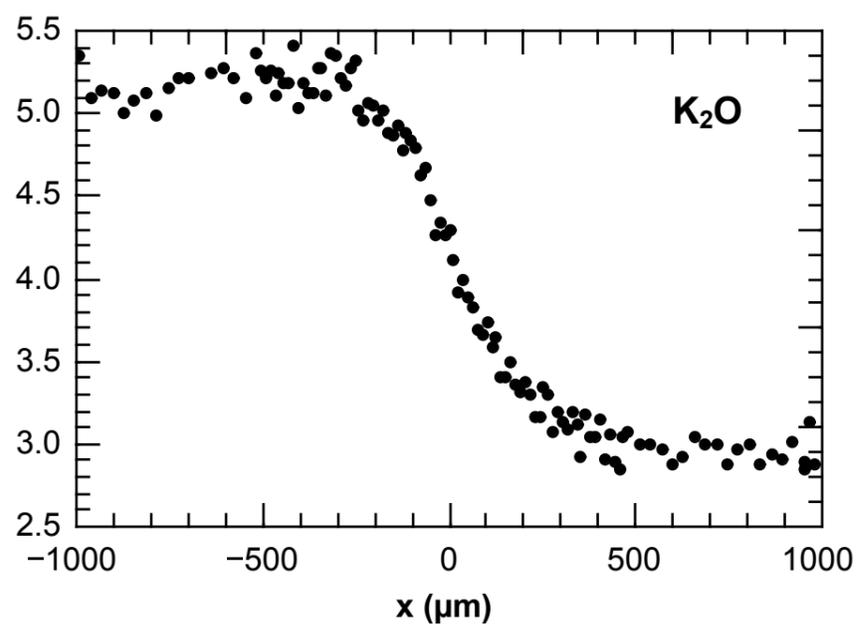

# Experiment P300-H2-0 (zero-time)

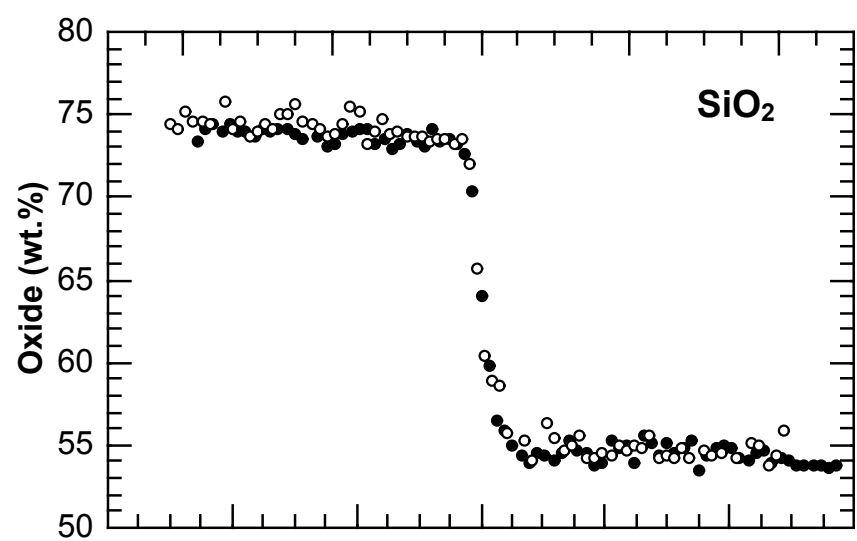
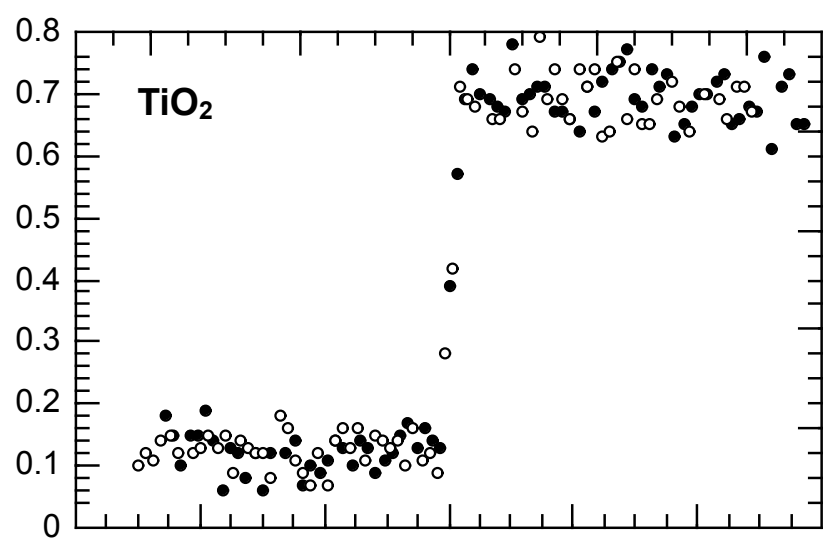
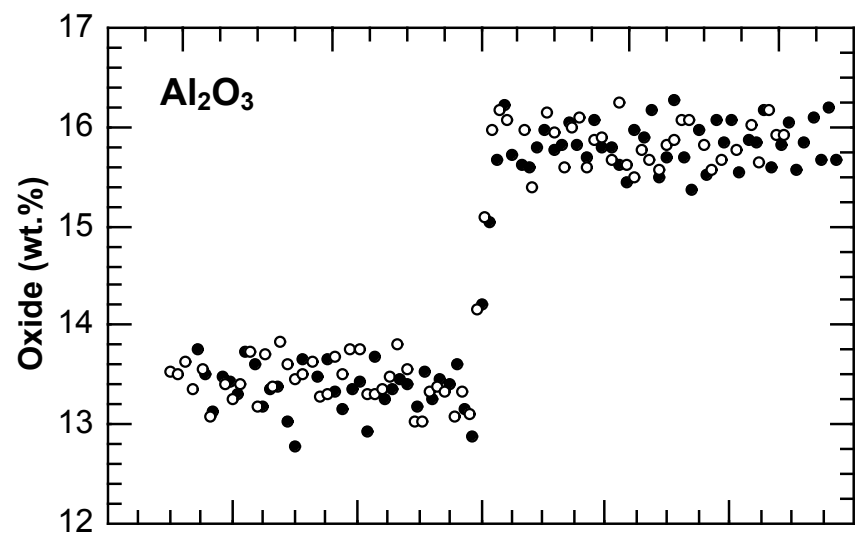
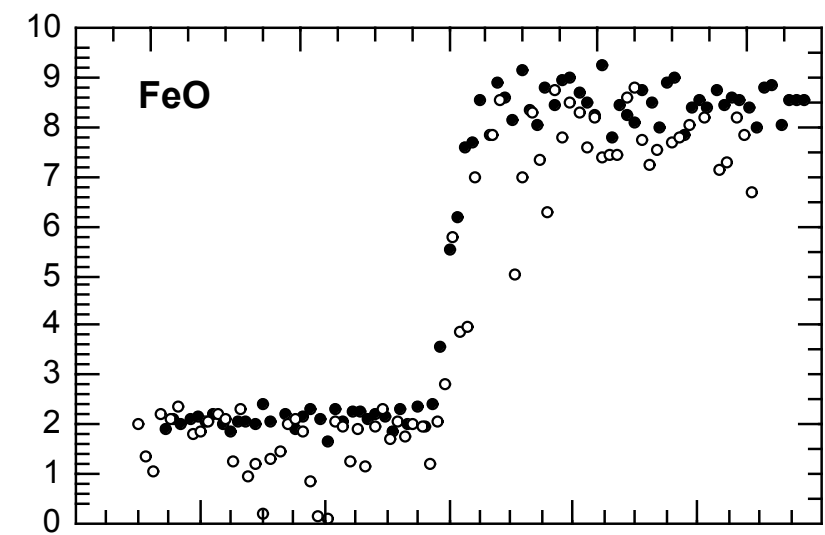
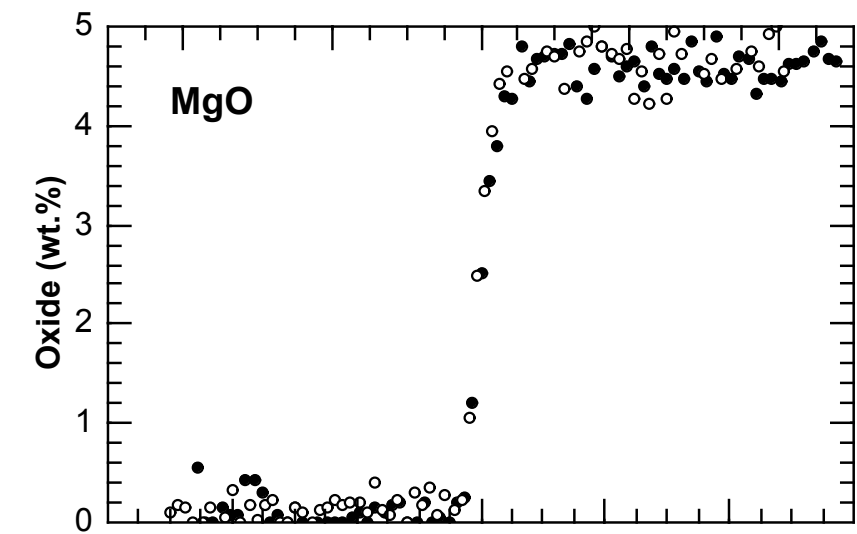
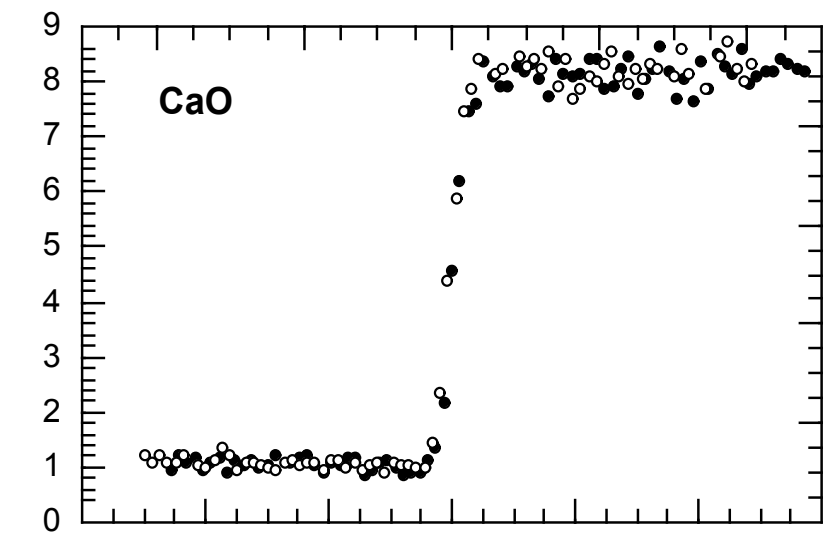
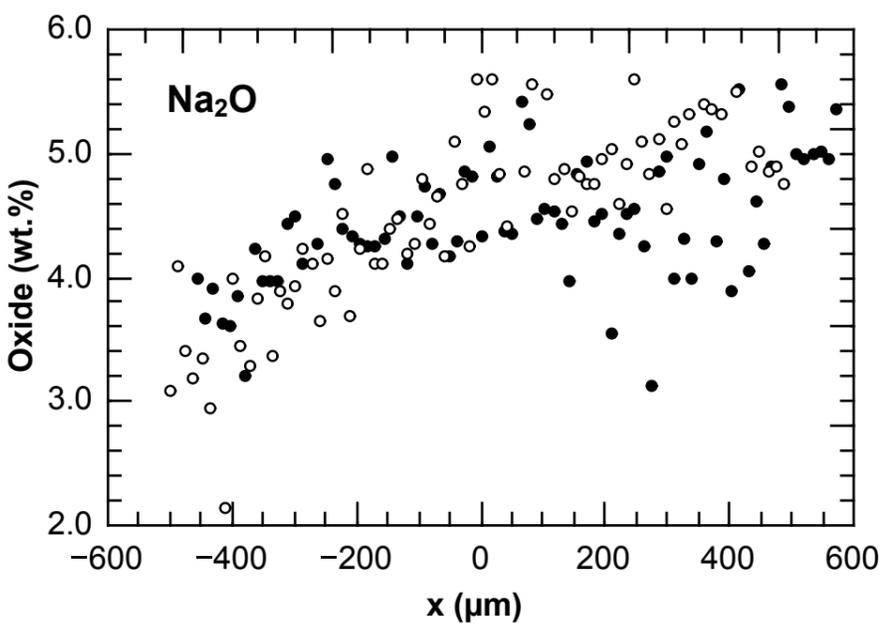
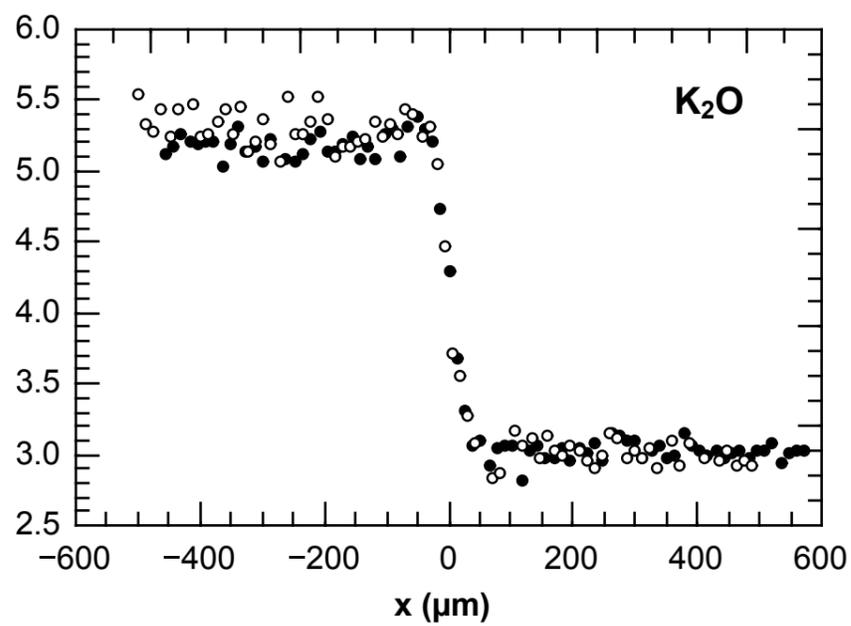

**Experiment P300-H2-1**

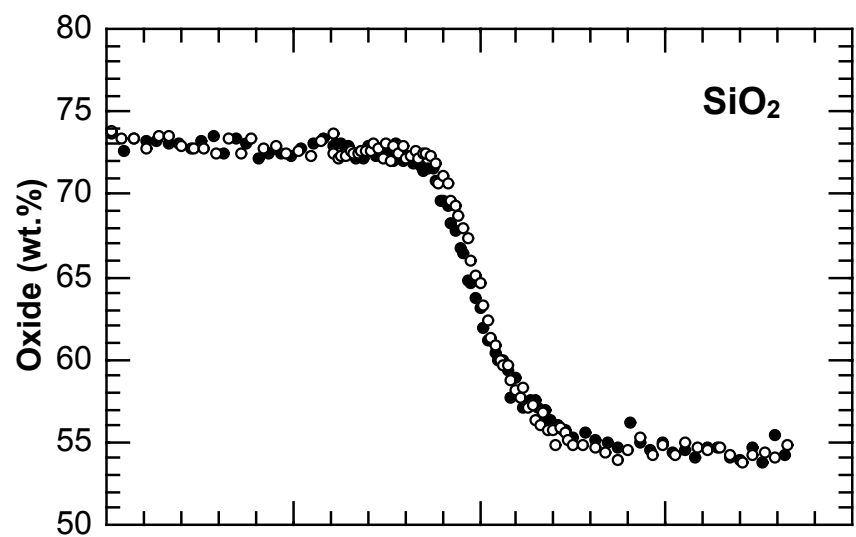
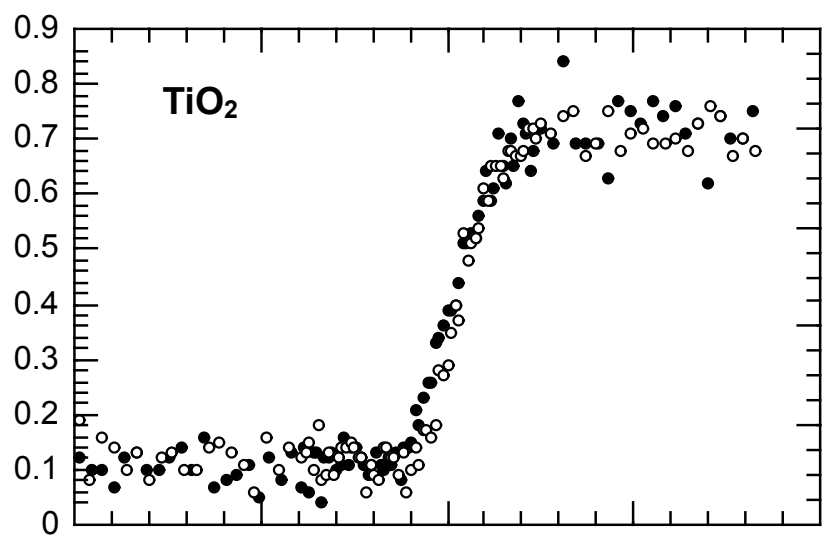
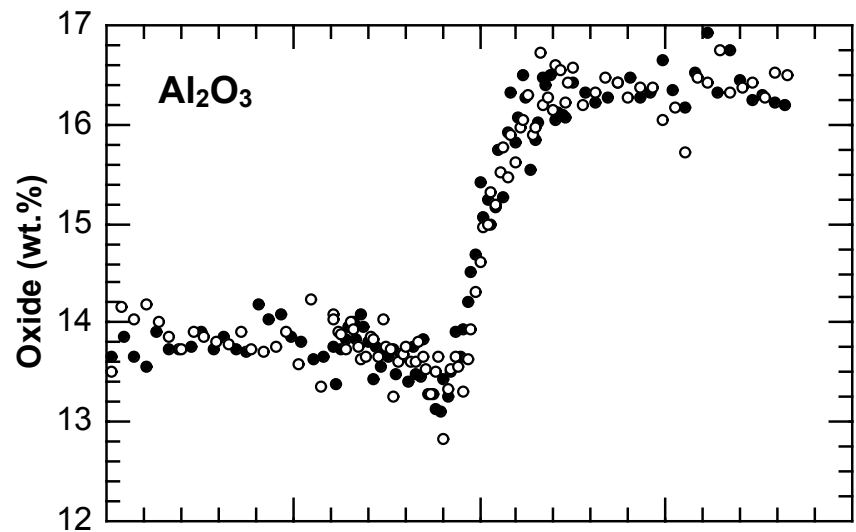
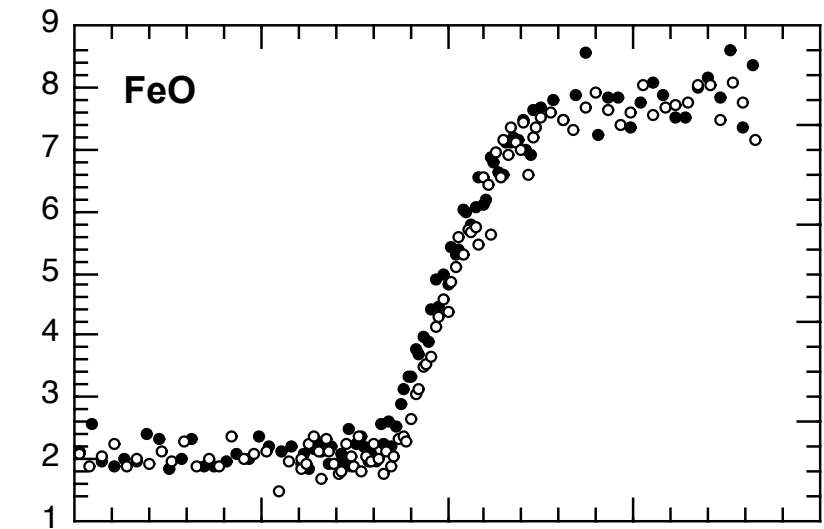
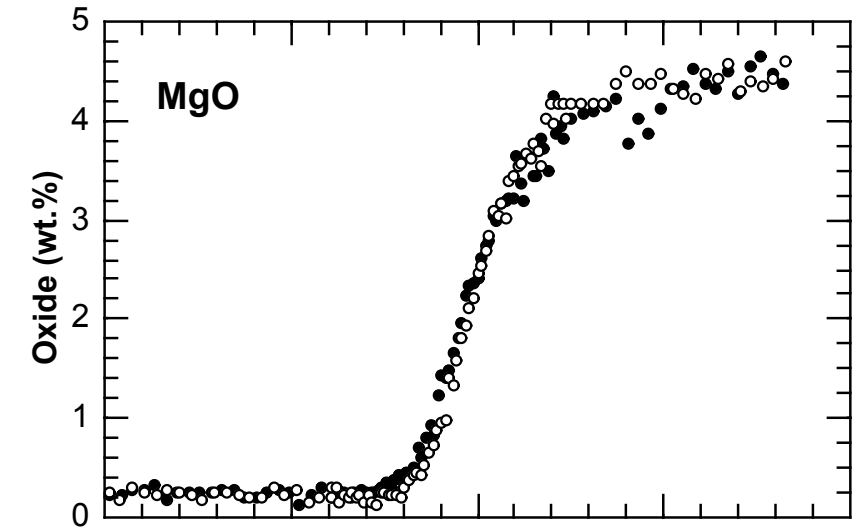
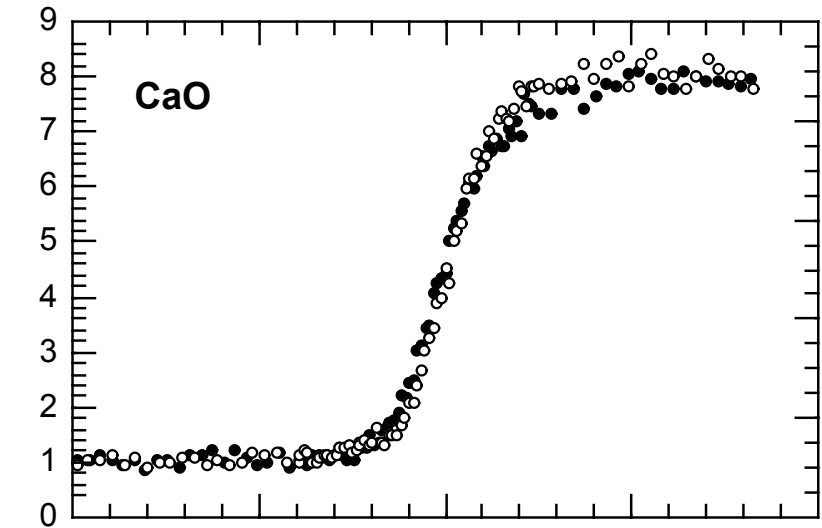
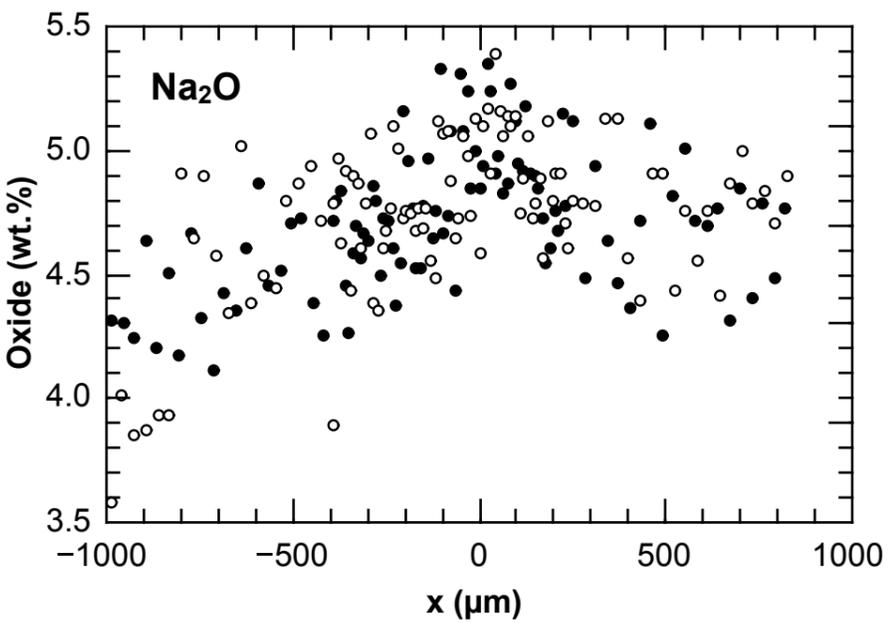
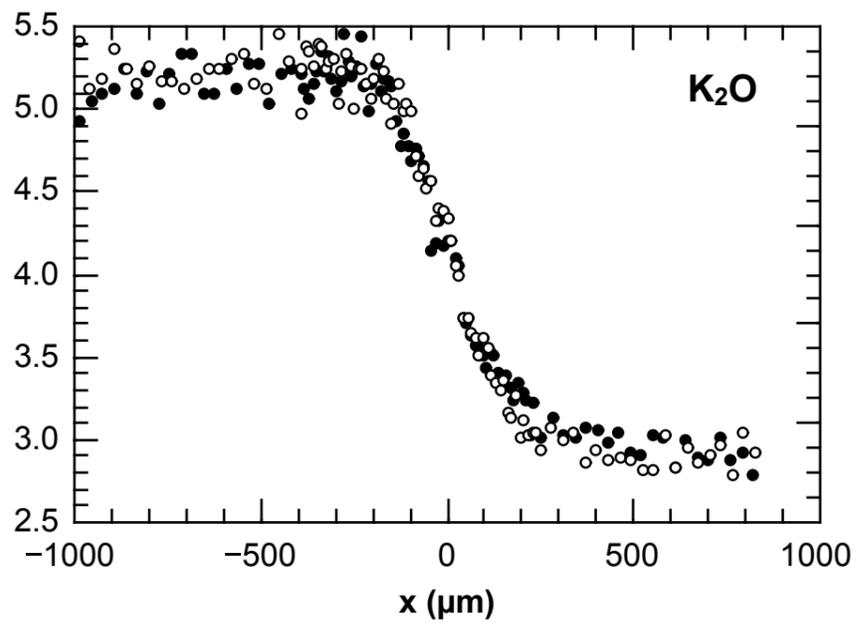

**Experiment P300-H2-4**

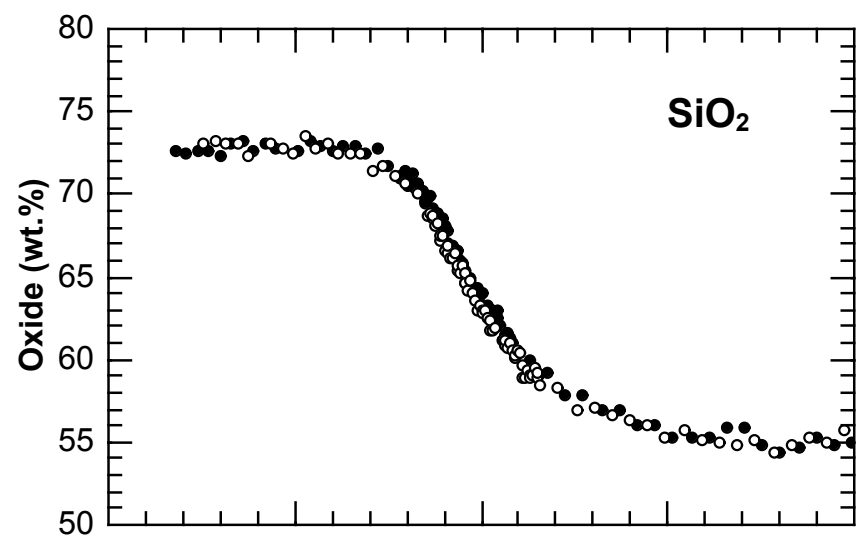
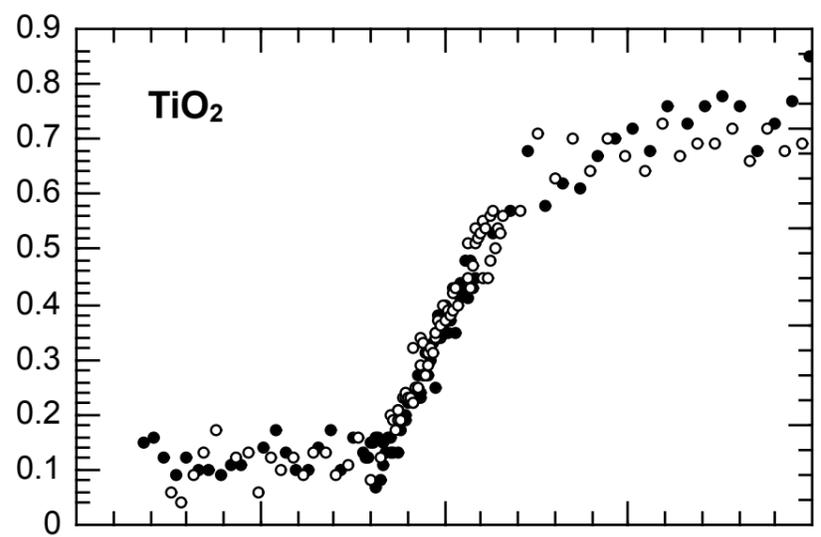
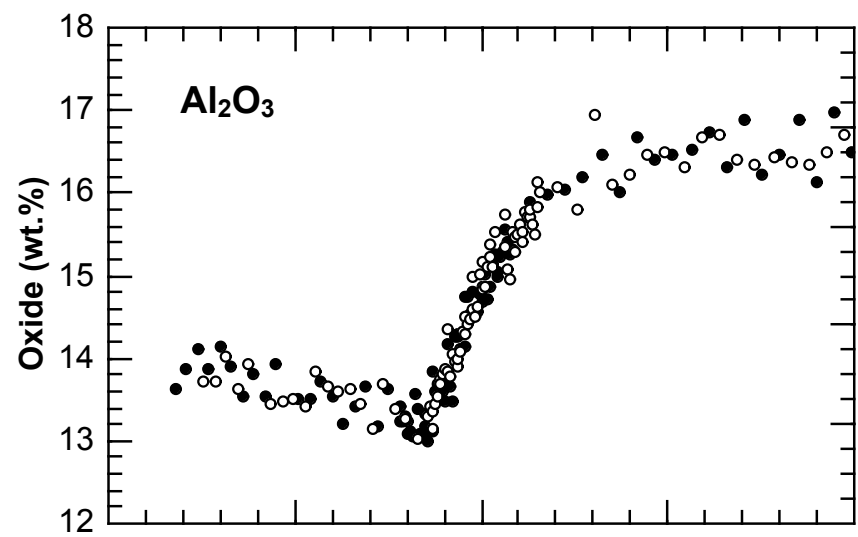
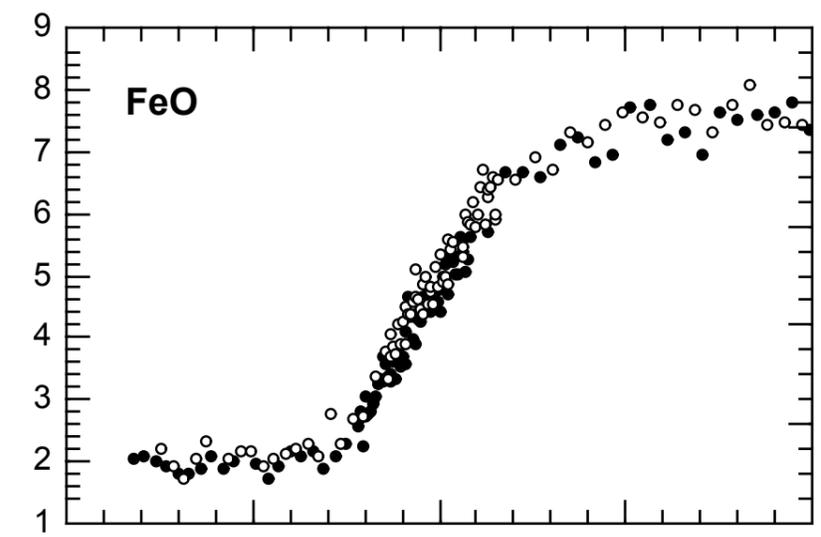
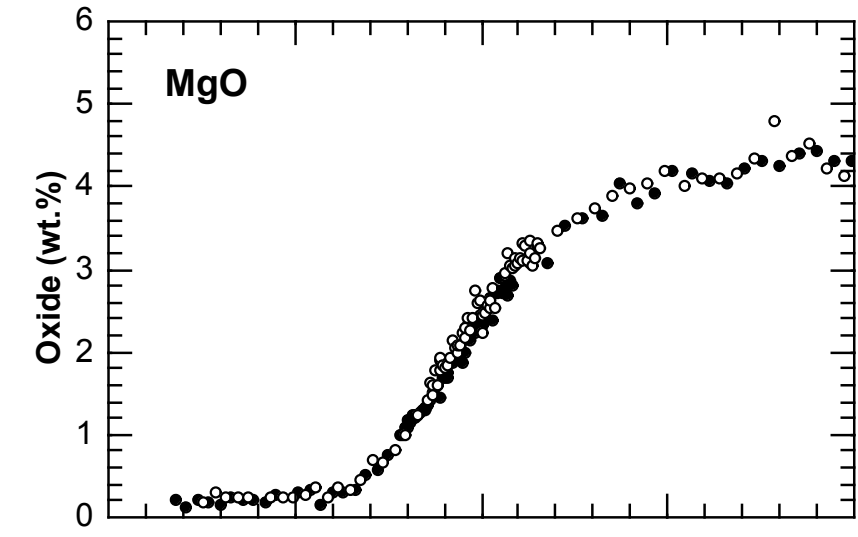
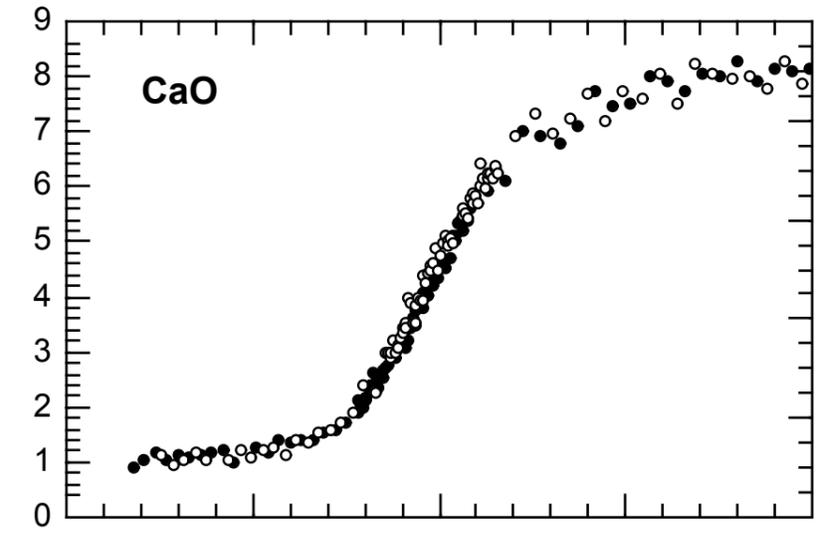
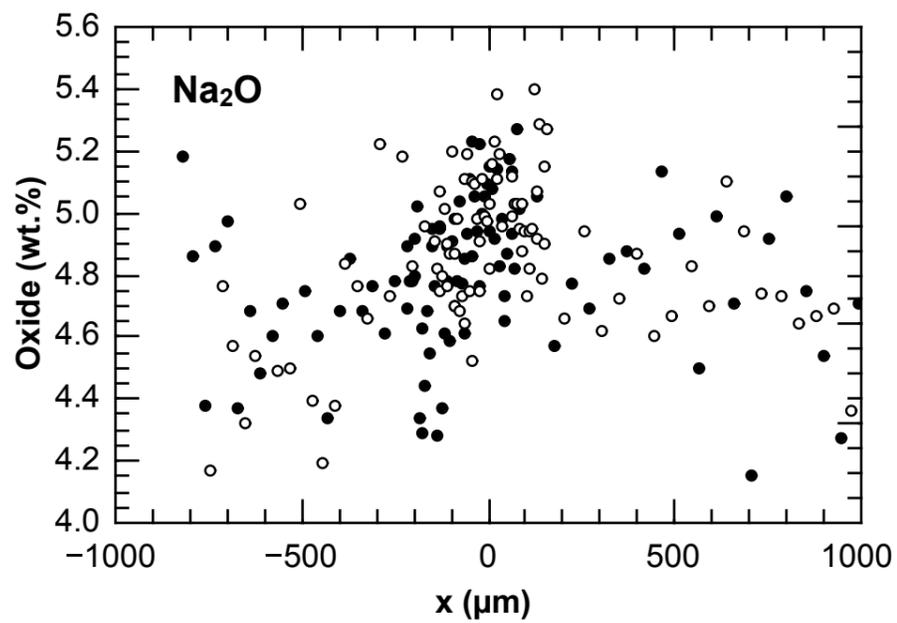
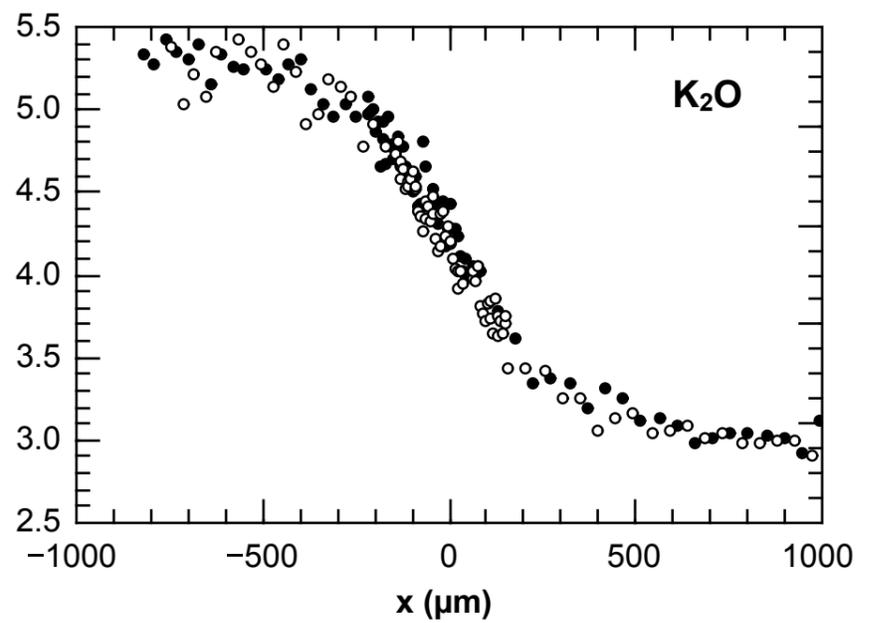

**Experiment P500-H2-4**

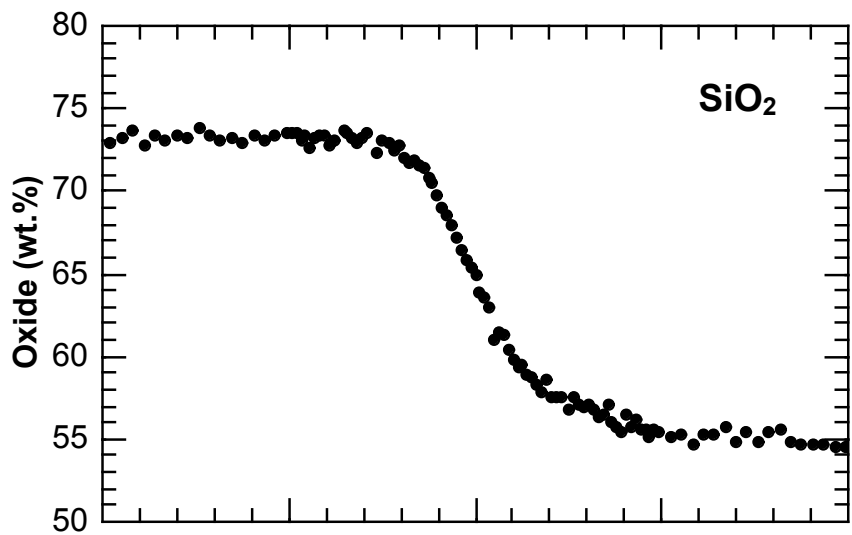
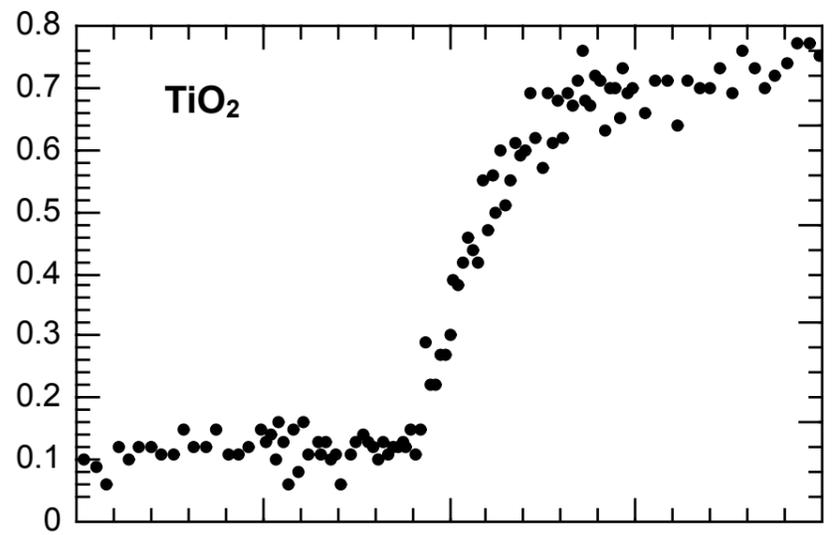
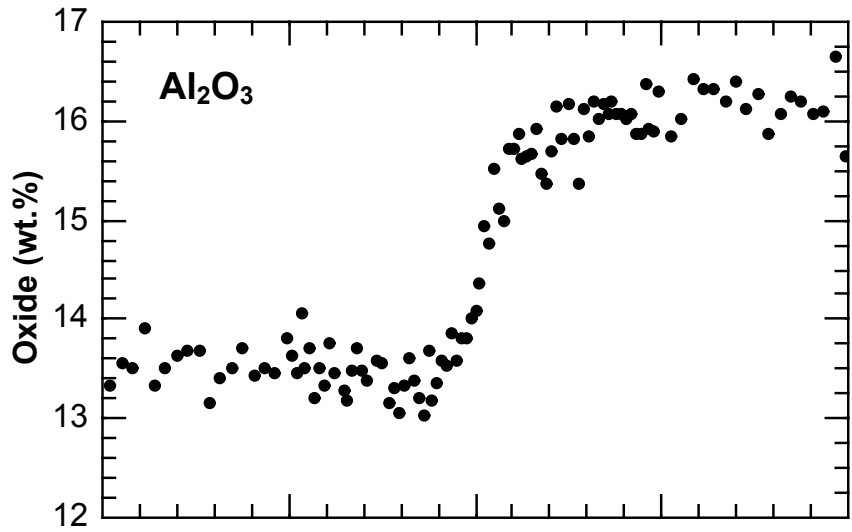
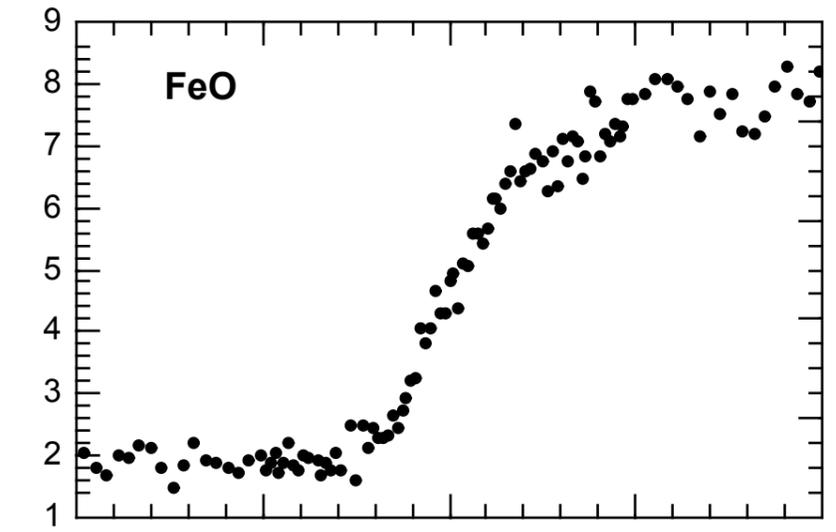
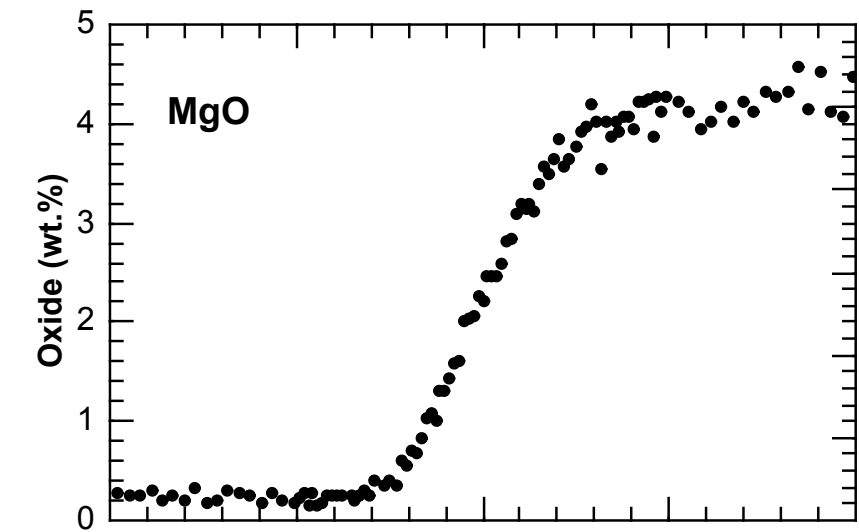
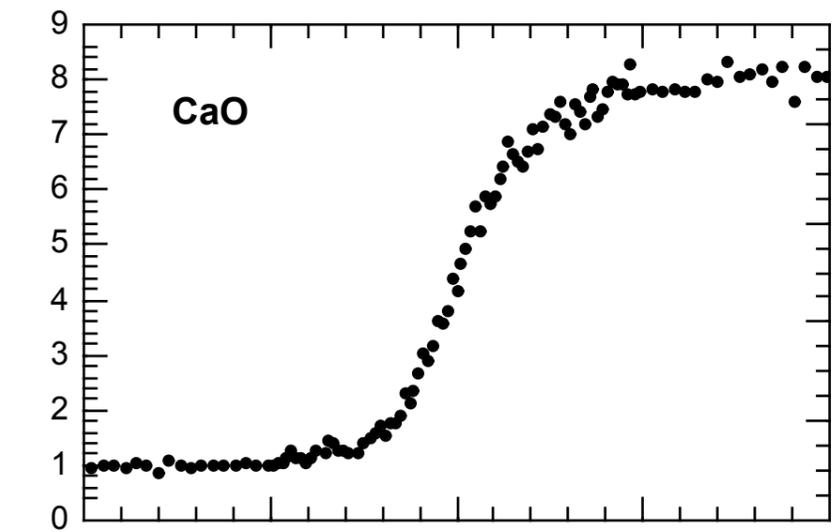
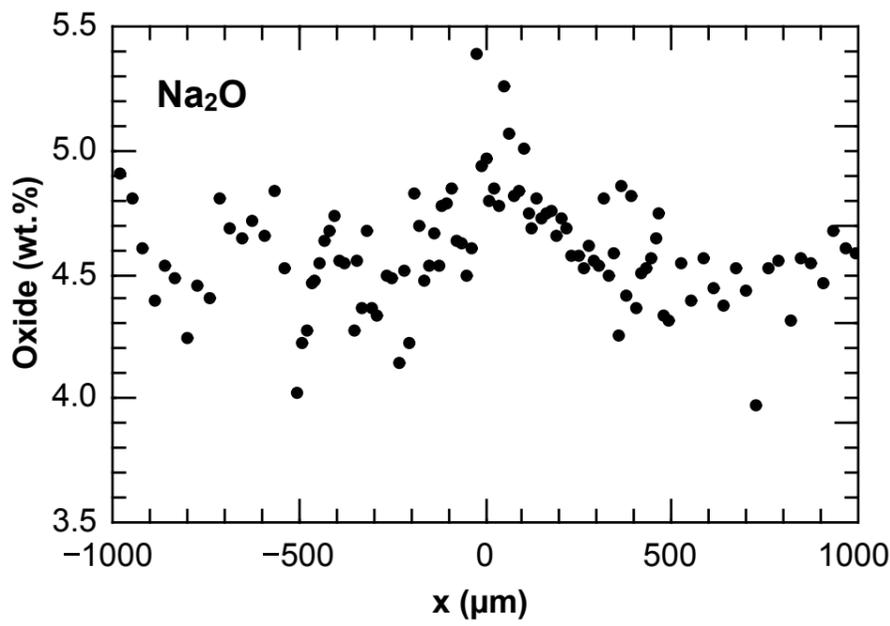
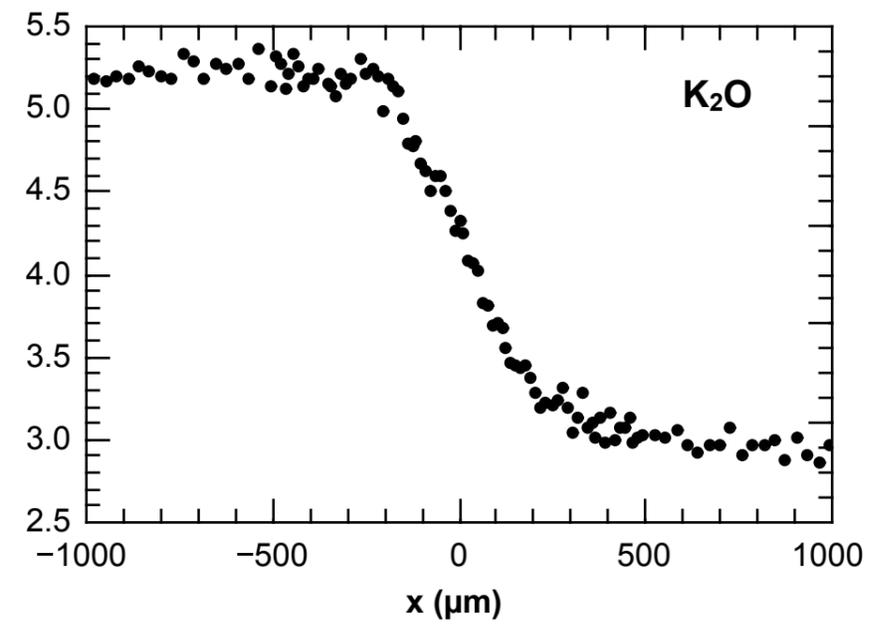

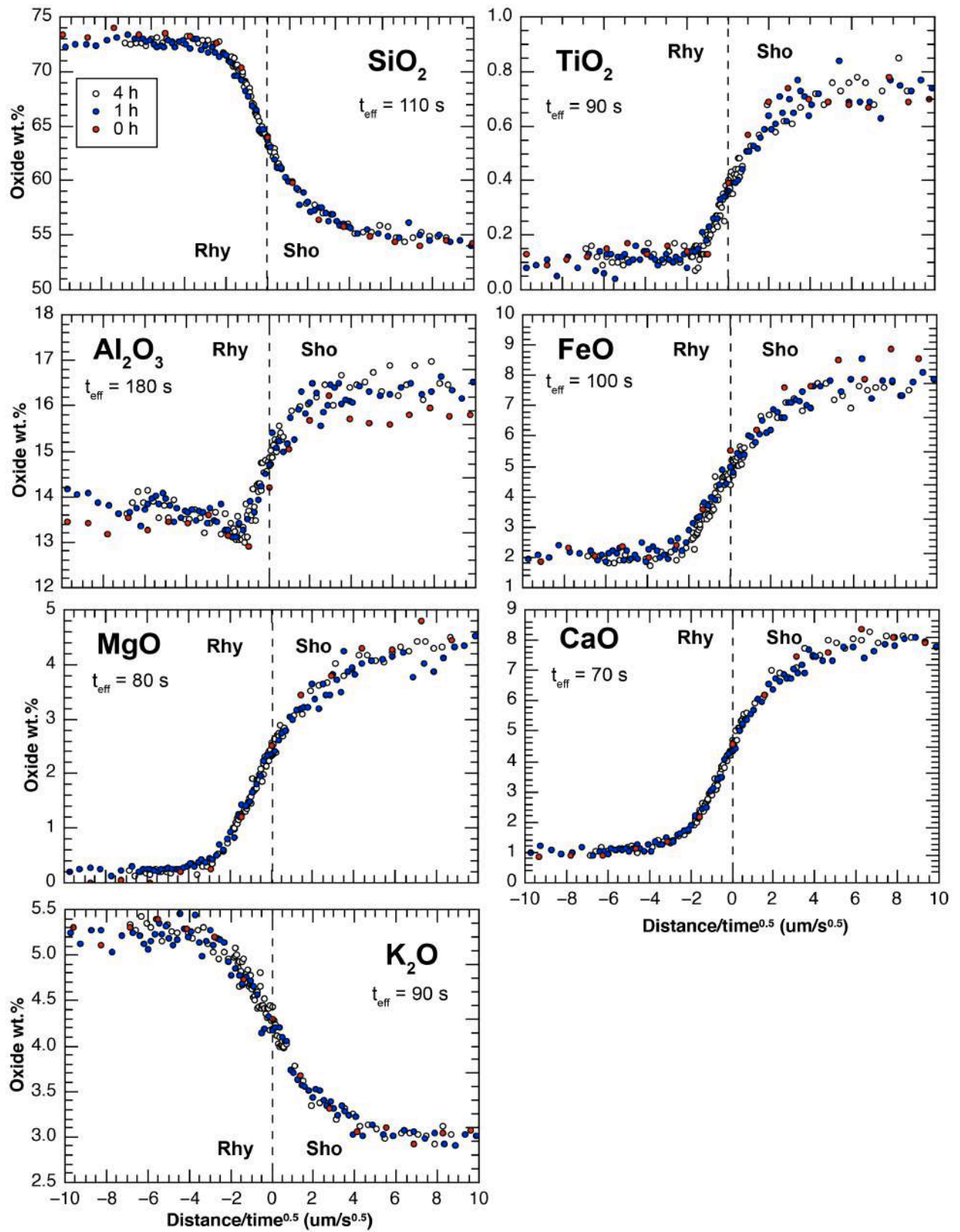

**Fig. S3.** Concentration-distance profiles of the 300 MPa, 2 wt.% $H_2O$ experiments, in which distance has been normalized by the square root of time. The effective durations used for the zero-time experiment are specified in each panel ($t_{eff}$). Effective durations are calculated by comparing the diffusive distances of the zero-time experiment with that of the long duration experiments (e.g. Zhang and Behrens, 2000).

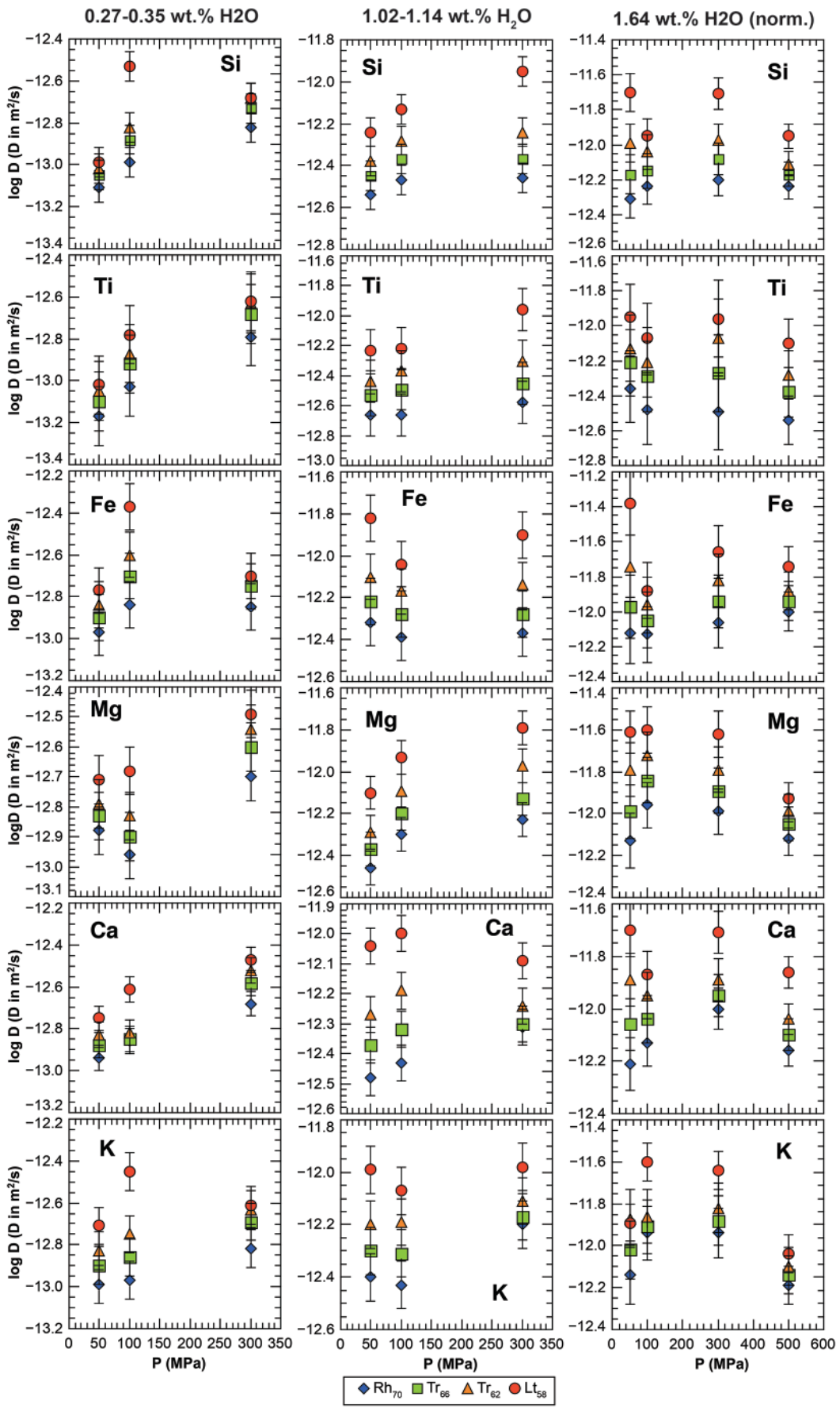

**Fig. S4**. Measured diffusivities (in $m^2/s$) of the six measured elements plotted against experimental pressure (MPa). The range of water content is indicated on the top of each row. The 2 wt.% nominal $H_2O$ series is strongly affected by differences in water content, and hence it has been normalized to 1.64 wt.% $H_2O$ (the water content of the 500 MPa experiment), in order to show possible water-independent diffusivity variations. Values were obtained using the equations given in **Table S2**.

Table S2: Parameters of linear equations relating water content and diffusivity for all six components (Si, Ti, Fe, Mg, Ca, K) at the four intermediate compositions (Lt$_{58}$ to Rh$_{70}$). Equations are in the form $log\ D = a*w + b$, where a and b are equation parameters, D is the diffusion coefficient in m$^2$/s, and w is the water content in wt.%. Errors (one standard deviation) of both parameters are also given in italics. Linear fits are plotted in Fig. 10.

| | | 50 Mpa | | | | 100 Mpa | | | | 300 Mpa | | | |
|---|---|---|---|---|---|---|---|---|---|---|---|---|---|
| | Comp. | a | err_a | b | err_b | a | err_a | b | err_b | a | err_a | b | err_b |
| Si | Rh$_{70}$ | 0.57 | 0.07 | -13.24 | 0.08 | 0.52 | 0.06 | -13.09 | 0.08 | 0.48 | 0.05 | -12.98 | 0.08 |
| Si | Tr$_{66}$ | 0.63 | 0.07 | -13.20 | 0.08 | 0.51 | 0.06 | -12.98 | 0.08 | 0.51 | 0.05 | -12.91 | 0.08 |
| Si | Tr$_{62}$ | 0.75 | 0.07 | -13.22 | 0.08 | 0.54 | 0.06 | -12.93 | 0.08 | 0.54 | 0.05 | -12.85 | 0.08 |
| Si | Lt$_{58}$ | 0.95 | 0.07 | -13.26 | 0.08 | 0.40 | 0.06 | -12.61 | 0.08 | 0.70 | 0.05 | -12.85 | 0.08 |
| Ti | Rh$_{70}$ | 0.48 | 0.15 | -13.27 | 0.17 | 0.39 | 0.13 | -13.12 | 0.16 | 0.35 | 0.11 | -12.93 | 0.16 |
| Ti | Tr$_{66}$ | 0.59 | 0.15 | -13.24 | 0.17 | 0.44 | 0.13 | -13.01 | 0.16 | 0.37 | 0.11 | -12.83 | 0.16 |
| Ti | Tr$_{62}$ | 0.71 | 0.15 | -13.24 | 0.17 | 0.45 | 0.13 | -12.94 | 0.16 | 0.37 | 0.11 | -12.74 | 0.16 |
| Ti | Lt$_{58}$ | 0.74 | 0.15 | -13.18 | 0.17 | 0.47 | 0.13 | -12.84 | 0.16 | 0.43 | 0.11 | -12.66 | 0.16 |
| Fe | Rh$_{70}$ | 0.59 | 0.12 | -13.09 | 0.13 | 0.50 | 0.10 | -12.96 | 0.12 | 0.60 | 0.08 | -13.05 | 0.12 |
| Fe | Tr$_{66}$ | 0.66 | 0.12 | -13.04 | 0.13 | 0.47 | 0.10 | -12.81 | 0.12 | 0.62 | 0.08 | -12.96 | 0.12 |
| Fe | Tr$_{62}$ | 0.78 | 0.12 | -13.03 | 0.13 | 0.45 | 0.10 | -12.70 | 0.12 | 0.67 | 0.08 | -12.91 | 0.12 |
| Fe | Lt$_{58}$ | 0.99 | 0.12 | -13.01 | 0.13 | 0.35 | 0.10 | -12.44 | 0.12 | 0.74 | 0.08 | -12.87 | 0.12 |
| Mg | Rh$_{70}$ | 0.55 | 0.08 | -13.04 | 0.10 | 0.70 | 0.07 | -13.12 | 0.09 | 0.53 | 0.06 | -12.86 | 0.09 |
| Mg | Tr$_{66}$ | 0.63 | 0.08 | -13.02 | 0.10 | 0.75 | 0.07 | -13.07 | 0.09 | 0.53 | 0.06 | -12.76 | 0.09 |
| Mg | Tr$_{62}$ | 0.75 | 0.08 | -13.03 | 0.10 | 0.78 | 0.07 | -13.00 | 0.09 | 0.54 | 0.06 | -12.67 | 0.09 |
| Mg | Lt$_{58}$ | 0.81 | 0.06 | -12.95 | 0.07 | 0.75 | 0.07 | -12.83 | 0.09 | 0.61 | 0.06 | -12.62 | 0.09 |
| Ca | Rh$_{70}$ | 0.53 | 0.06 | -13.08 | 0.07 | 0.54 | 0.05 | -13.01 | 0.07 | 0.53 | 0.05 | -12.87 | 0.07 |
| Ca | Tr$_{66}$ | 0.59 | 0.06 | -13.04 | 0.07 | 0.58 | 0.05 | -12.98 | 0.07 | 0.51 | 0.05 | -12.79 | 0.07 |
| Ca | Tr$_{62}$ | 0.69 | 0.06 | -13.02 | 0.07 | 0.59 | 0.05 | -12.93 | 0.07 | 0.52 | 0.05 | -12.73 | 0.07 |
| Ca | Lt$_{58}$ | 0.75 | 0.06 | -12.93 | 0.07 | 0.48 | 0.05 | -12.66 | 0.07 | 0.60 | 0.05 | -12.70 | 0.07 |
| K | Rh$_{70}$ | 0.60 | 0.10 | -13.13 | 0.11 | 0.77 | 0.08 | -13.21 | 0.10 | 0.65 | 0.07 | -13.00 | 0.10 |
| K | Tr$_{66}$ | 0.63 | 0.10 | -13.05 | 0.11 | 0.69 | 0.08 | -13.05 | 0.10 | 0.61 | 0.07 | -12.88 | 0.10 |
| K | Tr$_{62}$ | 0.69 | 0.10 | -13.00 | 0.11 | 0.64 | 0.08 | -12.91 | 0.10 | 0.61 | 0.07 | -12.82 | 0.10 |
| K | Lt$_{58}$ | 0.65 | 0.10 | -12.84 | 0.11 | 0.54 | 0.08 | -12.61 | 0.10 | 0.64 | 0.07 | -12.78 | 0.10 |